\documentclass{JHEP3}

\usepackage{amssymb}
\usepackage{amsfonts}
\usepackage{amsbsy}
\usepackage{amsmath}
\usepackage{amsthm}
\usepackage{graphicx}
\usepackage[vcentermath]{youngtab}
\usepackage{multirow}
\usepackage{latexsym}
\usepackage{array}
\usepackage{epsfig}
\usepackage{arydshln}

\newcommand{\cT}{\mathcal{T}}
\newcommand{\cV}{\mathcal{V}}

\newcommand{\cC}{\mathcal{C}}
\newcommand{\cD}{\mathcal{D}}

\newcommand{\cK}{\mathcal{K}}
\newcommand{\cN}{\mathcal{N}}

\newcommand{\cB}{\mathcal{B}}
\newcommand{\cF}{\mathcal{F}}
\newcommand{\cI}{\mathcal{I}}

\newcommand{\nn}{\nonumber}
\newcommand{\beq}{\begin{equation}}
\newcommand{\eeq}{\end{equation}}
\newcommand{\be}{\begin{equation}}
\newcommand{\ee}{\end{equation}}
\newcommand{\bea}{\begin{eqnarray}}
\newcommand{\eea}{\end{eqnarray}}   
\newcommand{\ben}{\begin{eqnarray*}}
\newcommand{\een}{\end{eqnarray*}}                  
\newcommand{\ba}{\begin{aligned}}
\newcommand{\ea}{\end{aligned}}
\newcommand{\bt}{\begin{tabular}}
\newcommand{\et}{\end{tabular}}
\newcommand{\bc}{\begin{center}}
\newcommand{\ec}{\end{center}}

\title{ F-theory fluxes, Chirality and Chern-Simons theories 
}

\author{Thomas W.~Grimm$^{1}$ and Hirotaka Hayashi$^2$\\
$^1$Max Planck Institute for Physics\\
          F\"ohringer Ring 6  \\
          Munich, 80805, Germany \\

$^2$School of Physics\\ Korea Institute for Advanced Study\\ 
Seoul 130-722, Korea\\
\\
\\
{\tt grimm} {\rm at} {\tt mppmu.mpg.de},
{\tt hayashi} {\rm at} {\tt kias.re.kr}
}

\preprint{MPP-2011-118\\
          KIAS-P11065}

\abstract{We study the charged chiral 
matter spectrum of four-dimensional F-theory compactifications on 
elliptically fibered Calabi-Yau fourfolds by using the dual M-theory description.
A chiral spectrum can be induced by M-theory 
four-form flux on the fully resolved Calabi-Yau fourfold. In M-theory 
this flux yields three-dimensional Chern-Simons couplings in the Coulomb 
branch of the gauge theory. In the F-theory compactification on an 
additional circle these couplings are only generated by one-loop corrections with charged fermions 
running in the loop. This identification allows us to 
infer the net number of chiral matter fields of the four-dimensional effective theory.   
The chirality formulas can be evaluated by using the intersection numbers
and the cones of effective curves of the resolved fourfolds. We argue 
that a study of the effective curves also allows to follow the resolution 
process at each co-dimension. To write simple chirality formulas 
we suggest to use the effective curves involved in the resolution process 
to determine the matter surfaces and to connect with the 
group theory at co-dimension two in the base. 
We exemplify our methods on examples with $SU(5)$ and $SU(5)\times U(1)$ 
gauge group. 
}

\begin{document}

\newpage

\section{Introduction}

In the last years there has been vast progress in understanding 
four--dimensional F-theory compactifications on elliptically fibered 
Calabi-Yau fourfolds. These 
compactifications can admit non-Abelian gauge groups which arise from stacks of 
7-branes. Geometrically this corresponds to a degeneration of the elliptic 
fiber over divisors in the base wrapped by the seven-branes \cite{Denef:2008wq,Weigand:2010wm}. 
At the intersection of two such divisors matter 
fields are localized. The presence of a 7-brane flux can lift
part of the matter spectrum such that a net number of chiral
fields remain in the $\cN=1$ low energy effective supergravity theory.
The aim of this paper is to determine formulas for 
the net number of such fields depended on the global geometric data 
of resolved Calabi-Yau fourfolds and the specification of four-form fluxes.
To justify the proposed chirality formulas we will exploit the duality 
between M-theory and F-theory, including one-loop corrections to 
the Chern-Simons terms in the effective theories obtained from M-theory.

A first approach in the derivation of a chirality formula for the charged 
matter fields along the intersection
of two 7-branes was to use the local data of the geometry and gauge 
bundle \cite{Donagi:2008ca, Beasley:2008dc, Hayashi:2008ba}. 
The required data included the classes of the matter curves and the 
local two-form flux components on the 7-branes.
For these consideration the input mainly came from two directions. 
In \cite{Donagi:2008ca,Hayashi:2008ba} a
spectral cover construction, more familiar from the heterotic string \cite{Friedman:1997yq}, 
has been used. In contrast, the formulas in \cite{Beasley:2008dc}
intensively use the local 7-brane gauge theory and more closely resemble 
the Type IIB analogs known from a weak 
coupling picture with D-branes. Despite these successes it is in 
general hard to obtain a global picture 
and develop the tools to study more complicated 7-brane configurations. 
In addition to the construction of compact Calabi-Yau fourfolds a 
complete global treatment has also to capture the flux 
data, which requires a deep understanding of the local singular 
geometry near the 7-branes and its global embedding. An analysis of  
the global geometry is particularly crucial 
when additional $U(1)$ symmetries are present \cite{Hayashi:2010zp,Grimm:2010ez}. 
Extensions of the spectral cover techniques to 
global constructions have been proposed in \cite{Marsano:2009gv,Blumenhagen:2009yv,Grimm:2009yu,Marsano:2010ix,Marsano:2011hv}.
In particular, in \cite{Blumenhagen:2009yv,Grimm:2009yu,Marsano:2011hv} consistency checks have been given 
to argue for the applicability of the spectral cover techniques to specific 
compact settings. 
In the present work we will take a different route, 
since we will specify the flux data directly on the 
resolved Calabi-Yau fourfolds with no reference to a 
spectral cover construction. 
 
Let us summarize our general strategy to derive chirality formulas in F-theory. 
Firstly, note that to study F-theory compactifications one inevitably 
has to address the fact that 
non-Abelian gauge groups arise from singular Calabi-Yau geometries $X_4$ for which 
the geometrical data determining the spectrum and couplings 
cannot be determined directly. However, a natural way to deal with the 
singularities is to perform a resolution of the singularities and work with 
the fully resolved Calabi-Yau fourfold $\tilde X_4$. Such resolutions can be performed using toric 
methods as in \cite{Candelas:1996su,Candelas:1997eh,Blumenhagen:2009yv,
Grimm:2009yu,Cvetic:2010rq,Chen:2010ts,Krause:2011xj,Braun:2011ux}, 
or stepwise as recently shown in \cite{Esole:2011sm,Marsano:2011hv}. 
The physics induced by the resolution process can be  
addressed in the M-theory description of F-theory \cite{Denef:2008wq,Weigand:2010wm}. 
M-theory compactification on an resolved elliptically fibered 
Calabi-Yau fourfold yields a specific three-dimensional 
effective theory with a couplings which can arise from a four-dimensional 
$\cN=1$ supergravity compactified on a circle \cite{Grimm:2010ks}. 
For the resolved Calabi-Yau fourfolds the theory will be in the Coulomb 
branch of the three-dimensional gauge theory.  

In the M-theory picture of F-theory 
the 7-brane fluxes correspond to four-form fluxes $G_4$ of the field-strength
of M-theory three-form potential. Not all $G_4$ fluxes will lift to a four-dimensional 
F-theory compactification. Crucially one has to impose that the allowed 
fluxes preserve four-dimensional Poincar\'e invariance in the F-theory limit \cite{Dasgupta:1999ss}. 
Further restriction are imposed by demanding the existence of an unbroken four-dimensional 
gauge theory. It was argued in \cite{Marsano:2011hv} that there are $G_4$ fluxes 
that satisfy these conditions, and reproduce the four-dimensional chirality formulas 
known from the spectral cover construction of an $SU(5)$ model.
The allowed $G_4$ fluxes crucially involve the wedges of two-forms Poincar\'e-dual to the 
exceptional resolution divisors. 
In this work we will give a physical interpretation of this fact.

To link the $G_4$ flux with the four-dimensional 
chirality it is crucial to point out that 
the M-theory reduction on a Calabi-Yau fourfold with $G_4$ flux 
induces terms in the three-dimensional gauge theory which are 
not obtained by a classical circle reduction of a general 
four-dimensional gauge theory. This can be inferred from the 
explicit reduction of \cite{Haack:2001jz,Grimm:2010ks}. In fact, the M-theory reduction 
will induce Chern-Simons terms for the $U(1)$ gauge-fields in the Coulomb 
branch. We note that in the reduction of the four-dimensional 
theory such terms must arise from one-loop corrections with 
charged fermions running in the loop. This links the charged 
matter spectrum of the four-dimensional theory with the 
$G_4$-flux corrections of the M-theory reduction.  
We will show that this provides us with an interpretation 
how the chiral matter spectrum can be determined from the flux data.

If the flux indeed encodes the net number of chiral fermions on the 
intersection curves of two 7-branes, the chiral index has to be of the form 
$\chi({\bf R}) = \int_{S_{\bf R}} G_4$, as already anticipated in \cite{Donagi:2008ca,Hayashi:2008ba}, and 
studied recently in \cite{Braun:2011zm,Marsano:2011hv,Krause:2011xj}. Here 
${\bf R}$ is the representation of the four-dimensional gauge group in which the 
fermions transform. The intersection curve of the two 7-brane in the base of $\tilde X_4$ 
will be called matter curve $\Sigma_{\bf R}$ if matter fields in the representation 
$\bf R$ are located along this curve. 
The difficulty in evaluating the expression for $\chi({\bf R})$ is 
to give a global and universal definition of the surface $S_{\bf R}$. 
Using heterotic/F-theory duality one expects that $S_{\bf R}$ 
is obtained by fibering the resolution 
$\mathbb{P}^1$'s over the matter curve. In fact, in the M-theory
picture the charged matter fields arise from M2-branes wrapping the 
$\mathbb{P}^1$-fibers of the resolved geometry. The group theory matches this 
geometric interpretation since the resolution $\mathbb{P}^1$'s over the 
matter curves can be associated 
to the weights of the representation~${\bf R}$ \cite{Intriligator:1997pq,Katz:1997eq}.
These states are massive on the resolved space $\tilde X_4$ and become massless 
in the singular F-theory limit.

To construct the matter surfaces for a given resolved Calabi-Yau fourfold we propose 
to exploit the data encoded by the cone of effective 
curves, i.e.~the Mori cone. It will be crucial to select a 
subcone of the full Mori cone, the 
relative Mori cone, consisting of curves in $\tilde X_4$ which 
shrink when going to the singular space $X_4$. This cone will be 
completed into the extended relative Mori cone by including 
other effective curves in the elliptic fiber, which intersect the 
exceptional resolution divisors. In simple cases this simply amounts to 
including the pinched elliptic fiber over the 7-brane. 
We will argue 
that the intersection of these curves with the exceptional 
divisors allows us to identify a pairing between generators 
of the extended relative Mori cone and weights of the four-dimensional 
gauge group. The exceptional divisors correspond to the simple 
roots of the gauge group. The identification of roots and weights 
with the geometric data has been proposed for local Calabi-Yau 
threefolds in \cite{Intriligator:1997pq,Katz:1997eq,Marsano:2011hv}. 
Note that a detailed analysis of 
which weights correspond to the effective curves in the extended relative Mori cone
also allows us to stepwise reconstruct the resolution process along the co-dimension two and three singularity 
loci in the base of $X_4$.
In this process, we make two assumptions. One is that the representation which can appear along the co-dimension 
two singularity loci are the same as the one of the matter fields localized along the curve. 
The second is that the degeneration of weights at the co-dimension three singularity points 
obeys the algebra $G_p$ when the singularity is enhanced to a type $G_p$. These assumptions 
are exploited already in \cite{Katz:1996xe,Donagi:2008ca,Beasley:2008dc,Hayashi:2009ge} and have been 
studied for compact settings in \cite{Marsano:2011hv, Krause:2011xj}. 
With these assumptions and the extended relative Mori cone at hand, we can generally 
determine the resolution process along the singularity loci. 

In this work we also include that case where additional geometrically 
massless $U(1)$ gauge fields are in the four-dimensional spectrum of the 
F-theory compactification. The methods to determine the resolution structure 
using the extended Mori cone naturally generalize to this situations, and 
one is able to explicitly construct the matter surfaces $S_{\bf R}$ also if 
distinguishing $U(1)$-charges of the representation $\bf R$ are present. However, 
one can generalize the Ansatz for the $G_4$ flux if one permits a 
gauging of the four-dimensional $U(1)$-symmetries. Such extra fluxes render the 
$U(1)$ massive, but allow to keep its global selection rules. 
An explicit example how such extra $U(1)$'s can be consistently induced in a Calabi-Yau
fourfold compactification was given in \cite{Grimm:2010ez}, and termed $U(1)$-restricted Tate model.    
The construction of fluxes in this model have been recently given in \cite{Braun:2011zm,Krause:2011xj}. 
In reference \cite{Braun:2011zm} a direct link to the 
chirality formulas for D7-branes and O7-planes was established. 
For $SU(5)$ models and their extensions it was shown in \cite{Krause:2011xj} 
that the chirality formula can be evaluated using 
the ambient fivefold geometry in which the Calabi-Yau fourfold is embedded. 
These techniques also allowed to
reduce the $G_4$ fluxes, using the ambient fivefold, to a two-form flux on the base $\cB$,
and reproducing the correct group theoretical factors as required for a 
valid chirality formula. In our formalism this detour is not required, and 
the $U(1)$ case appears as natural part of a more general construction.

To illustrate the derivation of the net chiralities we will consider two explicit 
examples of hypersurfaces in toric ambient spaces. The gauge theory will be $SU(5)$
and $SU(5) \times U(1)$ and we perform an explicit resolution of all co-dimension singularities 
as in \cite{Blumenhagen:2009yv,Grimm:2009yu,Chen:2010ts,Grimm:2010ez,Krause:2011xj} by modifying the toric ambient space. 
We compute the net chiralities induced
by a general $G_4$ flux compatible with four-dimensional Poincar\'e invariance and the 
preservation of the $SU(5)$ gauge symmetry in both cases. Our results are compared to the 
spectral cover and split spectral cover constructions \cite{Tatar:2009jk,Marsano:2009gv,Blumenhagen:2009yv}, 
and we find match of the chirality formulas for matter being localized near the 
$SU(5)$-brane as expected.

\section{F-theory chirality and three-dimensional Chern-Simons theories}

In this section we give a derivation of the F-theory 
chirality formulas by using one-loop corrections 
in a dual three-dimensional Chern-Simons theory. 
More precisely, we will exploit the description 
of F-theory via M-theory to show that a four-dimensional 
chiral spectrum can be induced by 
a special class of $G_4$-form fluxes 
on a resolved Calabi-Yau fourfold $\tilde X_4$.
This will lead to a derivation of formulas of the form
\beq \label{eq:chirality2}
 \chi({\bf R}) = n_{\bf R} - n_{\bf R^*} = \int_{S_{\bf R}} G_4 \ ,
\eeq
where $S_{\bf R}$ is a four-cycle in $\tilde X_4$.
Here we have denoted by $\chi({\bf R})$ the chiral index of $n_{\bf R}$ matter 
fields in the representation ${\bf R}$ minus $n_{\bf R^*}$ matter fields 
in the representation ${\bf R^*}$. 

In order to interpret chirality formulas involving 
$G_4$ we first have to summarize the properties 
of a fully resolved Calabi-Yau fourfold $\tilde X_4$ in 
section \ref{resolving-4folds}. In the compactification of 
M-theory on $\tilde X_4$ one can allow for $G_4$ fluxes 
in the reduction. We describe the M-theory and F-theory
 constraints on these fluxes in section \ref{introducingG4}. 
It is argued in section \ref{3dCS} that a certain class 
of M-theory fluxes induces Chern-Simons couplings 
in the three-dimensional effective theory. 
The matching these M-theory couplings with one-loop corrections of 
an F-theory setup compactified on a circle leads to 
chirality formulas of the form \eqref{eq:chirality2}. 

For completeness we establish a similar analysis for  
F-theory compactifications to six dimensions on elliptically 
fibered Calabi-Yau threefolds in appendix \ref{5dCS}. We include 
explicit formulas for the $SU(N)$ case. 
A more elaborated discussion of this duality including gravity 
can be found in \cite{BonettiGrimm}.

\subsection{Resolving Calabi-Yau fourfolds} \label{resolving-4folds}
 
Let us consider an elliptically fibered Calabi-Yau fourfold $X_4$ with fibers which 
can be singular over each complex co-dimension of the base $\cB$. We further demand 
that these singularities can be consistently resolved while still preserving the 
Calabi-Yau condition. Numerous Calabi-Yau three- and fourfold 
examples with various gauge groups have 
been constructed in refs.~\cite{Candelas:1996su,Candelas:1997eh,Blumenhagen:2009yv,Grimm:2009yu,Cvetic:2010rq,Chen:2010ts,Krause:2011xj} 
as hypersurfaces and complete intersections inside 
a toric ambient space. One can show that the singularities are resolved by adding 
new blow-up divisors to the ambient toric space. This can be done systematically 
as argued in \cite{Candelas:1996su,Candelas:1997eh}. Note that only on the resolved Calabi-Yau manifolds one can 
straightforwardly compute the topological data of the geometry. These are required 
to determine the spectrum and couplings of the F-theory compactification \cite{Grimm:2010ks}. 
The toric resolutions are equivalent, at least at co-dimension one and two relevant here, 
to the small resolutions performed for an $SU(5)$ gauge group in \cite{Esole:2011sm,Marsano:2011hv}.\footnote{We like to thank D.~Klevers for explicitly checking this equivalence.} 

For simplicity let us focus on geometries with a single gauge group $G$ over 
a divisor $S_{\rm b} = S\cdot \cB$ in the base $\cB$
of the Calabi-Yau manifold. Here the dot denotes the intersection of the divisors 
$S$ and $\cB$. The resolved Calabi-Yau fourfold will be named $\tilde X_4$ in the 
following.
We denote the set of inequivalent exceptional resolution divisors and the Poincar\'e-dual two-forms by 
\beq \label{def-E}
    D_i,\ \omega_i \ , \qquad i =1,\ldots, \text{rank}(G)\, .
\eeq
In addition we have divisors and Poincar\'e-dual two-forms 
\beq \label{def-omegaalpha}
    D_\alpha, \ \omega_\alpha \ , \qquad \alpha =1,\ldots, h^{1,1}(\cB)\, ,
\eeq 
The divisors $D_\alpha$ are obtained from divisors in the 
base $\cB$ as pre-image of the projection $\pi: X_4\rightarrow \cB $ if there 
is no gauge group located along divisors $D_\alpha \cdot \cB$. 
However, after the blow-up one has to modify the divisor $S$ in $X_4$ 
which hosts the gauge group $G$. 
One introduces the redefinition 
\beq
  S = \hat S + \sum_i a_i  D_i \ ,  
  \label{eq:shift}
\eeq
where $a_i$ are the Dynkin labels of 
the group $G$. Note that this modification has to be taken into account
when introducing a basis $D_\alpha$ on $\tilde X_4$. In such a basis one 
has the expansion
\beq \label{S-expansion}
  S = C^\alpha D_\alpha\ .
\eeq 
The simplest situation is that $S$ is one of the divisors $D_\alpha$.
Finally, if the elliptic fibration only has a single section, we introduce the two-form $\omega_0$ Poincar\'e-dual to the 
base $\cB$ itself. 

There are various generalizations to this setup. In particular, 
the geometry can induce additional $U(1)$ factors due to its fibration 
structure or additional singularities over curves in $\cB$. 
The number of extra $U(1)$'s is counted by 
\beq \label{def-nU(1)}
   n_{U(1)} = h^{1,1}(\tilde X_4) - h^{1,1}(\cB) - \text{rank}(G)\ .
\eeq
A particular example with an extra $U(1)$ is the $U(1)$-restricted Tate model discussed 
in~\cite{Grimm:2010ez}. The geometry is in this case restricted such that the discriminant locus develops an 
additional singularity over a curve, which after resolution induces a new two-form $\omega_X$.
In general, each extra $U(1)$ comes with a new element $\tilde \omega_m$ of $H^{1,1}(\tilde X_4)$, and 
can be represented by a divisor $\tilde D_m$. 
Note that the two-forms $\tilde \omega_m$ have intersection properties similar
to the $\omega_i$ introduced above. Hence, it will be useful to introduce the 
combined notation 
\beq \label{def-DomegaLambda}
   D_\Lambda = (D_i , \tilde D_m)\ , \quad \omega_\Lambda = (\omega_i,\tilde \omega_m) \ , \qquad \Lambda =1,\ldots, \text{rank}(G) +n_{U(1)}\, .
\eeq 
As we will recall below, the $D_\Lambda$ have to have special intersection properties 
such that the corresponding gauge-fields are well-defined in four dimensions. 
This will allow to select an appropriate basis for $D_\Lambda$.

It is important to stress that in F-theory the resolution $\tilde X_4$ is not physical. 
In fact, the F-theory compactification to four space-time dimensions has to be carried out on the 
singular space $X_4$ where the full non-Abelian gauge symmetry is present. However, the space 
$\tilde X_4$ can be used in the dual M-theory compactification. Recall that it is natural to 
describe F-theory via M-theory \cite{Denef:2008wq,Weigand:2010wm}. Starting with M-theory this 
interpretation requires to perform a T-duality along one of the one-cycles 
of the elliptic fiber of $X_4$ after going to Type IIA by shrinking the size of the elliptic fiber.
Note that in the dual Type IIB setup the shrinking of the elliptic fiber corresponds to 
a decompactification to four space-time dimensions. The compactification on the resolved space 
$\tilde X_4$ is thus only possible in the M-theory picture, before shrinking the sizes of the elliptic 
fiber and the resolution divisors. 
In such a generic point in the K\"ahler moduli space 
of $\tilde{X}_4$, one is in the Coulomb branch of the three-dimensional 
gauge theory obtained by the M-theory compactification.
The gauge group 
is 
\beq \label{Coulomb-Group}
   U(1)^{{\rm rank}(G)}\ \times\ U(1)^{n_{U(1)}}\ .
\eeq
The $U(1)$ gauge bosons arise from the expansion of the M-theory three-form $C_{3}$  
into the two-form $\omega_\Lambda$ introduced in \eqref{def-DomegaLambda} as 
\begin{equation} \label{C3expansion}
C_3 = A^\Lambda \wedge \omega_\Lambda  \ ,\qquad \quad \Lambda = 1,\ldots , \text{rank}(G)+n_{U(1)}\ .
\end{equation}
Only in the limit in which the exceptional divisors $D_{i}$ shrink 
to the holomorphic surface $S$ one recovers the non-Abelian gauge 
symmetry $G$ present in the four-dimensional F-theory compactification. 

Having a fully resolved Calabi-Yau fourfolds $\tilde{X}_{4}$, one can compute the 
complete set of intersection numbers, and other topological data such as Chern classes.
Let us here summarize the structure of intersection numbers. For a hypersurface or 
complete intersection in a toric ambient space they can be determined explicitly by inducing 
the intersection structure of the ambient space. The intersections depend on the `triangulation' as we will 
make more precise for the examples below. This implies that there will be various topological phases 
associated to an ambient space and its Calabi-Yau manifold \cite{Witten:1993yc}.  
We introduce the quadruple intersections as
\beq \label{def-KABCD}
    \cK_{ABCD} = \int_{\tilde X_4} \omega_A \wedge \omega_B \wedge \omega_C \wedge \omega_D\, ,
\eeq
where $\omega_A = (\omega_0,\omega_\alpha,\omega_\Lambda)$. 
For resolved elliptically fibered Calabi-Yau fourfolds one has 
several vanishing conditions for the intersection numbers. 
Firstly, recall that for four divisors inherited from the base $\cB$ one obviously has 
\beq \label{vanish_intersect1}
  \cK_{\alpha \beta \gamma \delta} =0 \, .
\eeq
More subtle are the vanishing intersections involving the blow-up divisors $D_\Lambda$. 
The following vanishing conditions hold:
\beq \label{vanish_intersect2}
 \cK_{\Lambda \alpha \beta \gamma} = 0 \, , \quad \cK_{0 \Lambda AB} = 0 \ , 
\eeq
where $A,B$ run over all possible indices as in \eqref{def-KABCD}.
To justify this recall that $\omega_\Lambda$ parameterizes the $U(1)$'s in \eqref{Coulomb-Group} through 
the expansion \eqref{C3expansion}.
However, these are three-dimensional gauge fields in an M-theory compactification 
on $\tilde X_4$. In order that they lift to four-dimensional gauge fields the two conditions 
\eqref{vanish_intersect2} have to be satisfied \cite{Grimm:2010ks}. In fact, for the 
explicit resolutions performed below, this condition is satisfied for an appropriate basis $D_\Lambda$.
Clearly, the conditions \eqref{vanish_intersect2} are consequences of the geometry of 
resolved elliptic fibrations.

Let us now turn to the non-vanishing intersections.
For a single gauge group $G$ with resolution divisors $D_i$ one finds
\beq \label{dynkin_intersect}
   \cK_{ij \alpha \beta} = - C_{ij} \,  C^\gamma \, \cK_{0 \alpha \beta \gamma}\ ,
\eeq
where $C^\alpha$ has been introduced in \eqref{S-expansion}. $C_{ij}$ is the Cartan matrix of the algebra associated to the gauge group $G$. Note that the conditions \eqref{vanish_intersect1}, \eqref{vanish_intersect2} and \eqref{dynkin_intersect} are independent of the phase, or triangulation, 
of the resolution part of $\tilde X_4$.\footnote{In might be necessary 
to reorder the divisors $D_i$ to keep the same form of \eqref{dynkin_intersect}.} 
Of crucial importance for the chirality formulas will be the intersection
numbers:
\beq
  \cK_{\alpha \Lambda \Sigma \Gamma} \ , \qquad \cK_{\Lambda \Sigma \Gamma \Delta}\ ,
\eeq 
with three or four exceptional divisors $D_\Lambda$ introcuded in \eqref{def-DomegaLambda}. 
These crucially depend on the phase as we will see below.
Let us note that the basis used for the computation of these intersection numbers is adapted to 
the structure of the elliptic fibration. A basis adapted to the K\"ahler cone, measuring positve volumes 
in the Calabi-Yau manifold, will be discussed in section \ref{KahlerMori}.

\subsection{$G_4$-form fluxes and their F-theory interpretation} \label{introducingG4}

In this section we introduce the $G_4$ fluxes on the resolved 
Calabi-Yau fourfold $\tilde X_4$. The $G_4$ fluxes have to be considered in the M-theory picture of 
F-theory and correspond to a non-trivial field strength of the M-theory 
three-form $C_3$. Together with the results of 
section \ref{3dCS}, this will allow us to 
find the set of fluxes which induce a net chiral matter spectrum 
along the intersection curves of the 7-branes in the F-theory limit.

Let us first summarize some of the key properties of $G_4$.
The flux is an element of the fourth cohomology group $H^{4}(\tilde X_4,\mathbb{R})$. It
can be split into a horizontal and vertical part $H^{4}_V \oplus H^{4}_H$, where $H^{4}_V$ is 
obtained by wedging two forms of $H^{2}(\tilde X_4,\mathbb{R})$, and $H^{4}_H$ are the four-forms 
which can be reached by a complex structure variation of the 
holomorphic $(4,0)$-form on $\tilde X_4$. 
In the following we will be concerned with fluxes in $H^{4}_V(\tilde X_4,\mathbb{R})$, which 
can be written as 
\beq
   G_4 = m^{AB} \omega_A \wedge \omega_B\ ,
\eeq
where $\omega_A$ is the basis introduced in section \ref{resolving-4folds}.
There are constraints on $G_4$, both from an M-theory and an F-theory perspective. 
Firstly, M-theory anomalies demand that $G_4$ is properly quantized \cite{Witten:1996md}
\beq \label{quantization}
   G_4 + \tfrac12 c_2(\tilde X_4)\ \in \ H^{4}_V(\tilde X_4,\mathbb{Z})\ .
\eeq
This condition is crucial for fluxes in $H^4_V$ since the second Chern class $c_2(\tilde X_4)$ is in this component of $H^4$. 
The quantization condition has recently been discussed in \cite{Collinucci:2010gz,Krause:2011xj} for specific 
gauge groups or specific geometries. 
However, let us stress that in general it is a hard question to determine a minimal 
integral basis of $H^4_V(\tilde X_4, \mathbb{Z})$.\footnote{In particular, even if one shows that a component of 
$c_2(\tilde X_4)$ can be written as $a\, \omega \wedge \tilde \omega$ for the effective $\omega, \tilde \omega$, the integrality 
of the coefficient $a$ does not imply that $\frac{a}{2} \omega \wedge \tilde \omega$ is non-integral. A fancy way to determine 
an integral basis is by using mirror symmetry \cite{Grimm:2009ef}.}

Let us now turn to the constraints on $G_4$ imposed in the F-theory perspective. 
In order that the M-theory fluxes $G_4$ actually lift 
to F-theory fluxes without breaking four-dimensional
Poincar\'e invariance and keeping the whole group $G$ unbroken, 
we have to enforce that various components
of $G_4$ vanish. In order to do that we define\footnote{Note that we changed the definition of $\Theta_{AB}$ 
compared with \cite{Grimm:2011tb,Grimm:2011sk}. The chosen definition will be convenient in the 
match with the four-dimensional result.} 
\beq\label{def-theta_gen}
 \Theta_{AB} = \int_{\tilde X_4} G_4 \wedge \omega_A \wedge \omega_B\ .
\eeq
The fluxes relevant for our F-theory compactifications have to satisfy
\bea 
  \Theta_{0\alpha} &=& 0 \ ,\qquad  \Theta_{\alpha \beta} =0\ , \nn\\
  \Theta_{i\alpha} &=& 0\ .  
  \label{eq:G-condition}
\eea
Let us comment on these various constraints.
The first two constraints are conditions on the existence of a Poincar\'e invariant 
four-dimensional theory. 
$\Theta_{0\alpha}$ correspond in the M-theory to F-theory limit 
to fluxes along the circle when performing the 4d/3d compactification 
as discussed in detail in \cite{Grimm:2011sk}. The fluxes $ \Theta_{\alpha \beta}$ 
are mapped to non-geometric fluxes in F-theory and make the 
existence of a four-dimensional effective theory questionable.
Note that the fluxes $\Theta_{\Lambda 0}, \Theta_{00}$ 
are automatically vanishing due to \eqref{vanish_intersect2}, and the fact that $\Theta_{00} = \Theta_{0\alpha} K^\alpha$
with a vector $K^\alpha$ parameterizing the first Chern class of $\cB$.
The second line in \eqref{eq:G-condition} are conditions on an unbroken gauge group $G$.
$\Theta_{i\alpha}$ is readily interpreted in the M-theory to F-theory limit. 
These fluxes have a four-dimensional interpretation 
and would induce gaugings of the axionic parts of the complexified K\"ahler 
moduli. This yields a breaking of the group $G$, which we demand to be
unbroken in our considerations.
In summary, we find that the only non-vanishing components of $\Theta_{AB}$
are given by 
\beq \label{def-theta}
   \Theta_{\Lambda \Sigma} = \int_{\tilde X_4} G_4 \wedge \omega_\Lambda \wedge \omega_\Sigma\ , \qquad 
   \Theta_{\alpha m} = \int_{\tilde X_4} G_4 \wedge \omega_\alpha \wedge \tilde \omega_m\ .
\eeq
where $\omega_i,\omega_j$ are the two-forms Poincar\'e dual to the 
resolution divisors, and $\tilde \omega_m$ are the forms parameterizing extra $U(1)$'s as 
introduced in \eqref{def-DomegaLambda}. 

Let us make some further comments on the significance of $\Theta_{\alpha m}$. 
In \eqref{eq:G-condition} we have demanded $\Theta_{\alpha i} = 0$ to prevent breaking the 
gauge group by a gauging involving the Cartan generators only. For the extra
$U(1)$'s such a gauging is precisely induced by $\Theta_{\alpha m}$, and we did not restrict to the 
case where it has to vanish. In fact the gauge invariant derivatives are 
\beq \label{DT-gauging}
  D T_{\alpha} = d T_\alpha + i \Theta_{\alpha m} A^m\ .
\eeq
Here $T_\alpha$ are the complexified K\"ahler volumes of the divisors in the base $\cB$.
The precise definition of $T_\alpha$ as well as the lift of \eqref{DT-gauging} from M-theory to 
F-theory can be found in \cite{Grimm:2010ks,Grimm:2011tb}. The presence of the gauging \eqref{DT-gauging} implies 
that the $U(1)$ can become massive by a Higgs effect. In fact, $A^m$ can `eat' the imaginary part of
$T_\alpha$ and gain a new degree of freedom as required for a massive $U(1)$. 
Due to supersymmetry such a gauging induces also a D-term, which gives a mass to the real part of $T_\alpha$.
This massive scalar appropriately combines with $A^m$ into a massive four-dimensional $\cN=1$ vector multiplet. 

\subsection{Four-dimensional chirality formula from three-dimensional loops} \label{3dCS}

Recall that in order to find a well-defined framework to 
deal with fluxes in F-theory we have used the fact that 
F-theory can be obtained as a limit of M-theory. In this limit four-dimensional F-theory compactifications on 
a singular Calabi--Yau fourfold $X_4$ are obtained from an M-theory compactification on the resolved fourfold 
$\tilde X_4$ in the limit of shrinking elliptic fiber and shrinking exceptional divisors. The two setups 
are best compared in three dimensions where the M-theory compactification on $\tilde X_4$ has 
to match a circle compactification of the four-dimensional F-theory effective action \cite{Grimm:2010ks}.
In the following we will argue that the M-theory compactification with $G_4$ induces additional 
Chern-Simons terms which are not induced by a classical Kaluza-Klein reduction of a four-dimensional 
$\cN=1$ gauge theory on a circle. The match is achieved 
only after including one-loop corrections with charged matter fermions running in the loop.  

Let us start by recalling some crucial facts about M-theory on a Calabi-Yau fourfold $\tilde X_4$ \cite{Haack:2001jz,Grimm:2010ks}. 
As in \eqref{Coulomb-Group} the three-dimensional gauge group is broken to $U(1)^{{\rm rk}G} \times U(1)^{n_{U(1)}}$
when performing the reduction on a resolved Calabi--Yau fourfold.
Hence,  the M-theory effective theory will be in the Coulomb branch in three-dimensional gauge theory. 
Note that the three-dimensional $\cN=2$ vector multiplets contain as bosonic fields 
\beq \label{N=2vectors}
   (\xi^\Lambda,A^\Lambda)\ , \qquad \Lambda = 1,\ldots, {\rm rk}G+n_{U(1)}\ ,
\eeq
where the $\xi^\Lambda$ are real scalars. 
The $A^\Lambda$ are the $U(1)$ gauge fields from the dimensional reduction 
of the M-theory three-form as in \eqref{C3expansion}, while 
the $\xi^\Lambda$ parameterize the size of the blow-ups in the M-theory compactification on $\tilde X_4$. 
The $\xi^\Lambda$ arise in the expansion 
of the normalized K\"ahler form $\tilde J= J\cdot \cV^{-1}$, where $\cV$ is the overall volume of $\tilde X_4$. 
Explicitly, one expands  
\beq \label{Kaehlerexpand}
   \tilde J = \xi^\Lambda \omega_\Lambda + L^\alpha \omega_\alpha + R \omega_0 \ ,
\eeq 
where $\omega_\alpha,\omega_\Lambda$ are the two-forms introduced in \eqref{def-omegaalpha}, 
\eqref{def-DomegaLambda}, and $\omega_0$ is the Poincar\'e dual to the base $\cB$. 

The key observation is that the inclusion of $G_4$ fluxes in the M-theory reduction 
induces a Chern-Simons term in the three-dimensional effective action. In particular, 
for the vector multiplets \eqref{N=2vectors} one finds a Chern-Simons term 
\beq \label{Chern-Simons3d}
 S^{(3)}_{\rm CS} = \frac14 \int_{\mathbb{M}^{2,1}} \Theta_{\Lambda \Sigma} \, A^\Lambda \wedge F^\Sigma
\eeq
where $\Theta_{\Lambda \Sigma}$ is given in terms of the $G_4$ flux as in \eqref{def-theta}, and $F^\Lambda$ is the field strength 
of $A^\Lambda$. Due to the $\cN=2$ supersymmetry of the three-dimensional theory $\Theta_{\Lambda \Sigma}$ has to be constant, which is 
consistent with \eqref{def-theta}.

We now turn to the F-theory picture, and consider a general $\cN=1$ gauge theory compactified to three dimensions 
on a circle. The four-dimensional theory is identified with the low energy effective theory obtained 
by reducing F-theory on a Calabi-Yau fourfold $X_4$. In the four-dimensional theory 
the charged fermions $\chi^s$ appear with a kinetic term \cite{Wess:1992cp}
\beq \label{ferm-kinetic-term}
    K_{r \bar s} \bar \chi^s \displaystyle{\not}{\cD} \chi^r \ ,
\eeq
where $\cD_\mu$ is the covariant derivative under the four-dimensional gauge group, and 
$K_{r \bar s}$ is the K\"ahler metric for the matter multiplets.
After compactification on $S^1$, the terms \eqref{ferm-kinetic-term} will induce a coupling of the fermions to 
the $S^1$-component of the four-dimensional vectors. These components are identified with the 
$\xi^\Lambda$ if one move to the Coulomb branch of the gauge theory \cite{Grimm:2010ks}, where the vector 
fields also span the Abelian group \eqref{Coulomb-Group}. The resulting three-dimensional coupling is 
a mass term for the $\chi^r$ with mass parameter $\xi^\Lambda$.
We aim to compare the three-dimensional theories of M-theory and F-theory 
after Kaluza-Klein reduction.
In a general framework of three-dimensional 
Abelian gauge theories, the quantum-corrected 
coupling of the Chen-Simons term $\frac12\int (k_{\Lambda \Sigma})_{{\rm eff}} A^\Lambda \wedge F^\Sigma$ 
can be written as \cite{Aharony:1997bx}
\begin{equation}
(k_{\Lambda \Sigma})_{{\rm eff}} = (k_{\Lambda \Sigma})_{{\rm class}} 
+ \frac{1}{2}\sum_{f}(q_f)_{\Lambda}(q_f)_{\Sigma}\ {\rm sign}
\Big(\sum_{\Gamma=1}^{{\rm rk}G}(q_{f})_{\Gamma} \xi^{\Gamma} + \tilde m_{f} \Big),
\label{eq:cs_3d}
\end{equation}
where the second term arises from a one-loop diagram with charged fermions running in the loop. 
Hence, $f$ runs through all the charged fermions, $(q_f)_\Lambda$ is a $U(1)_\Lambda$ charge and $\tilde m_f$
is the classical mass of the fermions.  
Since the three-dimensional theories we consider originate from the dimensional reduction of 
four-dimensional $\mathcal{N}=1$ supersymmetric gauge theories, the classical Chern-Simons term with 
these indices is absent \cite{Grimm:2010ks,Grimm:2011sk}, i.e.~$ (k_{\Lambda \Sigma})_{{\rm class}} =0$. 
Furthermore, since the fermions are massless in the F-theory limit $\xi^\Lambda \rightarrow 0$ one 
also has to set $\tilde m_f=0$.
Therefore, comparing the Chern-Simons couplings \eqref{Chern-Simons3d} of the M-theory reduction with the general 
one-loop expression \eqref{eq:cs_3d} we find the relation\footnote{In the match of M-theory 
with F-theory a factor $1/2$ has to be taken into account. This has been discussed in \cite{Grimm:2011tb,Grimm:2011sk}
for the gauge coupling function $f_{\rm M} = 1/2 f_{\rm F}$, $f_{\rm F}$ is the three-dimensional gauge coupling 
obtained after circle reduction.}
\begin{equation}
\Theta_{\Lambda \Sigma} = \frac{1}{2}\sum_{f}(q_f)_{\Lambda}(q_f)_{\Sigma}\ 
  {\rm sign}\Big(\sum_{\Gamma=1}^{{\rm rk}G}(q_{f})_{\Gamma}\xi^{\Gamma}\Big).
\label{eq:chirality3d}
\end{equation}
This expression gives the link between the $G_4$ fluxes on $\tilde X_4$ and the number 
of fermions running in the loop if the charges $(q_{f})_{\Gamma}$, and the sign-factors 
are given. We will now determine these data using the geometric M-theory setting.

To see how the right-hand side of \eqref{eq:chirality3d} can be written by the geometric data
we have to recall how the fermionic states arise in M-theory. Recall that in F-theory the matter 
fields arise from strings stretching between two 7-branes intersecting over a matter 
curve $\Sigma_{\bf R}$. Here we will indicate by ${\bf R}$ the representation of the four-dimensional gauge 
group in which the matter fields localized on this curve transform.  In the M-theory picture these 
string states correspond to M2-branes. More precisely, in the resolved phase $\tilde X_4$ 
the matter multiplets arise from M2-branes wrapped on the resolution $\mathbb{P}^1$'s  
fibered over the matter surface \cite{Weigand:2010wm}. Crucially, one can establish a map between 
the weights of the representation $\bf R$ and the resolution $\mathbb{P}^1$ fibered over 
the matter curves $\Sigma_{\bf R}$ \cite{Intriligator:1997pq,Katz:1997eq,Marsano:2011hv}. Therefore we
will denote the resolution curves associated to a weight ${\bf w}$ by $\cC_{\bf w}$. We will 
discuss this identification in much more detail in section \ref{Strategy}. Using this map 
one can give a geometric formula for the $U(1)_\Lambda$ charge of an 
M2-brane wrapping on a curve $\cC_{{\bf w}}$. 
The charge $(q_f)_\Lambda$ for the fields is given by  
\beq \label{def-qfi}
  (q_f)_\Lambda = q^{\bf w}_\Lambda = \int_{\cC_{{\bf w}}}\omega_\Lambda \ ,
\eeq 
where we have used that $U(1)$ charge of a fermion does only depend on the 
weight to which it corresponds.
The real scalar $\xi^i$ is obtained from the expansion of the K\"ahler form
as in \eqref{Kaehlerexpand}.  Hence, we can rewrite the sign part of \eqref{eq:chirality3d} as
\begin{equation}
{\rm sign} \sum_{k=1}^{{\rm rk}G}(q_{f})_{k}\xi^{k} = {\rm sign} \int_{\cC_{{\bf w}}} \tilde J \equiv {\rm sign} ({\bf w}) \ ,
\label{eq:sign_kahler}
\end{equation}
for a matter field in a weight ${\bf w}$. Here we have used the abbreviation ${\rm sign} ({\bf w})$ to 
indicate when a curve is positive or negative, i.e.~we introduce the notation 
\bea \label{wsmallbig}
  {\bf w} >0 \qquad  &\Leftrightarrow& \qquad  \int_{\cC_{{\bf w}}} \tilde J > 0 \ ,\\
  {\bf w} < 0 \qquad &\Leftrightarrow& \qquad  \int_{\cC_{{\bf w}}} \tilde J <  0 \ . \nn
\eea  
Motivated by the appearance of this sign-factor in \eqref{eq:chirality3d} we will in section \ref{Strategy} introduce 
in detail the notion of the relative Mori cone. Roughly speaking, the curves in the relative Mori cone 
are precisely the resolution curves $\cC_{{\bf w}}$ for which the sign \eqref{eq:sign_kahler} is positive.
Therefore, providing the techniques to determine the relative Mori cone of a compact 
Calabi--Yau fourfolds will determine the signs in \eqref{eq:chirality3d}.

Let us denote by $n_{\bf r}$ the number of fermions in the effective three-dimensional theory 
transforming in a representation ${\bf r}$. Using \eqref{def-qfi} and \eqref{eq:sign_kahler} we can rewrite 
\eqref{eq:chirality3d} as  
\beq \label{Theta_wweights}
   \Theta_{\Lambda \Sigma} = \frac12 \sum_{\bf r} n_{\bf r} \sum_{{\bf w} \in {W({\bf r})}} q^{\bf w}_\Lambda  q^{\bf w}_\Sigma\ {\rm sign}( {\bf w})\ ,
\eeq
where the sum runs over all representations for which $n_{\bf r}$ fermions appear in the spectrum. From the expression \eqref{Theta_wweights}, one can see that vector-like pairs drop off from the contribution to the Chern-Simons term. If there is a vector-like pair, we always have a pair of weight ${\bf w}$ and $-{\bf w}$ and their $U(1)$ charges are opposite $q_\Lambda^{-{\bf w}} = -q_\Lambda^{{\bf w}}$. Then, the contribution from the vector-like pair is 
\bea
q^{\bf w}_\Lambda  q^{\bf w}_\Sigma \ {\rm sign}( {\bf w}) + (q^{-{\bf w}}_\Lambda)  (q^{-{\bf w}}_\Sigma)\ {\rm sign}( -{\bf w}) &=& \nn \\
 q^{\bf w}_\Lambda  q^{\bf w}_\Sigma \ {\rm sign}( {\bf w}) + (-q^{{\bf w}}_\Lambda)  (-q^{{\bf w}}_\Sigma)\ (-{\rm sign}( {\bf w}))&=&0
\eea
Therefore, only the chiral indices $\chi({\bf R}) = n_{\bf R} - n_{{\bf R}^*}$ with some numerical factors appear in the right-hand side of \eqref{Theta_wweights}.
Clearly, for a given setup one can simply compute the $q_{\Lambda}^{\bf w}$ and determine the signs \eqref{eq:sign_kahler}. This allows to 
read off $\chi({\bf R})$. Formally, one can write this as  
\beq
  \chi({\bf R}) = t^{\Lambda \Sigma}_{\bf R} \Theta_{\Lambda \Sigma}\ ,
\eeq
where $t^{\Lambda \Sigma}_{\bf R}$ is a matrix associated to the representation $\bf R$. 
In fact, $t^{\Lambda \Sigma}_{\bf R}$ determines the matter surface $S_{\rm R}$ 
appearing in \eqref{eq:chirality2}. In the next section we will present a formalism to compute 
$t^{\Lambda \Sigma}_{\bf R}$ explicitly for a given Calabi-Yau geometry. 
Let us stress that in the evaluation of \eqref{Theta_wweights} one uses the three-dimensional 
fermion spectrum. However, due to the fact that this three-dimensional
is obtained as a $S^{1}$ compactification of a four-dimensional 
theory arising from an F-theory compactification, the zero mode spectrum in the three 
dimensional theories should match that in the four-dimensional theories. 
In other words, \eqref{eq:chirality2} equally determines the 
chirality in F-theory compactifications on the Calabi--Yau fourfold $X_4$. 
One could suspect that there are other modes running in the loop which 
arise from the Kaluza-Klein tower in the circle compactification. It will be 
shown in \cite{BonettiGrimm} that such modes generate other Chern-Simons couplings 
but do not contribute in \eqref{Theta_wweights}.

In summary, we need the following data in order to evaluate \eqref{Theta_wweights} and to determine the chiral index
\begin{itemize}
\item A detailed identification of the weights ${\bf w}$ with the resolution curves $\cC_{\bf w}$ for a given resolved 
compact Calabi-Yau fourfold. 
\item The information about of the sign the K\"ahler form $\tilde J$ integrated over the curves $\cC_{\bf w}$.
\end{itemize}
Both of these data will be induced in detail in section \ref{Strategy}, and evaluated for specific examples in section \ref{Examples}.

To end this section, let us point out that there is an elegant way to encode the match \eqref{eq:chirality3d} by a single 
auxiliary function $\mathcal{T}$.  One defines $\cT$ such that its second derivative with respect to $\xi^\Lambda$ 
will generate \eqref{eq:chirality3d}. Hence, one has 
\beq \label{ddT}
   \Theta_{\Lambda \Sigma} = 2\, \partial_{\xi^\Lambda} \partial_{\xi^\Sigma} \cT\ .
\eeq
From the M-theory perspective a natural definition of 
$\mathcal{T}$ is  
\begin{equation} \label{def-cT}
\mathcal{T} = \frac14 \int_{\tilde X_4} \tilde J \wedge \tilde J \wedge G\ ,
\end{equation}
which indeed satisfies \eqref{ddT}.
Let us isolate the part $\cT^{\rm c}$ of $\cT$ which encodes the data about the fermionic spectrum running in the loop by 
defining $\cT^{\rm c}  = \frac12 \xi^\Lambda \xi^\Sigma  \partial_{\xi^\Lambda} \partial_{\xi^\Sigma} \cT$. Then the condition 
\eqref{Theta_wweights} translates into 
\beq 
   \cT^{\rm c} = \frac{1}{8}\sum_{\bf r} n_{\bf r} \sum_{{\bf w} \in W({\bf r})} \int_{\cC_{\bf w}}\tilde J \big| \int_{\cC_{\bf w}}\tilde J  \big|\ .
\eeq
Note that the real function $\cT$
as defined in \eqref{def-cT} is well-known in the M-theory and F-theory reductions \cite{Haack:2001jz,Grimm:2010ks,Grimm:2011sk}. It encodes not 
only data about the spectrum, as argued here, but also encodes the three-dimensional scalar potential. 
In fact, after performing the F-theory limit also the four-dimensional D-terms can be read off from this 
real function \cite{Grimm:2010ks}. For example, in the complete expansion of \eqref{def-cT} also 
the components $\Theta_{m\alpha}$ appearing in the $U(1)$-gaugings \eqref{DT-gauging} are included and 
generate the corresponding D-terms.

\section{Strategy to derive chirality formulas on resolved fourfolds}
\label{Strategy}

In this section we will describe our strategy to explicitly evaluate 
the formula \eqref{eq:chirality2} to determine the four-dimensional chiral 
spectrum induced by non-trivial $G_4$ flux. The section is divided into several parts
which stepwise introduce the geometrical tools to perform the computations. A particular 
focus will be on the determination of the matter surfaces using the Mori cone generators of 
the resolved Calabi-Yau fourfold. In outlining the tools we will also explain how 
details of the resolution process at co-dimension two and three in the base $\cB$ 
can be inferred from the compact geometry using the Mori cone. 
The discussion of this section will be kept rather 
general. Examples for which these computations can 
be carried out explicitly are postponed to section~\ref{Examples}.

\subsection{The relative K\"ahler and Mori cone} \label{KahlerMori}

We have seen in the previous section \ref{3dCS} that the 
evaluation of the one-loop corrections \eqref{Theta_wweights} requires a detailed 
knowledge of the positivity of the resolution curve classes. 
In the following we want to formalize this further. We therefore 
endow the space of divisors of the resolved fourfolds $\tilde X_4$ 
with a cone structure by singling out positive K\"ahler forms. 
This will allow us to define the relative K\"ahler cone and relative 
Mori cone.

Recall that the K\"ahler cone is spanned by K\"ahler forms $J$ satisfying 
the positivity conditions 
$\int_{\Sigma^k} J^k > 0 $, where $\Sigma^k$ are $k$-dimensional 
holomorphic submanifolds of $\tilde X_4$. The 
K\"ahler cone can be spanned by a basis of two-forms or, equivalently, a basis of divisors. 
The cone dual to the K\"ahler cone is known as the Mori cone. It is spanned by a basis of 
effective curves combined with positive coefficients.
In the following we want to introduce the \textit{relative K\"ahler and Mori cones}, which 
parameterize fields which are driven to a special limit when blowing down the resolutions. 
Note that the exceptional divisors $D_{i}$ correspond 
to the simple roots of $G$. Since the weights are elements of the dual 
space to the simple roots, the weights correspond to holomorphic 
curves $\Sigma_{\cI}$ inside the Mori cone, such that
\beq
   D_i \ \Rightarrow \ \text{roots}\ ,\qquad \Sigma_\cI \ \Rightarrow \ \text{weights} \ .
\eeq 
The intersections $D_i \cdot \Sigma_{\cI}$ corresponds to the natural 
dual pairing of weights and roots. We will discuss the precise identification 
of a given curve with a weight in the next subsection. In the following we want 
to first give the definitions of the relative cones. 

In the shrinking limit of the exceptional divisors $D_{\Lambda}$, there are 
holomorphic curves $\Sigma$ which are contained in $D_{\Lambda}$ and map to 
points in $X_4$. We will call the space of all such shrinking 
curves the relative Mori cone:
\beq
   M(\tilde{X}_4/X_4)=\{ \Sigma \ |\  \Sigma\ \, {\rm effective\;curve\;mapping\;to\;a\;point\;in\;X_4}\}.
   \label{eq:relative_mori}
\eeq
In the M-theory interpretation of the F-theory compactification 
the charged matter fields arise from M2-branes 
wrapping on the holomorphic curves $\Sigma$ as discussed in section \ref{3dCS}.
We have seen in \eqref{Theta_wweights} that the evaluation of the chirality 
requires a knowledge about the positivity of the curves.
Later on we will also argue that the relative Mori cone plays a 
crucial role to identify the resolution process of higher co-dimension 
singularities.

The dual cone to the relative Mori cone is called 
the relative K\"ahler cone $K(\tilde{X}_4/X_4)$. Hence, the relative K\"ahler 
cone can be defined as
\beq
  K(\tilde{X}_4/X_4)=\{ D=\sum s_\Lambda D_\Lambda \ | \ D\cdot \Sigma > 0{\rm \;for\;all\;}\  \Sigma \in M(\tilde X_4/X_4) \}.
\eeq
Note that the relative K\"ahler cone for the Cartan generators of $G$ realized on 
a singular Calabi-Yau threefold was already introduced in \cite{Intriligator:1997pq}.
In this case the negative relative K\"ahler 
cone is identified with the sub-wedge of the Weyl chamber of $G$ in 
five dimensional gauge theories. 

Let us introduce a natural extension of the relative Mori 
cone. We will add additional generators to $M(\tilde X_4/X_4)$ which 
are effective curves in the elliptic fiber, 
i.e.~the Mori cone elements, and intersect the generators of 
the relative K\"ahler cone.
In simple cases this amounts to 
including the pinched elliptic fiber over the 7-brane. 
We will call the resulting cone $\widehat M(\tilde X_4/X_4)$
the \textit{extended relative Mori cone}. 
Clearly, we can similarly introduce the \textit{extended relative K\"ahler cone} $\widehat K(\tilde{X}_4/X_4)$
dual to $\widehat M(\tilde X_4/X_4)$. The cone $\widehat K(\tilde{X}_4/X_4)$ will contain 
one more generator $D_0 = \hat S$ which corresponds to the extended node of the Dynkin diagram of $G$.
This generator allows to extend \eqref{dynkin_intersect} to 
\beq 
  \cK_{IJ \alpha \beta} = - C_{IJ} \,  C^\gamma \, \cK_{0 \alpha \beta \gamma}\ ,
\eeq
where $I=(0,i)$ and $C_{IJ}$ is the extended Cartan matrix.

\subsection{Mori cone, singularity resolution, and connection with group theory}
\label{subsec:mori}

Having determined the relative Mori and K\"ahler cone, we now want to make contact 
with the group theory of the seven-brane gauge theory with gauge group $G$.
Our key point will be the precise association of some weights of a representation of the gauge group
with the elements of the relative Mori cone. 

\subsubsection{General discussion}

We start more general and introduce the charge vectors $\ell_{\cI, A}$ given 
by intersecting curves $\Sigma_\cI$ in the Mori cone with divisors $D_A$ in $\tilde X_4$ as
\beq \label{ell-def}
   \ell_{\cI,A} = \int_{\Sigma_\cI} \omega_A = \Sigma_\cI \cdot  D_A\ .
\eeq
We will determine the $\ell_{\cI,A}$ for specific examples in section \ref{Examples}. Let us 
make here some general comments, denoting henceforth by $\ell_\cI$ the vector with entries \eqref{ell-def}. For Calabi-Yau fourfold examples which are realized as 
hypersurfaces or complete intersections in a toric ambient space one determines the 
vectors $\ell_{\cI}$ in two steps \cite{Berglund:1995gd,Berglund:1996uy,Braun:2000hh}. 
Firstly, one uses the set of toric divisors $D_A$ of the 
ambient space and derives the $\ell_\cI$ using the Mori cone generators of the ambient space.
Since the ambient space can admit many triangulations, i.e.~topological phases connected by 
flop transitions, one obtains for a given geometry several sets of vectors $\ell_\cI^{\rm (I)},\ell_\cI^{\rm (II)}, \ell_\cI^{\rm (III)},\ldots $, each
set associated to a phase.  
Restricted to the Calabi-Yau manifold $\tilde X_4$ it can happen that different triangulations of the ambient 
space are connected by flops of curves which are not in $\tilde X_4$. This implies that several 
sets of the ambient space $\ell$-vectors have to be combined to describe the 
$\ell$-vectors of the Calabi-Yau manifold $\tilde X_4$.\footnote{By abuse of notation we have used 
the same symbols and indices for the $\ell$-vectors of the ambient space and the Calabi-Yau manifold $\tilde X_4$. Let us 
stress that even the number of $\ell$-vectors can differ for the two geometries.} Clearly, it will be our 
task to determine these vectors $\ell_\cI$ for $\tilde X_4$ itself in section  \ref{Examples}.
For completeness a brief account of the general procedure to determine the $\ell$-vectors for a 
Calabi-Yau hypersurface is given in appendix \ref{sec:mori_cone}.

Let us now make contact with the gauge theory on the 7-branes. We recall that we 
are working with the resolved fourfold $\tilde X_4$
and hence are on the Coulomb branch in the M-theory compactification to three dimensions. 
The geometrically massless gauge fields are then parameterizing the 
Abelian group \eqref{Coulomb-Group}, $U(1)^{\text{rank}(G)} \times U(1)^{n_{U(1)}}$.
In connection with this gauge group it will be crucial to 
analyze the $\ell$-vectors associated to $U(1)$-charges for the divisors $D_\Lambda$.
These are given by
\beq \label{U(1)charges}
  \ell_{\cI, \Lambda} = \Sigma_\cI \cdot D_\Lambda\ ,
\eeq
where $D_{\Lambda} = (D_i,\tilde D_m) $ as in \eqref{def-DomegaLambda}. 
In particular, for the Cartan $U(1)$'s in $G$ one has the Cartan charges $\ell_{\cI,i}=\Sigma_\cI \cdot D_i$,
where $D_i$ are the resolution divisors corresponding to the Cartan generators of $G$.
One realizes that a curve $\Sigma_\cI$ will be in the \textit{relative} Mori cone if 
it has negative intersection with one of the $D_\Lambda$:
\beq
  \ell_{\cI,\Lambda} < 0 \quad \Rightarrow \quad \Sigma_\cI \in M(\tilde X_4 /X_4)\ .
\label{eq:relative_mori_condition} 
\eeq
In fact, if the curve $\Sigma_\cI$ has the negative intersection 
with $D_\Lambda$, $\Sigma_\cI$ is contained in $D_\Lambda$ and shrinks to a point in $X_4$.
Note that if a curve $\Sigma_\cI$ is in the base $\cB$ itself, the intersection with the
$D_\Lambda$ vanishes due to the intersection structure \eqref{vanish_intersect2}. This is consistent with 
the fact that such curves have no $U(1)$-charges under the group \eqref{Coulomb-Group}, and 
are not in the relative Mori cone.

By computing the $U(1)$-charges of a curve $\Sigma_\cI$ with respect to the 
$D_\Lambda$, one can next determine a 
weight which reproduces the same $U(1)$ charges, and associate the weight to the curve $\Sigma_\cI$ or the $\ell$-vector $\ell_\cI$.
This leads to the identification
\beq
  \ell_\cI \  \cong \ \text{weight of a representation of}\ G\ .
\eeq 
Using this method one can associate a weight of $G$ to each generator of the relative 
Mori cone. This allows us to determine which weights $\bf{w}$ correspond to the effective curves and which weights do not. Since we know the weights which correspond to the generators of the relative Mori cone, other weights which correspond to effective curves should be realized by a linear 
combination of the weights in the relative Mori cone with positive integer coefficients. 
Applying this process, we determine the complete correspondence between the weights 
and the effective curves. 
Consistent with \eqref{wsmallbig} we use the following simplifying notation
\bea \label{notation_effective}
   {\bf w} >0 \quad &\Leftrightarrow& \quad {\bf w}\ \text{corresponds to an effective curve}\\ 
   {\bf w} <0 \quad &\Leftrightarrow& \quad {\bf w}\ \text{does not correspond to an effective curve}\ . \nn 
\eea
We argue in the following that details of the resolution process are contained in this information. 

Before turning to the resolution process, let us briefly 
comment on how one can also represent the roots as curves. 
In fact, due to the intersection numbers \eqref{dynkin_intersect}, we can always 
introduce a curve $\cC_{-\alpha_i}$ associated to the negative 
of a simple root $\alpha_i$ by the triple intersection of three divisors
\begin{equation}
\cC_{-\alpha_i} = D_i \cdot \tilde D \cdot \cD,
\label{simple-roots}
\end{equation}
where $\tilde D = v^\alpha D_\alpha$ and $\cD = s^\alpha D_\alpha$ are linear 
combinations of the divisors $D_\alpha$ inherited from the base. 
To ensure the correct normalization of the simple root $\alpha_i$ 
these divisors have to satisfy the condition
\begin{equation}
\cB \cdot S \cdot \tilde D \cdot \cD = 1 \ .
\end{equation}
Hence, for $D_i, \tilde D, \cD$ being holomorphic hypersurfaces of $\tilde X_4$ 
the curves which correspond to the negative simple roots 
are effective curves and elements in the relative Mori cone. 
The situation is different for the weights. 
Some of the weights do not correspond to effective curves. However, if one finds 
a weight which corresponds to an effective curve, one can 
construct other weights corresponding to effective curves from a linear combination of the 
original weight and the negative simple roots 
with positive integer coefficients. This does not 
mean that all the weights correspond to effective curves since some 
weights need negative coefficients for their construction.

The co-dimension one singularities in $\cB$ over the surface $S_{\rm b}$ 
determine the gauge group $G$ on the 7-branes.
Generically, there are also 
co-dimension two and co-dimension three singularities. 
Physically, charged matter fields are localized on the co-dimension two singularities 
and the Yukawa interaction between these fields is generated from co-dimension 
three singularities. Focusing on $G$ we realize that the 
Cartan divisors $D_i$ are $\mathbb{P}^{1}$-fibrations at the generic points 
in the surface $S_{\rm b}$. However, the $\mathbb{P}^{1}$-fibers may degenerate into 
smaller irreducible components along the singularity enhancement locus where the matter and 
Yukawa couplings are localized. 
The resolution of the co-dimension one singularity generates the extended Dynkin diagram of $G$. 
The resolution of the higher co-dimension singularity will 
generate another Dynkin diagram which may have a rank larger than  
${\rm rank}(G)$. 
We propose rules to determine the Dynkin diagrams from the resolution of the higher 
co-dimension singularity by exploiting the relative Mori cone. 

Let us consider a situation where the charged matter fields in the representation of ${\bf R}$ and ${\bf R}^{\ast}$ of $G$ 
are localized along the co-dimension two singularity enhancement locus $\Sigma_{\bf R}$. 
{}From the relative Mori cone, one can determine whether a weight of ${\bf R}$ or ${\bf R}^{\ast}$ corresponds to a effective curve or not. 
Then, the rule to determine the degeneration of $\mathbb{P}^{1}$ along $\Sigma_{{\bf R}}$ is that the negative of a simple root 
decomposes into a weight of ${\bf R}$ and a weight of ${\bf R}^{\ast}$ if both of them correspond to effective curves. 
If a curve corresponding to a weight lies in the relative Mori cone it is an effective curve.
In this decomposition process one has to use the generators of the extended relative Mori cone as much as possible.
In particular, one checks if the weight of ${\bf R}$ found in the decomposition can be further decomposed 
into a weight of ${\bf R}$ and the negative of a simple root, and if either of 
them is an element of the relative Mori cone or corresponds to the extended node. 
In this evaluation one should not mix with the weight of the other representation. Also, since 
the negative of a simple root is a generator of the relative Mori cone, it does not need to be decomposed further. 
By collecting all the irreducible components 
along $\Sigma_{{\bf R}}$ plus a curve corresponding to the extended Dynkin node, one can construct a Dynkin 
diagram generated from the resolution along the co-dimension two singularity locus $\Sigma_{\bf R}$. 
To make this algorithm more clear without introducing all the details of the global geometry
we give a simple $SU(5)$ example in subsection \ref{following_SU(5)}.

The co-dimension three singularity 
enhancement occurs at a point $p$ where at least two co-dimension two 
singularity loci intersect:
\beq \label{def-p}
   p=\Sigma_{\bf R} \cdot \Sigma_{\bf R'} \ \subset S_{\rm b}\ .
\eeq 
Here we suppose that the charged matter fields in the representation 
of ${\bf R}$ and ${\bf R}^{\ast}$ localized on one curve $\Sigma_{{\bf R}}$, 
and other charged matter fields in the representation of ${\bf R}^{\prime}$ and ${\bf R}^{\prime \ast}$ 
are localized along the other curve $\Sigma_{{\bf R}^{\prime}}$. 
Although the Dynkin diagram obtained from the resolution along the locus $\Sigma_{{\bf R}}$ consists of some of the weights of 
${\bf R}$, weights of ${\bf R}^{\ast}$ and the negative simple roots, 
the weights of ${\bf R}^{\prime}$ and the weights of ${\bf R}^{\prime \ast}$ can also form the nodes of the Dynkin diagram 
from the resolution at the co-dimension three singularity point $p$ obtained \eqref{def-p}. 
Hence, a weight of ${\bf R}, {\bf R}^{\prime \ast}$ or the negative of a simple root of $G$ further decomposes at $p$ if they 
are made of effective curves which correspond to any weights of ${\bf R}, {\bf R}^{\ast}, {\bf R}^{\prime}, {\bf R}^{\prime \ast}$ 
and the negative simple roots of $G$. When the singularity is enhanced to $G_{p} \supset G$ at $p$, this 
decomposition has to obey the algebra of $G_{p}$. From this decomposition rule, one can obtain all the weights 
and simple roots which form a Dynkin diagram obtained from the resolution of the co-dimension three singularity at $p$.

\subsubsection{A simple $SU(5)$ example} \label{following_SU(5)}

Since this explanation is rather abstract, let us illustrate the above procedure on a simple example 
with gauge group $SU(5)$. The representations are the ${\bf R} = {\bf 5}$, and ${\bf R} = {\bf 10}$
along enhancement curves $\Sigma_{\bf 5}$ and $\Sigma_{\bf 10}$ respectively.  
We will not introduce the complete geometry here, but rather focus on the 
determination of the Dynkin nodes over the enhancement loci. In other words we assume the 
the $\ell$-vectors have been determined for a given geometry, and the association of the generators 
of the relative Mori cone with the weights of $SU(5)$ has been performed.
We consider the following identification:
\begin{equation} \label{id_weightsandell}
\tilde \ell_{1} \cong -e_{3}, \qquad \tilde \ell_{2} \cong e_{3}+e_{4}, \qquad \tilde \ell_3 \cong -e_{1}+e_{2},\qquad \tilde \ell_4 \cong -e_{2}-e_{4}  \ ,
\end{equation}
where $\tilde \ell_i$ are the $\ell$-vectors generating the relative Mori cone, and $e_i$ 
are a orthonormal basis of $\mathbb{R}^5$ allowing 
to represent the roots and weights for $SU(5)$ and its representations. 
A compact Calabi-Yau fourfold which exactly yields the identification \eqref{id_weightsandell} 
can be found in section \ref{Example1}, see equation \eqref{eq:weight1}.

We first want to determine the weights which appear as curves in the relative Mori cone. 
For the weights of the {\bf 5} representation, $-e_3$ corresponds to an effective curve 
from the generators of the relative Mori cone \eqref{id_weightsandell}. 
Then, it is straightforward to see 
\begin{eqnarray}
e_4 &=& (-e_3) + (e_3+e_4),\\
-e_2 &=& (-e_2-e_4) + (-e_3) + (e_3+e_4),\\
-e_1 &=& (-e_1+e_2) + (-e_2-e_4) + (-e_3) + (e_3+e_4).
\end{eqnarray}
Hence, $e_4, -e_1$, and $-e_2$ correspond to effective curves. 
To determine $e_5$ or $-e_5$ corresponds to an effective curve, 
we use the fact that $e_1+e_2+e_3+e_4+e_5$ is a singlet of $SU(5)$. 
Then, we have
\begin{eqnarray}
e_5&=&e_1+e_3+e_5+ (-e_1) + (-e_3),\\
&=&(-e_2-e_4) + (-e_1) + (-e_3).
\end{eqnarray}
Therefore, $e_5$ corresponds to an effective curve. 
To summarize, the correspondence between the effective curves and the ${\bf 5}$ weights is given by 
\begin{equation} \label{effective5}
e_{1} < 0,\qquad e_2 < 0,\qquad e_3 < 0,\qquad e_4 > 0,\qquad e_5 > 0\ ,
\end{equation}
and has to be interpreted using the notation \eqref{notation_effective}.
A similar analysis can be carried out for the weights of the $\bf 10$ representation
\bea  \label{effective10}
 &&  e_1 + e_2 < 0,\qquad e_1+e_3<0,\qquad e_2 + e_3 < 0, \qquad e_2 +e_4 < 0, \\
 &&  e_3 +e_4 > 0, \qquad e_3 + e_5>0,\qquad e_4 + e_5 > 0. \nn
\eea
This concludes the identification of weights with effective curves.

Using this information we can now determine how the negative simple roots degenerate over
the enhancement curves $\Sigma_{{\bf 10}}$ and $\Sigma_{\bf 5}$.
Let us start with $\Sigma_{{\bf 10}}$, along which some of the negative simple
roots degenerate into ${\bf 10}$ and $\overline{{\bf 10}}$ weights. 
First, we consider the decomposition of the negative simple roots into smaller components along $\Sigma_{{\bf 10}}$
\begin{equation}
-({\rm simple}\;{\rm root}) = \overline{{\bf 10}}\;{\rm weight} + {\bf 10}\;{\rm weight}.
\label{eq:decomposition1-10}
\end{equation} 
Then, if both $\overline{{\bf 10}}$ weights and {\bf 10} weighs correspond to effective curves in the relative Mori cone, 
this degeneration occurs along $\Sigma_{{\bf 10}}$. 
The check of effectiveness of the curves corresponding to all the {\bf 10} weights 
was given in \eqref{effective10}. Hence, the degeneration of the negative simple roots along $\Sigma_{{\bf 10}}$ are
\begin{eqnarray}
-e_{2}+e_{3} &=& (-e_{2} - e_{4}) + (e_{3} + e_{4}),\label{eq:decomposition_D5-1}\\
 -e_{4}+e_{5} &=& (-e_{1}  - e_{4}) + (e_{1} + e_{5})
=(-e_{1} + e_{2}) + (-e_{2} -e_{4}) + (e_{1} + e_{5}).\label{eq:decomposition_D5-2} \nn 
\end{eqnarray}
To summarize, along $\Sigma_{\bf 10}$ the negative simple roots of $SU(5)$ plus the extended node $e_1-e_5$ split 
into 
\begin{equation}
e_{1}-e_{5},\quad e_{1}+e_{5},\quad -e_{1}+e_{2},\quad -e_{2}-e_{4},\quad -e_{3}+e_{4},\quad e_{3}+e_{4}.
\label{eq:D5gens_text}
\end{equation}
The resolution curves associated to the weights \eqref{eq:D5gens_text} form the extended Dynkin diagram of $D_{5}$ as depicted in Figure \ref{fig:Dynkin_phase1}.
Note that the well-known form of the $D_5$ Dynkin diagram is not directly visible by simply looking at the group-theoretic 
intersections of the elements \eqref{eq:D5gens_text}. However, this structure can be inferred by a local analysis as 
presented in appendix~\ref{sec:direct_comp}.  Let us stress that this information is not needed in the evaluation 
of the chirality formulas and hence will not play a major role in this work. 

We next turn to singularity enhancement locus 
$\Sigma_{{\bf \bar{5}}}$. In this case, some of the negative simple roots 
decompose into {\bf 5} weight and $\bar{{\bf 5}}$ weight,
\begin{equation}
-({\rm simple}\;{\rm root}) = \bar{{\bf 5}}\;{\rm weight} + {\bf 5}\;{\rm weight} .
\label{eq:decomposition1-5}
\end{equation}
Since $-e_3$ and $e_4$ corresponds to the effective curves from \eqref{effective5}, the decomposition of the negative simple roots along $\Sigma_{\bar{{\bf 5}}}$ is 
\begin{equation}
-e_{3} + e_{4} = (-e_{3}) + (e_{4}).
\end{equation}
Then, the negative simple roots of the $SU(5)$ plus the extended Dynkin node become
\begin{equation} \label{eq:A5gens_text}
e_{1}-e_{5},\quad -e_{1}+e_{2},\quad -e_{2}+e_{3},\quad -e_{3},\quad e_{4}, \qquad -e_{4}+e_{5}.
\end{equation}
The curves associated to these weights form the extended $A_{5}$ Dynkin diagram as depicted in 
the Figure \ref{fig:Dynkin_phase1}. Once again, we will only need in the derivation of the chirality formulas 
the identification of \eqref{eq:A5gens_text} with effective curves and not the precise match with the Dynkin diagram.

\subsection{Matter surfaces and the chiral index} 

Having discussed how the relative Mori cone can determine the resolution process of the higher co-dimension singularities
we next want to include the $G_4$ fluxes on $\tilde X_4$ and evaluate a chirality formula \eqref{eq:chirality2}. 
Recall from section \ref{introducingG4} that F-theory fluxes have to satisfy the conditions \eqref{quantization} 
and \eqref{eq:G-condition}. The non-vanishing components of $G_4$ are captured by the matrices $\Theta_{\Lambda \Sigma}$
and $\Theta_{m\alpha}$ introduced in \eqref{def-theta}.

Let us turn to the determination of the matter surfaces 
$S_{\bf R}$ appearing in \eqref{eq:chirality2} by using the extended 
relative Mori cone. As discussed in \cite{Donagi:2008ca,Hayashi:2008ba,Braun:2011zm,Marsano:2011hv,Krause:2011xj} 
this matter surface should be obtained by fibering the resolution $\mathbb{P}^1$'s over 
the matter curve $\Sigma_{\bf R}$ with representation ${\bf R}$ and ${\bf R^*}$. 
A relation of the matter surfaces with the 
weights of ${\bf R}$ was stressed in \cite{Marsano:2011hv}.
The fiber $\cC_{\bf w}$
corresponds to a weight ${\bf w}$ of the representation ${\bf R}$. The curves $\cC_{{\bf w}}$ 
are identical to the ones introduced in the resolution of the co-dimension 
two singularity locus $\Sigma_{{\bf R}}$. 
Hence, each curve $\cC_{\bf w}$ can be determined from the relative Mori cone
as discussed above. Such effective curves can be written by the triple 
intersection of divisors. We make the Ansatz 
\begin{equation} \label{P1ansatz}
\cC_{\bf w} = t^{A \Sigma}_{{\bf w}}\, D_A \cdot D_{\Sigma} \cdot \mathcal{D}\,
,\qquad \quad \mathcal{D}=s^{\alpha}D_{\alpha}\ ,
\end{equation}
with some real coefficients $t^{A \Sigma}_{{\bf w}} =  t^{A}_{{\bf w}} v^\Sigma$ which generally 
depend on the $s^\alpha$. Here $v^\Sigma D_{\Sigma}$ is an
exceptional divisor which contains the curve $\cC_{\bf w}$. 
Let us note that we checked this Ansatz for our examples, and 
showed that it can always be satisfied. This includes the observation 
that there exists a divisor $\cD$, that intersects the base $\cB$ in
a divisor, which can be separated as in \eqref{P1ansatz}. For weights of representations of the gauge group $G$ on $S_{\rm
b}$,
$\cD$ intersects $S_{\rm b}$ in a curve. It would be desirable to give a
geometric proof that
there always exists a representation of the class of $\cC_{\bf w}$ of the form
\eqref{P1ansatz}. 

At least for $SU(N)$ gauge theories with matter in
the fundamental and anti-symmetric representation, one
can show that the curve $\cC_{\bf w}$ can be generally written as
\eqref{P1ansatz} using a group theory argument. In the fundamental 
or anti-symmetric representation, all the Cartan charges
of the highest weight are non-negative. On the other hand, the other weights
have at least one negative Cartan charge. Hence, the 
Ansatz \eqref{P1ansatz} might only be impossible if the highest weight
appears as a generator of the relative Mori cone. For the other weights 
one can always choose a component $D_{\Lambda}$ in the Ansatz \eqref{P1ansatz} 
which has negative intersection number with the curve $\cC_{\bf w}$. If the highest weight is
a generator of the relative Mori cone, all the weights correspond to 
effective curves in the relative Mori cone since the negative simple
roots are always effective curves by \eqref{simple-roots}. When one sums up
all the weight $e_i,\; (i=1, \cdots, N)$ in the fundamental representation,
or the weights $e_i + e_j,\; (1 \leq i \neq j \leq N)$ in the anti-symmetric
representation, one has $N(e_1+ \cdots + e_N)$ or $(N-1)(e_1 + \cdots +
e_N)$ respectively. Namely, the singlet of $SU(N)$ corresponds to the
effective curve in the relative Mori cone. However, if the curve
corresponding to the singlet is in the relative Mori cone, the relative
K\"ahler cone cannot be defined since $\int_{\cC_{{\rm singlet}}}\sum
s^{i}D_{i} = 0$. Therefore, the highest weight cannot be the
generator of the relative Mori cone and the generators of the relative Mori
cone should have at least one negative Cartan charge. This negative Cartan
charge indicates that the curve is contained in an exceptional divisor.
Hence, one can always make the Ansatz \eqref{P1ansatz} for the $SU(N)$ gauge
theories with matter in the fundamental and anti-symmetric representations.

Since our final interest is in the matter surface $S_{{\bf R}}$ we still have to extract a surface out of the 
curve \eqref{P1ansatz}. In order to do that we propose to pull out the divisor $\cD$ which becomes a 
curve in $S_{\rm b}$.
In order that the normalization of the $t^{A \Sigma}_{{\bf w}}$ in \eqref{P1ansatz} is fixed we demand that the curve 
$\cD\cdot S \cdot B$  intersects exactly once with the matter curve $\Sigma_{{\bf R}}$ in $S_{\rm b}$. 
In other words we normalize $\cD$ such that 
\begin{equation}
    \Sigma_{{\bf R}} \cdot \mathcal{D} =1\ .
\label{eq:multiplicity}
\end{equation}
The condition \eqref{eq:multiplicity} fixes the normalization of $\cD$, 
and via \eqref{P1ansatz} the normalization of the $t^{A \Sigma}_{\bf w}$.
The class of the matter surface $S_{\bf R}$ is then fixed and given by
 \beq \label{matter-surfaces}
       S_{\bf R}  = t^{A \Sigma}_{{\bf w}}\, D_A \cdot D_\Sigma\, ,
 \eeq
with $t^{A \Sigma}_{\bf w} = t^{A}_{\bf w} v^\Sigma$ as in \eqref{P1ansatz}. For a fixed $\cD$ the parameters 
$t^{A \Sigma}_{{\bf w}}$ are 
determined from the intersection numbers between the curve $\cC_{\bf w}$ and the divisors 
$D_{A}$. The intersection numbers are already known as entries of the $\ell$-vectors. 
This procedure does not determine the parameters uniquely, but fixes the class of the curve $\cC_{\bf w}$ and 
the matter surface $S_{\bf R}$. Curves 
are in the same class if their intersection number with the divisors are identical.
Strictly speaking one should note that $S_{\bf R}$ depends on the chosen weight for the representation ${\bf R}$. 
However, as will become more clear momentarily, this ambiguity drops out 
from the chirality formula \eqref{eq:chirality2}.

Using the non-vanishing components of the $G_4$ flux \eqref{def-theta}, the 
chirality formula \eqref{eq:chirality2} together with the explicit 
form of the matter surfaces $S_{\bf R}$
\eqref{matter-surfaces} yields 
\beq
   \chi({\bf R}) = t^{A \Sigma}_{\bf w} \Theta_{A\Sigma}\ .
   \label{chirality}
\eeq 
{}From this expression one can infer that the chirality is indeed independent of the 
weight ${\bf w}$ and only depends on the representation ${\bf R}$. 
Suppose that a curve corresponding to another weight ${\bf w}^{\prime}$ of the 
representation ${\bf R}$ also appears along the locus $\Sigma_{{\bf R}}$. 
Since any two weights can be related by a linear combination of simple roots, one can 
write ${\bf w}^{\prime} = {\bf w} - \sum_{i}u^{i}\alpha_i$. Since the negative simple roots can be 
always written as \eqref{simple-roots}, we can expand $\tilde{D} = v^{\alpha}D_{\alpha}$, and identify $\cD$ 
of \eqref{simple-roots} and \eqref{P1ansatz}. 
Hence, adding or subtracting the negative simple roots to or from the weight ${\bf w}$ corresponds 
to adding or subtracting $D_i \cdot \tilde{D}$ to or from $t_{\bf w}^{A \Sigma }D_{A} \cdot D_{\Sigma}$. 
Then, the expression \eqref{chirality} evaluated for the two different weights ${\bf w}$ and ${\bf w}^{\prime}$ yields 
\begin{eqnarray} \label{independent_of_weight}
t_{{\bf w}^{\prime}}^{A \Sigma}\Theta_{A \Sigma} &=& t_{{\bf w}}^{A \Sigma}\Theta_{A\Sigma} + u^{i}v^{\alpha}\Theta_{i\alpha} \\
&=& t_{{\bf w}}^{A \Sigma}\Theta_{A\Sigma}\ ,\nn
\end{eqnarray}
where we use \eqref{eq:G-condition}. Therefore, the chirality formula \eqref{chirality} does not depend on the weight ${\bf w}$
but only on the representation ${\bf R}$. In geometrical terms this also implies that \eqref{chirality} does not 
depend on the topological phases of the $\tilde X_4$ distinguishing different Calabi-Yau resolutions of $X_4$. In 
fact, in different phases other weights of the 
same representation are associated to the matter curves, and \eqref{independent_of_weight} ensures independency of the 
resolution phase.



\section{Examples} \label{Examples}

In this section we discuss two illustrative examples of explicitly resolved Calabi-Yau hypersurfaces realized in a
toric ambient space. Our first example will admit an $SU(5)$ singularity over a divisor in the base, as in the 
torically realized GUT models of \cite{Blumenhagen:2009yv,Grimm:2009yu}. In the second example an additional $U(1)$ will be present, such 
that $n_{U(1)}=1$ in \eqref{def-nU(1)}. The toric construction will correspond to the $U(1)$-restricted Tate model of \cite{Grimm:2010ez}. 
The toric methods required to perform the computations of this subsection have been explained 
in~\cite{Candelas:1996su,Candelas:1997eh,Blumenhagen:2009yv,
Grimm:2009yu}, and where recently reviewed in~\cite{Knapp:2011ip}. The determination of the Mori cone can be performed using 
the methods of \cite{Berglund:1995gd,Berglund:1996uy,Braun:2000hh} as reviewed in appendix \ref{sec:mori_cone}.

\subsection{A Calabi-Yau hypersurface with $SU(5)$ gauge group} \label{Example1}

As the first example, we consider a Calabi--Yau fourfold $\tilde{X}_4$ which has a K3 fibration. 
The K3 fibration itself has an elliptic fibration such that it can be used  in an F-theory compactification. 
Such a Calabi--Yau fourfold can be obtained from a hypersurface in the ambient toric space whose points on edges of the polyhedron are
\begin{eqnarray}
\begin{array}{|ccccc|rl|} \hline
\multicolumn{5}{|c|}{\text{points}} &\ \text{divisor}& \hspace*{.1cm} \text{basis}\hspace*{.3cm} \\ \hline \hline
 -1 & 0 & 0 & 0 & 0 & D_{1} & \\
 0 & -1 & 0 & 0 & 0 & D_{2} &  \\
 3 & 2 & 0 & 0 & 0 & D_{3} &= \cB \\
 3 & 2 & 1 & 0 & 0 & D_{4} & =\hat S\\
 3 & 2 & -1 & 0 & 0 & D_{5} & \\
 3 & 2 & 0 & 1 & 1 & D_{6} & \\
 3 & 2 & 0 & -1 & 0 & D_{7} & \\
 3 & 2 & 0 & 0 & -1 & D_{8} & =H \\
 2 & 1 & 1 & 0 & 0 & D_{9} & =B_1 \\
 1 & 1 & 1 & 0 & 0 & D_{10} & =B_2\\
 1 & 0 & 1 & 0 & 0 & D_{11} & =B_3  \\
 0 & 0 & 1 & 0 & 0 & D_{12} & =B_4 \\
\hline
\end{array}
\label{eq:toric3}
\end{eqnarray}

Note that we have introduced a basis of independent toric divisors $\cB,\hat S,H$, and $B_i$. These $7$ divisors will span
a basis of independent divisors on a generic hypersurface $\tilde X_4$ embedded in the class $K = \sum_i D_i$, such that $h^{1,1}(\tilde X_4)=7$.\footnote{This is a consequence of 
the Lefschetz hyperplane theorem.}  
One realizes from \eqref{eq:toric3} that $S_{\rm b} = \cB \cdot \hat S$ is the $\mathbb{P}^2$  base 
of the K3-fibration. The normal bundle $N_{S_{\rm b}|\cB}$ is trivial $N_{S|\cB} = \mathcal{O}_{\mathbb{P}^{2}}$. 
In \eqref{eq:toric3} we also introduced the blow up 
divisors $B_i$ for the resolution of an $A_{4}$ singularity over $S_{\rm b}$. The to $S_{\rm b}$ associated divisor $\hat S$ in $\tilde X_4$ is 
given by $S = \hat S + B_1 + B_2 + B_3 + B_4$ as an $SU(5)$ version of \eqref{eq:shift}.  
The Hodge numbers of the Calabi--Yau fourfold can be computed as  
\begin{equation}
h^{1,1}(\tilde{X}_{4}) = 7,\qquad h^{2,1}(\tilde{X}_{4})=0,\qquad h^{3,1}(\tilde{X}_{4})=2148,\qquad \chi(\tilde{X}_{4}) = 12978.
\end{equation}
%

\subsubsection{Mori cone, resolutions and group theory}
\label{sec:degeneration}

The generators of the Mori cone for the Calabi--Yau fourfold $\tilde{X}_{4}$ can be obtained by the method described in \cite{Berglund:1995gd,Berglund:1996uy,Braun:2000hh}. 
Note that the generators of the Mori cone for a toric ambient space is generally different from the generators of the 
Mori cone for a Calabi--Yau fourfold hypersurface. In general, some of the triangulations for the ambient space are 
connected by a flop of a curve which is not inside the Calabi--Yau fourfold. In our case, we have 54 star-triangulations of the polyhedron \eqref{eq:toric3} from the origin. However some of them are connected by the flops of the curves which are not 
included in $\tilde{X}_{4}$. If the defining equation of a curve and the Calabi--Yau hypersurface $\tilde{X}_4$ cannot be 
satisfied simultaneously because they are elements of the Stanley-Reisner ideal, the flop of the curve is not a 
true flop in the Calabi--Yau fourfold $\tilde{X}_4$. This can be confirmed from the distinct intersection 
numbers of $\tilde{X}_4$ since different phases give different intersection numbers. 
In the derivation of the intersection numbers we use the star-triangulations ignoring the interior points in the facets. 
The presence of these points indicates the existence of point-like singularities in the ambient space.
Since the Calabi--Yau hypersurface does generically not intersect these singularities, 
a star-triangulation of the points in the polyhedron \eqref{eq:toric3}  yields a smooth Calabi--Yau hypersurface.
In our example \eqref{eq:toric3}, we find that the true number of the triangulations for $\tilde{X}_{4}$ is three.
The generators of the Mori cone for the three phases are: \begin{eqnarray}
\begin{array}{|ccc;{2pt/2pt}cccc|c|ccc;{2pt/2pt}cccc|c|ccc;{2pt/2pt}cccc|} \cline{1-7} \cline{9-15} \cline{17-23}
\multicolumn{7}{|c|}{\text{phase I}} & &\multicolumn{7}{c|}{\text{phase II}} & & \multicolumn{7}{c|}{\text{phase III}}\\
\cline{1-7} \cline{9-15} \cline{17-23}
\ell_{1} &\ell_{2} &\ell_{3} & \ell_{4} & \ell_{5} & \ell_{6} &  \ell_{7}  & &
\ell_{1} &\ell_{2} &\ell_{3} & \ell_{4} &\ell_{5} &\ell_{6} &\ell_{7}  & &
\ell_{1} &\ell_{2} &\ell_{3} & \ell_{4} &\ell_{5} &\ell_{6} &\ell_{7}  \\
\cline{1-7} \cline{9-15} \cline{17-23}
 0 & 0 & 0 & 1 & 0 & 0 & 0 && 0 & 0 & 0 & 1 & 0 & 0 & 0 && 0 & 0 & 0 & 1 & 0 & 0 & 0 \\
 0 & 0 & 0 & 0 & 1 & 0 & 0 && 0 & 0 & 0 & 0 & 1 & 0 & 0 && 0 & 0 & 0 & 0 & 1 & 0 & 0 \\
 -2 & -3 & 1 & 0 & 0 & 0 & 0 && -2 & -3 & 1 &0 &0 &0 & 0 && -2 & -3 & 1 &0 &0 & 0 & 0 \\
 1 & 0 & -2 & 0 & 0 & 1 & 0 && 1 & 0 & -2 & 0 &0 & 1 & 0 && 1 & 0 &-1 & 1 & 0 & 1 &-1 \\
 1 & 0 & 0 & 0 & 0 & 0 & 0 && 1 & 0 & 0 & 0 & 0 & 0 & 0 && 1 & 0 & 0 & 0 & 0 & 0 & 0 \\ 
 0 & 1 & 0 & 0 & 0 & 0 & 0 && 0 & 1 & 0 & 0 & 0 & 0 & 0 && 0 & 1 & 0 & 0 & 0 & 0 & 0 \\
 0 & 1 & 0 & 0 & 0 & 0 & 0 && 0 & 1 & 0 & 0 & 0 & 0 & 0 && 0 & 1 & 0 & 0 & 0 & 0 & 0 \\
 0 & 1 & 0 & 0 & 0 & 0 & 0 && 0 & 1 & 0 & 0 & 0 & 0 & 0 && 0 & 1 & 0 & 0 & 0 & 0 & 0 \\
 0 & 0 & 1 & 0 & 0 & -2 & 1 && 0 & 0 & 1 & 1 & 1 & -1 &-1 && 0 & 0 & 0 & 0 & 1 &-2 & 1 \\
 0 & 0 & 1 & 0 & 1 & 0 & -1 && 0 & 0 & 1 & -1 &0 & -1 & 1 && 0 & 0 & 0 & -2 & 0 & 0 & 1 \\
 0 & 0 & 0 & 1 & -1 & 1 & -1 && 0 & 0 & 0 & 0 &-2 & 0 & 1 && 0 & 0 & 0 & 0 & -2 & 1 & 0 \\
 0 & 0 & 0 & -1 & 0 & 0 & 1 && 0 & 0 & 0 & 0 & 1 & 1 &-1 && 0 & 0 & 1 & 1 & 1 & 0 &-1 \\
\cline{1-7} \cline{9-15} \cline{17-23}
\end{array} 
\hspace*{.5cm}
\label{Mori-generators1}
\end{eqnarray}
In the following we will indicate the phase by writing $\ell^{\rm (I)}_i$, $\ell^{\rm (II)}_i$, and $\ell^{\rm (III)}_i$
for the Mori vectors of the three phases respectively. Note that
\beq
   \ell^{\rm (I)}_1 = \ell^{\rm (II)}_1 = \ell^{\rm (III)}_1\ ,\qquad \ell^{\rm (I)}_2 = \ell^{\rm (II)}_2 = \ell^{\rm (III)}_2\ ,\qquad  \ell^{\rm (I)}_3 = \ell^{\rm (II)}_3 =\ell^{\rm (III)}_3 + \ell^{\rm (III)}_7 \ .
\eeq
{}From this identification we already realize that the phase III will be special, since its $\ell$-vectors appear more non-trivially in the last identification.

A subset of the generators of the Mori cone for each phase corresponds to effective curves and can be identified with weights in a representation 
of $SU(5)$ in the way described in the section \ref{subsec:mori}. 
One can read off the Cartan matrix $C_{ij}$ from the intersection numbers \eqref{dynkin_intersect}. 
By comparing it with the Cartan matrix of $SU(5)$, one can deduce that the blow up divisors $B_1, B_2, B_3, B_4$ 
correspond to the simple roots $e_{1}-e_{2}, e_{4}-e_{5}, e_{2}-e_{3}, e_{3}-e_{4}$ respectively. 
Here $e_{i}$ denotes the orthonormal basis of $\mathbb{R}^{5}$. Then, we can identify 
the generators of the Mori cone \eqref{Mori-generators1} 
with the weights of some representations of $SU(5)$ from the Cartan charges \eqref{U(1)charges}. 
Since we are interested in the extended relative Mori cone, the relevant generators for the phase I 
is $\ell_3^{\rm (I)}, \ell_4^{\rm (I)}, \ell_5^{\rm (I)}, \ell_6^{\rm (I)}, \ell_7^{\rm (I)}$. Among them the 
generators of the relative Mori cone \eqref{eq:relative_mori} have negative intersection 
numbers with the Cartan divisors. Hence, they are $\ell_4^{\rm (I)}, \ell_5^{\rm (I)}, \ell_6^{\rm (I)}, \ell_7^{\rm (I)}$ 
and $\ell_3^{\rm (I)}$ corresponds to the extended Dynkin node. The weights of the 
generators of the relative Mori cone for the phase I are 
\begin{equation}
\ell_{4}^{\rm (I)} \cong -e_{3}, \qquad \ell_{5}^{\rm (I)} \cong e_{3}+e_{4}, \qquad \ell_{6}^{\rm (I)} \cong -e_{1}+e_{2},\qquad \ell_{7}^{\rm (I)} \cong -e_{2}-e_{4}.
\label{eq:weight1}
\end{equation}
The extended node $\ell_3^{\rm (I)}$ corresponds to $e_1-e_5$. For the phase II, the generators of the relative Mori cone are $\ell_4^{\rm (II)}, \ell_5^{\rm (II)}, \ell_6^{\rm (II)}, \ell_7^{\rm (II)}$. They correspond to the following weights
\begin{equation}
\ell_{4}^{\rm (II)}  \cong e_{1}+e_{5}, \qquad \ell_{5}^{\rm (II)}  \cong-e_{2}+e_{3}, \quad  \ell_{6}^{\rm (II)}  \cong-e_{1} - e_{4}, \qquad \ell_{7}^{\rm (II)} \cong e_{2}+e_{4}.
\label{eq:weight2}
\end{equation}
Similarly, $\ell_{3}^{\rm (II)}$ corresponds to the extended Dynkin node $e_{1}-e_{5}$, and  $\ell_3^{\rm (II)}, \ell_4^{\rm (II)}, \ell_5^{\rm (II)}, \ell_{6}^{\rm (II)}, \ell_7^{\rm (II)}$ are 
the generators of the extended relative Mori cone. For the phase III, 
the generators of the relative Mori cone and their correspondence to the weights are 
\begin{equation}
\ell_{4}^{\rm (III)}  \cong -e_{4}+e_{5},  \qquad \ell_{5}^{\rm (III)}  \cong-e_{2}+e_{3},  \qquad \ell_{6}^{\rm (III)} \cong -e_{1}+e_{2}, \qquad \ell_{7}^{\rm (III)} \cong e_{1}+e_{4}.
\label{eq:weight3}
\end{equation} 
In this phase, we have a generator $\ell_{3}^{\rm (III)}$ which corresponds 
to a weight $-e_4 - e_5$ and does not shrink to a point in $X_4$. 
Hence $\ell_3^{\rm (III)}, \ell_4^{\rm (III)}, \ell_{5}^{\rm (III)}, \ell_6^{\rm (III)}, \ell_7^{\rm (III)}$ 
are the generators of the extended relative Mori cone. 

\begin{figure}[tb]
\begin{center}
\begin{tabular}{c}
\includegraphics[width=100mm]{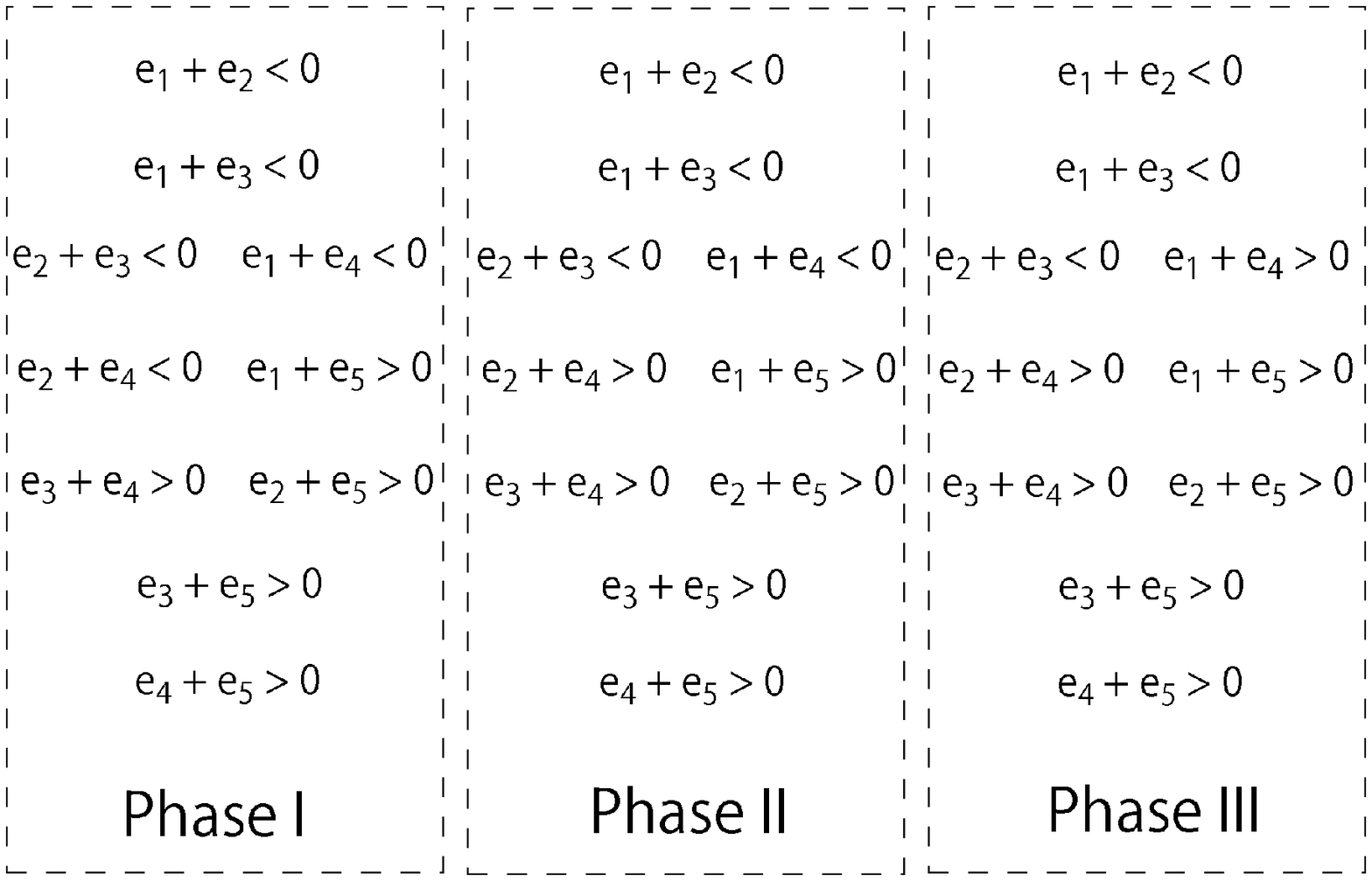} \\ 
\end{tabular}
\caption{The effective curves corresponding to the weights of {\bf 10} representation. The negative sign means that the negative of the weight corresponds to a effective curve.}
\label{fig:10phase1}
\end{center}
\end{figure}

So far we have determined the generators of the extended 
relative Mori cone. This implies that the weights \eqref{eq:weight1}--\eqref{eq:weight3} correspond to effective curves. 
Using the strategy of section \ref{subsec:mori} one can also determine whether other weights correspond to effective curves from
the relative Mori cone. Comparing \eqref{eq:weight1} and \eqref{id_weightsandell} we note that section~\ref{following_SU(5)} precisely discusses 
the phase I of the resolved Calabi-Yau fourfold \eqref{eq:toric3}. The identification of the effective curves was given in \eqref{effective5}. 
Following the same strategy also for phases II and III, one shows that for all the three phases one has 
\begin{equation}
e_{1} < 0,\qquad  e_2 < 0,\qquad e_3 < 0,\qquad e_4 > 0,\qquad e_5 > 0. \label{eq:5phase1}
\end{equation}
where we use the notation \eqref{notation_effective}. In section \ref{following_SU(5)} also the determination of the effectiveness of the $\bf{10}$ weights has 
been given for phase I. The result was given in \eqref{effective10}. Repeating the same analysis for the phases II and III one 
finds the result summarized in Figure~\ref{fig:10phase1}.

As also explained in section \ref{subsec:mori}, one next determines the weights which describe the curves fibering over the matter 
curves $\Sigma_{\bf 5}$ and $\Sigma_{\bf 10}$. This allows to determine the degeneration structure of the curves 
in the Cartan resolution divisors $D_i$ from the relative Mori cone. 
One considers the splits 
\bea
\Sigma_{\bf 10}: &\qquad& -({\rm simple}\;{\rm root}) = \overline{{\bf 10}}\;{\rm weight} + {\bf 10}\;{\rm weight}\  ,\\
\Sigma_{\bf 5}: &\qquad& -({\rm simple}\;{\rm root}) =  \bar{{\bf 5}}\;{\rm weight} + {\bf 5}\;{\rm weight}\ .
\eea 
For the phase I this was explained in detail in section \ref{following_SU(5)}. The result was that the weights corresponding 
to the matter curves $\Sigma_{\bf 10}$ and $\Sigma_{\bf 5}$ are 
\bea
\Sigma_{\bf 10}: &\quad &e_{1}-e_{5},\;e_{1}+e_{5},\;-e_{1}+e_{2},\;-e_{2}-e_{4},\;-e_{3}+e_{4},\;e_{3}+e_{4} \ ,    \label{eq:D5gens}\\
\Sigma_{\bf 5}:  &\quad& e_{1}-e_{5},\;-e_{1}+e_{2},\;-e_{2}+e_{3},\;-e_{3},\;e_{4}\, -e_{4}+e_{5},  \label{eq:A5gens}
\eea
as shown in \eqref{eq:D5gens_text} and \eqref{eq:A5gens_text}. 
The weights \eqref{eq:D5gens} form the extended Dynkin diagram of $D_{5}$ in Figure \ref{fig:Dynkin_phase1}. 
Similarly, the weights \eqref{eq:A5gens} form the extended $A_{5}$ Dynkin diagram also depicted 
in Figure \ref{fig:Dynkin_phase1}. The intersection numbers 
for both Dynkin diagrams can be calculated from the direct computation in appendix \ref{sec:direct_comp}. 
We will not need these intersection numbers in the following.

\subsubsection{Yukawa couplings at co-dimension three}

\begin{figure}[tb]
\begin{center}
\begin{tabular}{c}
\includegraphics[width=100mm]{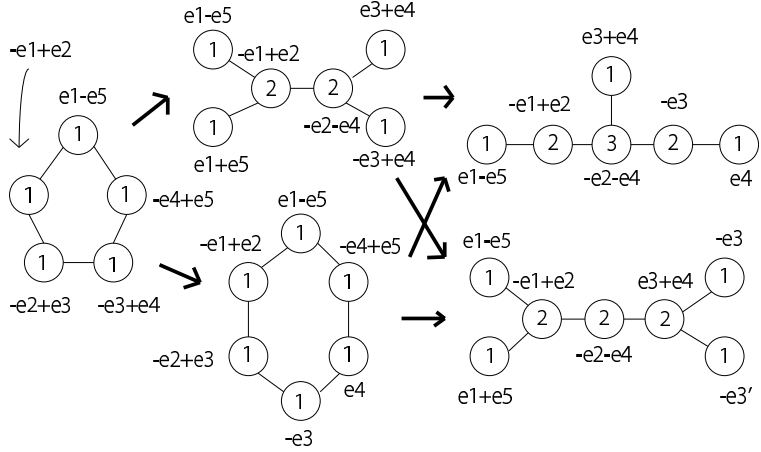} \\ 
\end{tabular}
\caption{The chain of the Dynkin diagrams for the phase I. The number in the nodes denotes the multiplicity. The intersection structure cannot be inferred by 
simple group theoretic arguments about the weights, but requires an inspection of the resolution geometry.}
\label{fig:Dynkin_phase1}
\end{center}
\end{figure}

We have studied the degeneration along the co-dimension-two 
singularity locus. The singularity further enhances along the 
$E_{6}$ and $D_{6}$ Yukawa points. At the Yukawa points, the curves 
of \eqref{eq:D5gens} and \eqref{eq:A5gens} further degenerate 
into smaller irreducible components. In this case, the degeneration 
generates the curves corresponding to {\bf 10}, $\overline{{\bf 10}}$ weights 
and also {\bf 5}, $\bar{{\bf 5}}$ weights from one Yukawa point. 
In general, when the singularity is enhanced to $G_p$ at the 
co-dimension-three point, our proposal is that the degeneration of the curves 
obeys the algebra of $G_p$. 
Namely, the further degeneration is possible only if the decompositions of the weights at $E_6$ and $D_6$ points obey the $E_6$ and $D_6$ algebra respectively. First, let us see the degeneration of the extended $D_{5}$ Dynkin diagram at the $E_{6}$ enhancement point. Since $e_{1}-e_{5},\; -e_{1}+e_{2},\; -e_{2}-e_{4},\; e_{3}+e_{4}$ correspond to the generators of the relative Mori cone, they do not degenerate further. Since ${\bf 5}$ or $\bar{{\bf 5}}$ weights can appear at the $E_6$ enhancement point, the negative simple root $-e_3+e_4$ in \eqref{eq:D5gens} can decompose as
\begin{equation}
-e_{3} + e_{4} = (-e_{3}) + (e_{4}).
\label{eq:decomposition_E6_2}
\end{equation}
Moreover, we have the decomposition
\begin{eqnarray}
e_{1} + e_{5} &=& -e_{2} - e_{3} - e_{4},\nn \\
&=& (-e_{2}-e_{4}) + (-e_{3}).
\label{eq:decomposition_E6}
\end{eqnarray}
We use the fact that $e_{1} + e_{2} + e_{3} + e_{4} + e_{5}$ is a singlet in $SU(5)$. The decomposition \eqref{eq:decomposition_E6_2} obviously obeys the $E_6$ algebra since the adjoint weight decomposes into a vector-like pair. In order to see that the decomposition \eqref{eq:decomposition_E6} obeys the algebra of $E_{6}$ but does not obey the algebra of $D_{6}$, one can consider the following decomposition
\begin{eqnarray}
E_{6} &\supset& SU(5) \times U(1)_{1} \times U(1)_{2}  \label{eq:E6}\\
{\bf 78} &\rightarrow& {\bf 1}_{0,0} + {\bf 1}_{0,0} + {\bf 1}_{-5,-3} + {\bf 1}_{5,3} + {\bf 24}_{0,0} \nn \\
 &&+ {\bf 5}_{-3,3} + \bar{\bf 5}_{3,-3} + {\bf 10}_{-1,-3} + \overline{\bf 10}_{1,3} + {\bf 10}_{4,0} + \overline{\bf 10}_{-4,0}, \nn \\[.1cm]
D_{6} &\supset& SU(5) \times U(1)_{1} \times U(1)_{2}  \label{eq:SO(12)}\\
\bf{66} &\rightarrow& {\bf 1}_{0,0} + {\bf 1}_{0,0} + {\bf 24}_{0,0}  \nn \\
&&+ {\bf 5}_{2,2} + {\bf 5}_{2,-2} + \bar{\bf 5}_{-2,2} + \bar{\bf 5}_{-2,-2} + {\bf 10}_{4,0} + \overline{\bf 10}_{-4,0}. \nn 
\end{eqnarray}
{}From the $E_{6}$ decomposition \eqref{eq:E6}, one can associate the $E_{6}$ algebra 
\begin{equation}
{\bf 10}_{4,0} \rightarrow \overline{\bf 10}_{1,3} + \bar{\bf 5}_{3,-3}
\end{equation}
to \eqref{eq:decomposition_E6}. However, one cannot associate the $D_{6}$ algebra 
to \eqref{eq:decomposition_E6} since the $U(1)$ charges are not conserved under the 
decomposition. Hence, this degeneration corresponds to the $E_{6}$ enhancement point. 
To summarize, we have the weights 
\begin{equation}
e_{1} - e_{5},\quad -e_{1}+e_{2},\quad -e_{2}-e_{4},\quad e_{3}+e_{4},\quad -e_{3},\quad e_{4},
\label{eq:E6weights}
\end{equation}
at the $E_6$ Yukawa point. The weights \eqref{eq:E6weights} form the $E_{6}$ Dynkin diagram depicted in 
Figure \ref{fig:Dynkin_phase1}. 
As noted in \cite{Esole:2011sm}, \eqref{eq:E6weights} does not form the `extended' 
$E_{6}$ diagram and the rank does not enhances at the $E_{6}$ Yukawa point. 

One can also do the same analysis for the degeneration of the extended $D_{5}$ 
Dynkin diagram at the a $D_{6}$ enhancement point. At the $D_{6}$ Yukawa point, 
the degeneration of \eqref{eq:decomposition_E6} is impossible since it does not 
satisfy the $D_{6}$ algebra. Hence, $e_{1} + e_{5}$ remains to be irreducible at 
the $D_{6}$ Yukawa point. On the other hand, $-e_{3} + e_{4}$ can decompose 
differently to \eqref{eq:decomposition_E6_2} as
\begin{equation}
-e_{3} + e_{4} = (-e_{3}) + (-e_{3}^{\prime}) + (e_{3} + e_{4}).
\label{eq:decomposition_D6}
\end{equation}
The decomposition \eqref{eq:decomposition_D6} is possible for the $D_{6}$ 
enhancement point since one has two $\bar{{\bf 5}}$ representations with 
different charges under $U(1)_{1}\times U(1)_{2}$ in the decomposition of 
$SO(12)$ as displayed in \eqref{eq:SO(12)}. The U(1) charge conservation corresponding to \eqref{eq:decomposition_D6} becomes
\begin{equation}
{\bf 24}_{0,0} \rightarrow \bar{\bf 5}_{-2,2} + \bar{\bf 5}_{-2,-2} + {\bf 10}_{4,0}.
\end{equation}
This is not allowed in the $E_6$ algebra. Note that our proposal is that one has to decompose the weights into the 
generators of the extended relative Mori cone as much as possible. 
Therefore, the degeneration at the $D_6$ Yukawa point should not stop 
at \eqref{eq:decomposition_E6_2} but proceeds further to \eqref{eq:decomposition_D6} since both $-e_3$ and $e_3+e_4$ are the generators of the extended relative Mori cone.  
To summarize, we have the weights
\begin{equation}
e_{1}-e_{5},\quad e_{1}+e_{5},\quad -e_{1}+e_{2},\quad -e_{2}-e_{4},\quad e_{3}+e_{4},\quad -e_{3},\quad -e_{3}^{\prime}.
\label{eq:D6gens}
\end{equation}
at the $D_6$ Yukawa point. The curves associated to \eqref{eq:D6gens} form the extended $D_{6}$ 
Dynkin diagram in Figure \ref{fig:Dynkin_phase1}.

The chains of the Dynkin diagrams for the phase II and III can be computed in a similar manner and they 
are depicted in Figures \ref{fig:Dynkin_phase2} and \ref{fig:Dynkin_phase3}.

\subsubsection{$G_4$-flux and chirality }

In this section we test the chirality formula \eqref{eq:chirality2} for the matter fields 
in the ${\bf 10}$ and $\bar{{\bf 5}}$ representation from the F-theory compactifications on 
the Calabi--Yau fourfold \eqref{eq:toric4}. 
The necessary information is the $G_4$ flux and the mater surfaces $S_{{\bf R}}$ 
for the ${\bf 10}$ and $\bar{{\bf 5}}$ matter fields. Hereafter, we also focus on the phase I. 
A construction of the $G_4$ flux and matter surfaces for $SU(5)$ examples
can be also be found in \cite{Marsano:2011hv}.

First we determine $G_4$ flux. 
We consider the $G_4$ flux constructed from the intersection of the divisors 
of $\tilde{X}_4$. In order to preserve four-dimensional Poincar\'e invariance 
and the $SU(5)_{GUT}$ symmetry, the $G_4$ flux should satisfy the conditions \eqref{eq:G-condition}. 
We find such $G_4$ flux from the expansion by the general intersection of the divisors. 
Without any constraint, we have $7\times 8/2 = 28$ generators of such surfaces. 
However, not all of them are independent. First of all, we have the constraints from the 
Stanley-Reisner ideal of the toric ambient space for $\tilde X_{4}$. 
For the phase I, the Stanley-Reisner ideal is 
\begin{eqnarray}
SR&=&\{D_{2}D_{10}, D_{3}D_{9}, D_{3}D_{10}, D_{3}D_{11}, D_{3}D_{12}, 
  D_{4}D_{5}, D_{4}D_{11}, D_{4}D_{12}, D_{1}D_{9}, D_{1}D_{11}, D_{5}D_{9}, \nonumber\\
&&D_{5}D_{10}, D_{5}D_{11}, D_{5}D_{12},D_{9}D_{12}, 
D_{1}D_{2}D_{3}, D_{1}D_{2}D_{4}, D_{6}D_{7}D_{8}\}.
\label{eq:SR1}
\end{eqnarray} 
Hence we have 15 constraints for the surfaces and all of them are independent. 
We have another constraints coming from the incompatibility between the Stanley-Reisner 
ideal and the Calabi--Yau hypersurface equation. 
Those constraints are 
\begin{equation}
D_{1}D_{3}, D_{1}D_{4}, D_{2}D_{3}, D_{2}D_{4}, D_{2}D_{9}.
\label{eq:SR2}
\end{equation} 
However, not all the constraints from \eqref{eq:SR1} and \eqref{eq:SR2} are independent. 
There are actually 19 independent constraints in total, so the number of the 
true generators for the expansion of surfaces is $28 - 19 = 9$. 
We choose
\begin{equation}
B_4^{2},\;B_3 \cdot B_4,\; B_3^{2},\; B_2\cdot B_3,\; B_2^{2},\; B_4\cdot H, H^{2},\; H\cdot \hat{S},\; \cB\cdot H
\end{equation}
Then, the general expansion of the $G_4$ flux by the nine independent surfaces is  
\begin{equation}
G_{4} = \alpha_{1}B_4^{2} + \alpha_{2}B_3\cdot B_4 + \alpha_{3}B_3^{2} + 
        \alpha_{4}B_2\cdot B_3 + \alpha_{5}B_2^{2} + \alpha_{6}B_4\cdot H  + 
        \alpha_{7}H^{2} + \alpha_8 H \cdot \hat{S} + \alpha_{9}\cB\cdot H.
\label{eq:G}
\end{equation}
The condition \eqref{eq:G-condition} reduces the nine parameters to one parameter,
\begin{equation}
G_{4}=\beta(8B_2 \cdot B_3 -4 B_3\cdot B_3 - 2 B_3 \cdot B_4 +3 B_4^{2} + 9 B_4 \cdot H),
\label{eq:G-1}
\end{equation} 
where $3\beta=\alpha_2$. We choose $\beta$ such that all the coefficients are integers. 

Let us turn to the matter surface $S_{{\bf 10}}$ and $S_{\overline{{\bf 10}}}$. 
{}From Figure \ref{fig:Dynkin_phase1}, the matter surface $S_{{\bf 10}}$ 
corresponds to the weights $e_{1}+e_{5}, e_{3}+e_{4}$, and  
$S_{\overline{{\bf 10}}}$ corresponds to the weight $-e_{2}-e_{4}$. 
The class of the curves corresponding to $e_{1}+e_{5}, e_{3}+e_{4}$ and $-e_{2}-e_{4}$ 
can be determined from the intersection numbers \eqref{Mori-generators1}. 
For example, let us consider a curve $\ell_7$ corresponding to the weight $-e_{2}-e_{4}$. 
Since our final interest is the matter surface, we have to pull out a divisor, which 
intersects $S_b$ in a curve, from the triple intersection representing the curve. 
We make the Ansatz $(\mu H + \nu S )$ for this divisor. Furthermore, when $\ell_i$ has a negative 
intersection number with a divisor $B_i$, the triple intersection representing 
$\ell_i$ has a component $B_i$. Hence, we can make a general Ansatz for 
$\ell_7$, $\ell_7=a B_2 \cdot B_3 \cdot (\mu H + \nu S)$. The parameter $a$ can be determined 
from the intersection numbers and the result is
\begin{equation}
  {\bf\overline{10}}: \quad  \ell_7^{\rm (I)} \cong -e_{2} - e_{4}\quad \rightarrow \quad \tfrac{1}{3\mu} B_2 \cdot B_3 \cdot (\mu H + \nu S).
\end{equation}
Note that we solve only for $a$ and not for $\mu$ and $\nu$ in this process. 
In order to obtain a matter surface corresponding to the weight $-e_{2}-e_{4}$, 
one has to pull out a divisor in $\cB$ with a correct multiplicity. The correct 
multiplicity can be determined from the intersection with the matter curve 
$\Sigma_{{\bf 10}}$ \eqref{eq:multiplicity}. Since the {\bf 10} matter curve lies in the class $c_{1}(\cB)|_{S_{\rm b}}$, 
\eqref{eq:multiplicity} becomes 
\begin{equation}
S_{\rm b} \cdot_{\cB} c_{1}(\cB) \cdot_{\cB} \mathcal{D} = 1.
\label{eq:pullout1}
\end{equation}
In this case, the divisor can be chosen to be $\mathcal{D}=\frac{1}{3\mu}(\mu H + \nu S)$. 
Hence the matter surface $S_{{{\bf \overline{10}}}}$ corresponding to the weight $-e_{2}-e_{4}$ is 
\begin{equation}
S_{{\bf \overline{10}}} = B_2 \cdot B_3\ . 
\label{eq:matter_surface10-1}
\end{equation}
{}From this expression of the matter surface $S_{\overline{{\bf 10}}}$, and 
the $G_4$ flux \eqref{eq:G-1} one can compute the chirality for 
the ${\bf 10}$ matter fields. The chirality formula \eqref{eq:chirality2} becomes 
\beq
 \chi(\overline{{\bf 10}}) = \int_{S_{ \overline{\bf 10}}} G_{4} = 3 \cdot 54 \beta.
\label{eq:chirality2-ex1}
\eeq
where we have used the intersection numbers of $\tilde X_4$.

For the chirality formula of the $\bar{{\bf 5}}$ matter, one has to determine 
the matter surface $S_{\bar{{\bf 5}}}$. We consider a curve corresponding to the weight $-e_{3}$ 
which is one of the generators of the relative Mori cone for the phase I. From the intersections 
one can determine the class of the curve $\ell_4$ corresponding to the weight $-e_{3}$. We 
also make an Ansatz such that the triple intersection has a component $(\mu H + \nu S )$ 
which is then being dropped in the determination of the matter surface $S_{\bar{{\bf 5}}}$. Moreover, since $\ell_4$ has a 
negative intersection number with $B_4$, we can make a general Ansatz 
$\ell_4 = (\sum_A a^A D_A) \cdot B_4 \cdot (\mu H + \nu S)$, where $\sum_A a^A D_A$ is a general linear combination 
of the divisors in \eqref{eq:toric3}. The parameters $a_i$ can be determined from the 
intersection numbers, such that $\ell_4$ is represented by
\begin{equation} 
{\bf\overline{5}}: \quad  \ell_4^{\rm (I)} \cong -e_{3} \quad \rightarrow \quad
\tfrac{1}{24 \mu}\big( B_2 -B_3 + 9H \big)\cdot B_{4} \cdot (\mu H + \nu S),
\end{equation}
where we choose a special representative in the solutions just for the simplicity of the expression. 
In order to obtain the matter surface $S_{\bar{{\bf 5}}}$, 
one has to pull out a divisor $\cD$ which satisfies the condition
\begin{equation}
S \cdot_{\cB} (8c_{1}(\cB) - 5S) \cdot_{\cB} \mathcal{D} = 1,
\label{eq:pullout2}
\end{equation}
where $(8c_{1}(\cB) - 5S)|_{S}$ is a class of the $\bar{{\bf 5}}$ matter curve. 
By using the condition \eqref{eq:pullout2}, one finds $\mathcal{D}$ being of the form 
$\mathcal{D}=\frac{1}{24\mu}(\mu H + \nu S)$. 
Hence the matter surface $S_{\bar{{\bf 5}}}$ is 
\begin{equation} \label{Sbar5}
S_{\bar{{\bf 5}}} = (B_2 - B_3 + 9 H)\cdot B_{4}.
\end{equation}
Therefore, the chirality formula for the $\bar{{\bf 5}}$ matter becomes
\beq
 \chi({\bf \bar 5}) = \int_{S_{\bf \bar 5}} G_{4} = - 3 \cdot 54 \beta \ ,
\label{eq:chirality2-ex1-2}
\eeq
where we have inserted the matter surface \eqref{Sbar5} and the flux \eqref{eq:G-1}, 
and used the intersection numbers of $\tilde X_4$.
{}From \eqref{eq:chirality2-ex1} and \eqref{eq:chirality2-ex1-2} we find 
the relation $\chi({{\bf 10}}) = -\chi(\overline{{\bf 10}}) = \chi({{\bf \bar 5}})$, 
which is consistent with the anomaly conditions for $SU(5)$ gauge theories. 
Note that we did not discuss the quantization of $\beta$ appearing 
in \eqref{eq:chirality2-ex1} and \eqref{eq:chirality2-ex1-2}. 
This can be done by investigating the integrality properties 
of the basis used in \eqref{eq:G-1}, and satisfying the constraint \eqref{quantization}.

\subsubsection{Relation to three-dimensional Chern-Simons term}

One can also see that the chirality \eqref{eq:chirality2-ex1} and \eqref{eq:chirality2-ex1-2} 
can be obtained from the formula \eqref{eq:chirality3d}. 
The sign in \eqref{eq:sign_kahler} is determined from the relative Mori cone. 
The effectiveness of the curves corresponding to {\bf 10} weights is depicted in the 
first column of Figure \ref{fig:10phase1}. The effectiveness of the curves corresponding to ${\bf 5}$ 
weights is \eqref{eq:5phase1}. The $U(1)_i$ charges $(q_f)_i$ of each weight can be determined from \eqref{U(1)charges}. 
By inserting all the information, the formula \eqref{eq:chirality3d} for the 
Calabi--Yau fourfold \eqref{eq:toric3} in the phase I is 
\begin{align} \label{eq:3d-ex1}
&\Theta_{23} = -\chi({\bf 10}),& \qquad  &
\Theta_{24} = \tfrac{1}{2} \chi({\bf 10}) + \tfrac{1}{2}\chi(\bar{{\bf 5}}), &\\
&\Theta_{33} = \chi(\bar{{\bf 5}}),&\qquad &
\Theta_{44} = -\chi({\bf 10}),& \nn \\
&\Theta_{13} = \tfrac{1}{2}\chi({\bf 10}) - \tfrac{1}{2}\chi(\bar{{\bf 5}}),&\qquad & 
\Theta_{34} = \tfrac{1}{2}\chi({\bf 10}) - \frac{1}{2}\chi(\bar{{\bf 5}}) & \nn  \\
&\Theta_{11} = -\chi({\bf 10}) + \chi(\bar{{\bf 5}}),& \qquad &
\Theta_{22} = \chi({\bf 10}) - \chi(\bar{{\bf 5}})& \nn
\end{align} 
and the other components are zero. Using the intersection numbers with the 
$G_4$-flux \eqref{eq:G-1}, one can explicitly compute the components $\Theta_{ij}$. 
One finds 
\begin{equation}
\Theta_{23} = 3 \times 54 \beta,\qquad
\Theta_{24} = -3 \times 54 \beta, \qquad
\Theta_{33} = -3 \times 54 \beta, \qquad
\Theta_{44} = 3 \times 54 \beta, \label{eq:theta-ex1-1}
\end{equation}
with all the others being zero. By inserting the explicit numbers \eqref{eq:theta-ex1-1} into \eqref{eq:3d-ex1}, one obtains
\begin{equation}
\chi({\bf 10}) = -3 \times 54 \beta,\qquad \chi(\bar{{\bf 5}}) =  -3 \times 54 \beta.
\label{eq:3d_matching1}
\end{equation}
This precisely matches the chirality obtained from integrating $G_4$-fluxes over the matter surfaces \eqref{eq:chirality2-ex1} and \eqref{eq:chirality2-ex1-2}.

\subsubsection{Comparison with spectral cover}

We compare the results of the proposed chirality formula \eqref{eq:chirality2-ex1} 
and \eqref{eq:chirality2-ex1-2} with the spectral cover computation. 
The chirality formula for the matter in {\bf 10} representation 
is \cite{Curio:1998vu, Diaconescu:1998kg}
\begin{equation}
n_{{\bf 10}} = -\lambda \eta \cdot (5K_{S_{\rm b}} + \eta),
\label{eq:chirality}
\end{equation}
where $S_{\rm b}$ is a surface wrapped by the $SU(5)$ brane. 
The divisor $\eta$ in $S_{\rm b}$ is related to a normal 
bundle $N_{S_{\rm b}|\cB}$ by the equation
\begin{equation}
c_{1}(N_{S_{\rm b}|\cB}) = 6 K_{S_{\rm b}} + \eta\, .
\label{eq:normal_eta}
\end{equation}
Finally $\lambda$ is related to the $G_4$ flux 
and takes values in $\mathbb{Z}+\frac{1}{2}$ \cite{Curio:1998bva}. 
Note that the chirality formula \eqref{eq:chirality} is a 
local expression on the surface $S_{\rm b}$, In constrast, the chirality formula \eqref{eq:chirality2} 
is defined by integration over the whole resolved Calabi--Yau fourfold $\tilde{X}_4$. In the current example, 
we chose $S_{\rm b} = \mathbb{P}^{2}$ and $N_{S_{\rm b}|\cB} = \mathcal{O}_{\mathbb{P}^{2}}$. Hence, 
$\eta = 18 H_{\mathbb{P}^{2}}$ where $H_{\mathbb{P}^{2}}$ is a hyperplane class of $\mathbb{P}^{2}$. 
The chirality formula \eqref{eq:chirality} becomes 
\begin{equation}
  n_{{\bf 10}} = - \lambda (18H_{\mathbb{P}^{1}} \cdot_{\mathbb{P}^{2}} 3H_{\mathbb{P}^{2}}) = -54 \lambda
\label{eq:chirality1-ex1}
\end{equation}
One also finds the same number for the matter in the $\bar{{\bf 5}}$ representation. 
Comparing the spectral cover result \eqref{eq:chirality1-ex1} with the 
result \eqref{eq:chirality2-ex1}, we find agreement if we identify $3\beta = \lambda$. 
Since the chirality for the {\bf 10} matter fields and the chirality of $\bar{{\bf 5}}$ 
matter fields are the same, the same identification is 
satisfied for the comparison between \eqref{eq:chirality1-ex1} and \eqref{eq:chirality2-ex1}.

\subsection{A $U(1)$-restricted hypersurface with $SU(5) \times U(1)$ gauge group}

In the previous subsection we have exemplified the use of the Mori cone generators and 
their connection to a non-Abelian gauge group $SU(5)$ on a single stack of 7-branes. 
We now present a second example where an additional geometrically massless $U(1)$ is 
present. In an $SU(5)$ model this $U(1)$ can be identified with the $U(1)_X$ in $SO(10)$. 
The geometric construction presented here corresponds to the  
$U(1)$ restricted Tate model introduced in~\cite{Grimm:2010ez}. 

The Calabi-Yau fourfold which we will construct has a base which is a $\mathbb{P}^3$
blown up along a curve into a surface $S_{\rm b}$. The gauge group on $S_{\rm b}$ is 
engineered to be $SU(5)$.  
The points on edges of the polyhedron for the $U(1)_{X}$ restricted Tate model for such a setup
are:  
\begin{eqnarray}
\begin{array}{|ccccc|rl|} \hline
\multicolumn{5}{|c|}{\text{points}} &\ \text{divisor}& \hspace*{.1cm} \text{basis}\hspace*{.3cm} \\ \hline \hline
 -1 & 0 & 0 & 0 & 0 & D_{1} & \\
 0 & -1 & 0 & 0 & 0 & D_{2} &  \\
 3 & 2 & 0 & 0 & 0 & D_{3} &= \cB \\
 3 & 2 & 1 & 1 & 1 & D_{4} & =H \\
 3 & 2 & -1 & 0 & 0 & D_{5} & \\
 3 & 2 & 0 & -1 & 0 & D_{6} & \\
 3 & 2 & 0 & 0 & -1 & D_{7} & \\
 3 & 2 & 1 & 1 & 0 & D_{8} & =\hat{S} \\
 2 & 1 & 1 & 1 & 0 & D_{9} & =B_1 \\
 1 & 1 & 1 & 1 & 0 & D_{10} & =B_2\\
 1 & 0 & 1 & 1 & 0 & D_{11} & =B_3  \\
 0 & 0 & 1 & 1 & 0 & D_{12} & =B_4 \\
 -1 & -1 & 0 & 0 & 0& D_{13} & =X \\
\hline
\end{array}
\label{eq:toric4}
\end{eqnarray}
Note that the inclusion of the last vertex corresponding to $X= D_{13}$ 
enforces $a_6 = 0$ in the standard Tate constraint. Here $a_6$ is the coefficient of 
the $z^6$ term, with $z=0$ being the base $\cB$. This method can be quite generally  
applied to obtain a geometrically massless $U(1)$'s in 
the four-dimensional spectrum \cite{Grimm:2010ez}.

In \eqref{eq:toric4} we have introduced the independent divisors $\cB,H,\hat S,B_i$ and $X$.
Note that $B_{1}, \cdots, B_{4}$ are the exceptional divisors resolving the $A_{4}$ singularity. 
$X$ originates from the resolution of an $SU(2)$ singularity along a curve outside $S_{\rm b}$.
As in \eqref{eq:shift} we introduce a divisor $S = \hat{S} + B_1 + B_2 + B_3 +B_4$. 
The Hodge numbers of the Calabi--Yau fourfold $\tilde{X}_{4}$ are 
\begin{equation}
h^{1,1}(\tilde{X}_{4})=8,\;\; h^{2,1}(\tilde{X}_{4})=0,\;\; h^{3,1}(\tilde{X}_{4})=1020,\;\; \chi(\tilde{X}_{4})=6216.
\end{equation}
In the following we will determine the Mori vectors and follow the resolution process. This will 
allow us to determine the net chirality induced by an F-theory compatible flux.

\subsubsection{Mori cone, resolutions and group theory}
\label{sec:degeneration2}

We begin our analysis with the determination of the Mori cone and 
its relation to the group theory of $SU(5) \times U(1)$.
Restricting to star-triangulations of the polyhedron \eqref{eq:toric4} including the origin, and 
using the method described in appendix \ref{sec:mori_cone}, 
we find twelve phases for the hypersurface. 
As in the previous example we use the star-triangulations 
ignoring the interior points in the facets.
For simplicity we will focus on one phase in the following. 
The generators of the Mori cone for the phase are 
\begin{eqnarray}
\begin{array}{|c c c;{2pt/2pt} c c c c c|}
\hline
\ell_1 & \ell_2 & \ell_3 & \ell_4 & \ell_5 & \ell_6 & \ell_7 & \ell_8\\
\hline
0 & 0 & 0 & 0 & 0 & 0 & 0 & 1 \\
0 & 0 & 0 & 1 & 0 & 0 & -1 & 1 \\
-3 & -1 & 1 & 0 & 0 & 0 & 0 & 0 \\
0 & 1 & 0 & 0 & 0 & 0 & 0 & 0 \\ 
1 & 0 & 0 & 0 & 0 & 0 & 0 & 0 \\
1 & 0 & 0 & 0 & 0 & 0 & 0 & 0 \\
0 & 1 & 0 & 0 & 0 & 0 & 0 & 0 \\
1 & -1 & -2 & 0 & 0 & 1 & 0 & 0 \\
0 & 0 & 1 & 0 & 1 & -2 & 0 &0 \\
0 & 0 & 1 & 1 & -1 & 0 & 0 & 0 \\
0 & 0 & 0 & -1 & -1 & 1 & 1 & 0 \\
0 & 0 & 0 & 0 & 1 & 0 & -1 & 0 \\
0 & 0 & 0 &0 & 0 & 0 & 1 & -1 \\
\hline
\end{array}
\label{eq:mori4}
\end{eqnarray}

The singular limit $\tilde{X}_4 \rightarrow X_4$ is the limit in which $B_{1},\ldots, B_{4}$ collapse 
to the surface $S_{\rm b}$ and $X$ collapses to a curve. 
In this limit, each curve $\ell_{4}, \ell_{5}, \ell_{6}, \ell_{7}, \ell_{8}$ shrinks to a point in $X_4$. 
Hence, they are the generators of the relative Mori cone. 
Note that the Cartan charges of $SU(5)$ are the same as the ones of \eqref{Mori-generators1}. 
Furthermore, $\ell_3$ intersects with the Cartan divisors. 
Hence, the generators of the extended relative Mori cone are $\ell_{3}, \ell_{4}, \ell_{5}, \ell_{6}, \ell_7, \ell_{8}$.

Some of the weights are charged under the new $U(1)$ which originates from the 
reduction along the Poincar\'e dual two-form of $X$. As for the 
Cartan divisor of the other $U(1)$, we use the divisor \cite{Grimm:2010ez}
\begin{equation}
B_{5} = X - B -[c_1(\cB)],
\label{eq:U(1)}
\end{equation}
where $[c_1(\cB)] = H + D_5 + D_6 + D_7 + S$.\footnote{Here we use $[c_1(\cB)]$ including $S$ instead of $\hat S$. More precisely, 
one could also write $\pi^* c_{1}(\cB)$. Any modification by a shift with blow-up divisors will just result in 
a change of basis in the following discussion.} This redefinition is required to ensure that the intersection 
numbers satisfy the vanishing condition \eqref{vanish_intersect2} in this basis.
The Poincar\'e dual two-from $\omega_5$ of $B_5$ is used for the dimensional 
reduction to obtain the $U(1)_X$ gauge field $A_X$ as $C_3 = A_X \wedge \omega_5$. 
From the intersection between $B_i$ with $i=1,\cdots, 4$, and $B_\Lambda$ with $\Lambda=1,\cdots, 5$, 
one obtains a part of the Cartan matrix of $SO(10)$ as \cite{Krause:2011xj}
\begin{eqnarray}
B_i \cdot B_\Lambda \cdot D_\alpha \cdot D_\beta = \left(
\begin{array}{ccccc}
-2 & 1 & 0 & 0 & 0 \\
1 & -2 & 1 & 0 & 0 \\
0 & 1 & -2 & 1 & 1 \\
0 & 0 & 1 & -2 & 0
\end{array}
\right)
\cB \cdot S \cdot D_\alpha \cdot D_\beta \ .
\label{eq:SO(10)_Cartan}
\end{eqnarray}
Since $B_5 \cdot B_5$ does not localize on the surface $S$, the component $C_{55}$ of \eqref{dynkin_intersect} does not reproduce the $5-5$ 
component of the $SO(10)$ Cartan matrix. This is consistent with a four-dimensional theory 
with gauge group $SU(5)_{GUT} \times U(1)_X$ rather then $SO(10)$, and 
matter representations originating from an underlying $SO(10)$.
This precisely occurs in the $U(1)_X$ restricted model as 
a global realization of the $(4+1)$ split 
spectral cover model considered in \cite{Marsano:2009gv, Blumenhagen:2009yv}.
Indeed, we can identify the weights corresponding to the generators of the Mori cone \eqref{eq:mori4} 
with weights in the representation of $SO(10)$ when we identify the root corresponding to the 
Cartan divisor $B_5$ with $e_4+e_5$ which is one of the simple roots of $SO(10)$. 

Motivated by the intersections \eqref{eq:SO(10)_Cartan}, we identify the Cartan divisors $B_{\Lambda}, (\Lambda = 1, \cdots, 5)$ with $e_1-e_2, e_4-e_5, e_2-e_3, e_3-e_4, e_4+e_5$, which are the simple roots of $SO(10)$. From the Cartan charges in \eqref{eq:mori4}, one can also identify the curves $\ell_{4}, \ell_{5}, \ell_{6}, \ell_{7}, \ell_{8}$ with weights in the representation of $SO(10)$
\begin{eqnarray}
\ell_4 &\cong& -\tfrac{1}{2}e_1-\tfrac{1}{2}e_2 + \tfrac{1}{2}e_3+\tfrac{1}{2}e_{4}-\tfrac{1}{2}e_5 = +e_3+e_4-e_{\Sigma},\nonumber\\
\ell_5 &\cong& \tfrac{1}{2}e_1-\tfrac{1}{2}e_2 + \tfrac{1}{2}e_3-\tfrac{1}{2}e_{4}+\tfrac{1}{2}e_5 =-e_2-e_4+e_{\Sigma},\nonumber\\
\ell_6 &\cong& -e_1+e_2,\label{eq:gen_mori4}\\
\ell_7 &\cong& \tfrac{1}{2}e_1+\tfrac{1}{2}e_2 - \tfrac{1}{2}e_3+\tfrac{1}{2}e_{4}+\tfrac{1}{2} e_5 =-e_3+e_{\Sigma},\nonumber\\
\ell_8 &\cong& -\tfrac{1}{2}e_1-\tfrac{1}{2}e_2 - \tfrac{1}{2}e_3 - \tfrac{1}{2}e_{4}-\tfrac{1}{2}e_5 =-e_{\Sigma},\nonumber
\end{eqnarray}
where $e_{\Sigma}$ denotes $\tfrac{1}{2}(e_1+e_2+e_3+e_4+e_5)$. In terms of $SU(5)$, $\ell_4$ corresponds to a weight in the ${\bf 10}$ representation, $\ell_5$ corresponds to a weight in the ${\bf \overline{10}}$ representation, and $\ell_7$ corresponds to a weight in the ${\bf \bar{5}}$ representation. They are identical to the ones in the phase I of the first example \eqref{eq:weight1}. In the $U(1)_X$ restricted model, we can further understand their $SO(10)$ origins. The weights corresponding to $\ell_4$ and $\ell_7$ come from the ${\bf 16}^{\prime}$ representation and the weight corresponding to $\ell_5$ comes from the ${\bf 16}$ representation of $SO(10)$. We can also consider a weight corresponding to a curve $\ell_7 + \ell_8$
\begin{equation}
\ell_7 + \ell_8 \cong -e_3.
\end{equation}
In $SU(5)$ this curve corresponds to a weight of the ${\bf \bar{5}}$ representation. Its $SO(10)$ origin 
is a weight of the ${\bf 10}$ representation since $\pm e_a \; (a=1, \cdots, 5)$ are the weights 
of the ${\bf 10}$ representation of $SO(10)$. Therefore, we have two types of ${\bf \bar{5}}$
 representations originating from the ${\bf 16}^{\prime}$ and ${\bf 10}$ representation of $SO(10)$. We also have a 
 singlet field associated with the weight $-e_{\Sigma}$ which corresponds to the curve $\ell_8$. In addition, the generator $\ell_3$ corresponds to the extended weight $e_1-e_5$ up to a 
term $e_{\Sigma}$ which is a singlet in $SU(5)$. This curve does not shrink 
in the singular limit and is an additional generator for the extended relative Mori cone.

{}From the relative Mori cone, one can determine the resolution 
structure. Since the $SU(5)$ Cartan charges of the 
$\ell_3, \ell_{4}, \ell_{5}, \ell_{6}, \ell_{7}$ are the 
same as the ones of the phase I in the first
example \eqref{eq:weight1}, the resolution of the chains 
$A_4 \rightarrow D_5 \rightarrow E_6, D_6$ and 
$A_4 \rightarrow A_5 \rightarrow E_6, D_6$ are essentially the same except for the singlet term $e_{\Sigma}$. 
However, we have other chains $A_4 \rightarrow A_5 \rightarrow A_6$ and $A_4 \rightarrow A_5^{\prime} \rightarrow A_6$. 
The $A_6$ singularity enhancement appears as a point where the two $A_5$ and $A_5^{\prime}$ 
singularity enhancement loci meet. Let us focus on the resolution along these chains. 
From the generators of the relative Mori cone, we have identified the 
two ${\bf \bar{5}}$ matter fields with $\ell_7$ and $\ell_7+\ell_8$. 
Therefore, the decompositions of the negative simple roots along the two $A_5$  
and $A_5^{\prime}$ singularity enhancement loci are 
\begin{eqnarray}
-e_3 + e_4 &=& \big(-e_3 + e_{\Sigma}\big) + \big(e_4 - e_{\Sigma}\big),\label{eq:SU(6)_1}\\
-e_3+e_4 &=& (-e_3) + (e_4). \label{eq:SU(6)_2}
\end{eqnarray}
Indeed, one can reconstruct the weights $e_4 - e_{\Sigma}$ and $e_4$ from the generators of the relative Mori cone
\begin{eqnarray}
e_4 - e_{\Sigma} &\cong& \ell_4 + \ell_7 + \ell_8,\\
e_4 &\cong& \ell_4 + \ell_7
\end{eqnarray} 
Hence, both weights correspond to the effective curves in the relative Mori cone. 
 
\begin{figure}
\begin{center}
\begin{tabular}{c}
\includegraphics[width=100mm]{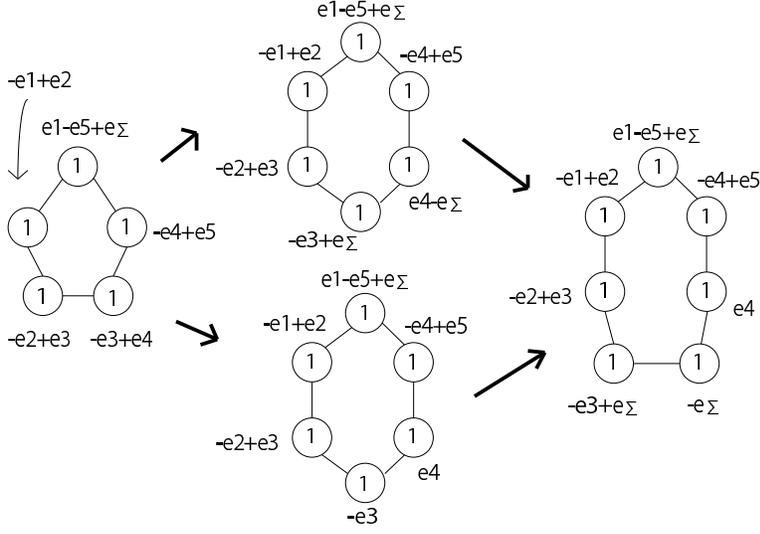} \\ 
\end{tabular}
\caption{The chain of the Dynkin diagrams for $A_4 \rightarrow A_5 \rightarrow A_6$ and 
$A_4 \rightarrow A_5^{\prime} \rightarrow A_6$. $e_\Sigma$ denotes the singlet weight $\frac{1}{2}(e_1+e_2+e_3+e_4+e_5)$. 
The number in the node denotes the multiplicity.}
\label{fig:Dynkin_A7-1}
\end{center}
\end{figure}

At the $A_6$ singularity enhancement points a further degeneration occurs in 
accord with the $SU(7)$ algebra. The ${\bf 5}$ and ${\bf \bar{5}}$ weights arising from 
different $SO(10)$ representations are located at the $A_6$ points. From the 
decomposition \eqref{eq:SU(6)_1}, $-e_3 + e_{\Sigma}$ cannot further 
decompose into smaller pieces since it is already a generator of the relative 
Mori cone. However, $e_4 - e_{\Sigma}$ can decompose as 
\begin{equation}
e_4 - e_{\Sigma} = (e_4) + \big(- e_{\Sigma} \big)\ .
\label{eq:SU(7)_1}
\end{equation}
The consistent degeneration requires that \eqref{eq:SU(6)_2} decomposes as 
\begin{equation}
-e_3 = (-e_3 + e_{\Sigma}) + \big(-e_{\Sigma} \big)\ , \label{eq:SU(7)_2}
\end{equation}
and $e_4$ remains unchanged. We can see that the decompositions \eqref{eq:SU(7)_1} 
and \eqref{eq:SU(7)_2} also obey the $SU(7)$ algebra. To summarize, we 
have the curves corresponding to the weights 
\begin{equation}
-e_1+e_2,\qquad -e_2+e_3, \qquad -e_3+e_{\Sigma},\qquad -e_{\Sigma}, \qquad e_4, \qquad -e_4+e_5,
\end{equation}
at the $A_6$ singularity enhancement points. The degeneration chain is depicted in Figure \ref{fig:Dynkin_A7-1}.

\subsubsection{G$_4$-flux and chirality}

In the $U(1)_X$ restricted model, we turn on a $G_4$-flux of the following form\footnote{In general, one can also include 
an additional flux $G_4 = m^{\Sigma \Lambda} \omega_{\Sigma} \wedge \omega_\Lambda$, we have checked that \eqref{eq:G-condition} restricts this Ansatz
to a one-parameter family. It order to keep the analysis simple, we will not include this flux in the following.}
\begin{equation}
G_4 = F_X \wedge \omega_{\tilde{X}}\, ,
\label{eq:G4_res}
\end{equation}
where $F_X = n^\alpha \omega_\alpha$, and $\omega_{\tilde{X}}$ is the Poincar\'e dual two-form to a linear 
combination of the exceptional divisors $B_{\Lambda},\;(\Lambda=1,\cdots, 5)$. 
This means that we turn on a gauge flux in the direction of $\omega_{\tilde{X}}$. 
Since we preserve the $SU(5)_{GUT}$ symmetry for all $n^\alpha$, the condition \eqref{eq:G-condition}, given by 
$\Theta_{i\beta}=0$, reduces to  
\begin{equation}
   \int_{\tilde{X}_4} \omega_\alpha \wedge \omega_{\tilde{X}} \wedge \omega_{i} \wedge \omega_{\beta} = 0.
\label{eq:orthogonality2}
\end{equation}
Then, $\omega_{\tilde{X}}$ can be determined up to an overall constant. The Poincar\'e dual to $\omega_{\tilde{X}}$ is 
\begin{equation}
B_{\tilde{X}} = \alpha(-2 B_1 -4 B_3 -6 B_4 -3 B_2 -5 B_{5}),
\label{eq:U(1)_X}
\end{equation}
We later fix the overall constant $\alpha$ by requiring that the $U(1)$ direction $\omega_{\tilde{X}}$ matches the $U(1)_X$ considered in  the $(4+1)$ split spectral cover model. Furthermore, the $G_4$-flux \eqref{eq:G4_res} should satisfy the condition \eqref{eq:G-condition}. The condition can be satisfied when $F_X$ is an element of $H^{1,1}(\cB)$ because of \eqref{vanish_intersect2}. Hence the Poincar\'e dual of $F_X$ can be 
\begin{equation}
F_X = a H + b S.
\end{equation}
The first constraint of \eqref{eq:G-condition} is also satisfied 
due to the requirement \eqref{eq:orthogonality2}.

Let us determine the charges for matter fields under the $U(1)_X$ obtained along \eqref{eq:U(1)_X}. 
The Cartan charges can be computed by \eqref{U(1)charges}. We have seen in 
the previous subsection that the curve corresponding to the $\overline{{\bf 10}}$ 
matter field is $\ell_5$ and the curves corresponding to the 
two $\bar{{\bf 5}}$ matter fields are $\ell_7$ and 
$\ell_7 + \ell_8$. One can determine their curve classes from the intersection 
numbers \eqref{eq:mori4}. Note that the curve classes in the Calabi--Yau 
fourfold can be expressed as the triple intersection of the divisors. 
When $\ell_i$ has a negative intersection number with $B_i$, $B_i$ can be 
chosen as a component in the triple intersection. Hence, we  
make the Ans\"atze 
\begin{align}
 &\ell_5 = a B_2 \cdot B_3 \cdot (\mu H + \nu S)\; ,&   \quad 
  &\ell_7=(\sum_A a^A D_A) \cdot B_4 \cdot (\mu H + \nu S)\;,&\\ 
 & \ell_7+\ell_8=(\sum_A b^A D_A) \cdot B_4 \cdot (\mu H + \nu S)\;,& \quad &
 \ell_8 = (\sum_A c^A D_A) \cdot X \cdot (\mu H + \nu S)\;, & \nn 
\end{align}
where $\sum_A a^A D_A$ etc.~are general linear combinations of the divisors in \eqref{eq:toric4}. 
The parameters $a, a_A$, $b_A$ $c_A$ are determined from the intersection 
numbers \eqref{eq:mori4}. Then, the curve classes 
for $\ell_5, \ell_7, \ell_7+\ell_8$ and $\ell_8$ are
\begin{eqnarray}\label{eq:curve10_res}
\overline{{\bf 10}}&:&\quad \ell_5 \cong -e_2-e_4 + e_{\Sigma} \quad \rightarrow \quad  \tfrac{1}{3\mu - 2 \nu}B_2\cdot B_3 \cdot (\mu H + \nu S),\\
\bar{{\bf 5}}&:&\quad \ell_7 \cong -e_3 + e_{\Sigma} \quad \rightarrow \quad \tfrac{1}{(\mu -2\nu)(7\mu - 2\nu)}\Big(2\mu B_2 + (\mu+2\nu)B_3 \nonumber \\
&&\hspace{4 cm}+ (\mu + 2\nu)B_4 - (\mu -2\nu) B_5 \Big) \cdot B_4 \cdot (\mu H + \nu S),  \label{eq:curve5bar1_res}\\
\bar{{\bf 5}}&:&\quad \ell_7+ \ell_8 \cong - e_3 \quad \rightarrow \quad \tfrac{1}{4(\mu-2 \nu)(3\mu -\nu)} (2(3\mu-2\nu) B_2+ (5\mu - 2\nu) B_3 \nonumber \\
&&\hspace{4 cm}+ 2(3\mu -2\nu) B_4 + (\mu -2\nu) B_5) \cdot B_4 \cdot (\mu H + \nu S), \label{eq:curve5bar2_res}\\
{\bf 1}&:&\quad \ell_5 \cong -e_{\Sigma} \quad \rightarrow \quad \frac{1}{63\mu + 94\nu}(6B_4 + 8H + B_5 ) \cdot X \cdot (\mu H + \nu S),\label{eq:curvesinglet_res}
\end{eqnarray}
where we choose special representatives in the solutions just for simplicity. 
From the explicit forms \eqref{eq:curve10_res}--\eqref{eq:curvesinglet_res}, one can determine the charge under $U(1)_X$ for each weight
\begin{equation}
\ell_5 \rightarrow \overline{{\bf 10}}_{-\alpha}\, ,\qquad 
\ell_7 \rightarrow \bar{{\bf 5}}_{-3\alpha}\, , \qquad 
\ell_7+ \ell_8 \rightarrow \bar{{\bf 5}}_{2\alpha} \qquad 
\ell_8 \rightarrow {\bf 1}_{5\alpha}.
\end{equation}
For the comparison to the $(4+1)$ split spectral cover model, we choose the direction 
of $U(1)$ from $\omega_{\tilde{X}}$ to be the $U(1)_X$ in the $(4+1)$ split spectral cover model. 
Hence, we take $\alpha=1$ hereafter. Note that we have determined the matter representation 
curves directly from the generators of the extended relative Mori cone. This approach is 
different from the one discussed in \cite{Krause:2011xj}. In \cite{Krause:2011xj}, the 
matter representation curves are determined from the direct computation of the degeneration 
of the Tate form as done in the appendix \ref{sec:direct_comp} by exploiting the Stanley-Reisner ideal.

The other information necessary for the computation of the chirality \eqref{eq:chirality2} 
are the matter surfaces $S_{{\bf 10}}$ and $S_{\bar{{\bf 5}}}$. They can be determined from the curves 
in the extended relative Mori cone. The curves are already obtained as 
\eqref{eq:curve10_res}--\eqref{eq:curve5bar2_res}. Then, we pull out $\mu H + \nu S$ 
with the correct multiplicity. The correct multiplicity can be computed from the 
consideration of the intersection between $\mu H + \nu S$ and the matter curve 
$\Sigma_{{\bf 10}}$, $\Sigma_{\bar{{\bf 5}}}$. The condition for the $\Sigma_{{\bf 10}}$ matter 
curve is the same as \eqref{eq:pullout1}. However, the $\Sigma_{\bar{{\bf 5}}}$ curve 
splits into two components $a_{3,2}=w=0$ and $a_1a_{4,3}-a_{2,1}a_{3,2}= w=0$, where $w=0$ defines the surface $S_{\rm b}$
and the definitions 
of the $a_{i,j}$ can be found in \eqref{eq:Tate}, \eqref{eq:SU(5)}. The matter fields 
in the $\bar{{\bf 5}}_{-3}$ and $\bar{{\bf 5}}_{2}$ representation are localized along $a_{3,2}=w=0$ 
and $a_1a_{4,3}-a_{2,1}a_{3,2}= w=0$, respectively. The matter fields in 
the singlet ${\bf 1}_{5}$ are localized along the curve $a_{3,2}=a_{4,3}=0$.  Hence, the 
condition \eqref{eq:multiplicity} becomes
\begin{eqnarray}
\cB \cdot S \cdot c_{1}(\cB) \cdot  (\mu H + \nu S)&=& 3\mu - 2\nu,\\
\cB \cdot S \cdot (3c_{1}(\cB) - 2c_{1}(N_{S|\cB})) \cdot   (\mu H + \nu S) &=& 7\mu -2\nu,\\
\cB \cdot S \cdot (5c_1(\cB) -3 c_1(N_{S|\cB})) \cdot  (\mu H + \nu S) &=& 12\mu - 4\nu,\\
\cB \cdot (3c_1(\cB) -2c_1(N_{S|\cB})) \cdot (4c_1(\cB) - 3c_1(N_{S|\cB}))) \cdot  (\mu H + \nu S) &=& 63\mu + 94\nu.
\end{eqnarray}
Therefore, the matter surfaces are 
\begin{eqnarray}
S_{\overline{{\bf 10}}_{-1}} &=& B_2 \cdot B_3,\label{eq:matter10_res}\\
S_{\bar {{\bf 5}}_{-3}} &=&  \tfrac{1}{\mu -2\nu} (2\mu B_2 + (\mu+2\nu)B_3+ (\mu + 2\nu)B_4 - (\mu -2\nu) B_5 \big) \cdot B_4,\label{eq:matte5bar1_res}\\
S_{\bar {{\bf 5}}_{2}} &=& \tfrac{1}{\mu-2 \nu} (2(3\mu-2\nu) B_2+ (5\mu - 2\nu) B_3+ 2(3\mu -2\nu) B_4 + (\mu -2\nu) B_5) \cdot B_4, \label{eq:matter5bar2_res}\\
S_{{\bf 1}_{5}} &=& (6B_4 + 8H + B_5)  \cdot X. \label{eq:mattersinglet_res}
\end{eqnarray}

With the $G_4$ flux \eqref{eq:G4_res} and the matter surfaces 
\eqref{eq:matter10_res}--\eqref{eq:mattersinglet_res}, we compute the chirality 
of the matter fields in each representation by using \eqref{eq:chirality2}. The intersection between 
the matter surfaces \eqref{eq:matter10_res}-\eqref{eq:mattersinglet_res} and the $G_4$ flux \eqref{eq:G4_res} yields the numbers
\begin{eqnarray}
\chi(\overline{{\bf 10}}_{-1}) &=& \int_{S_{\overline{{\bf 10}}_{-1}}} \hspace*{-.3cm} G_4 = - 3a + 2b\,, \qquad \quad 
\chi(\bar{{\bf 5}}_{-3}) = \int_{S_{\bar{{\bf 5}}_{-3}}} \hspace*{-.3cm} G_4 = -21a+6b\, , \nn \\
\chi(\bar{{\bf 5}}_{2}) &=& \int_{S_{\bar{{\bf 5}}_{2}} } \hspace*{-.1cm} G_4 = 24a -8b\,,\qquad  \qquad  
\chi({\bf 1}_{5}) = \int_{S_{{\bf 1}_{5}}} \hspace*{-.1cm} G_4 = 315a+470b.\label{eq:chi_ex2}
\end{eqnarray}
Note that consistent with anomaly cancellation in the four-dimensional gauge theory one has $\chi({\bf 10}_{1}) = \chi(\bar{{\bf 5}}_{-3})+\chi(\bar{{\bf 5}}_{2})$.

It is important to stress that we did not address the quantization of the real scalars $a,b$
defining the $G_4$ flux \eqref{eq:G4_res}. In order to do that one has to evaluate the 
condition \eqref{quantization}. It is a trivial task to compute $c_2(\tilde X_4)$ for a torically realized 
hypersurface. The complication lies in the question if the base elements chosen in \eqref{eq:G4_res}
are actually part of a minimal integral basis of $H^{4}_{\rm V}(\tilde X_4,\mathbb{Z})$. Furthermore, \eqref{quantization} can 
imply that we have to switch on another component of $G_4$ to compensate for the half-integrality 
of $c_2(\tilde X_4)/2$.

\subsubsection{Relation to three-dimensional Chern-Simons term}

The chirality \eqref{eq:chi_ex2} can be obtained from the three-dimensional 
Chern-Simons term by using \eqref{eq:chirality3d}. Since the $SU(5)$ Cartan charges of 
the generators of the relative Mori cone for the $U(1)_X$ restricted model are the 
same as the ones in the first example, the effectiveness of the curves corresponding 
to the ${\bf 10}$ weights and the ${\bf 5}$ weights are essentially the same. However, 
they are actually ${\bf 16}$ representation or ${\bf 10}$ representation of $SO(10)$ and the 
Cartan charge for the $U(1)$ obtained from $B_5$ is affected by the singlet term $e_{\Sigma}$. 
Hence, we also have to take into account the singlet term to compute the 
formula \eqref{eq:chirality3d}. Since the ${\bf 10}$ weights of $SU(5)$ come from the 
${\bf 16}^{\prime}$ weights of $SO(10)$, the weight $e_i + e_j$ is indeed the weight $e_i+e_j -e_{\Sigma}$. 
For the ${\bf \bar{5}_{-3}}$ weights coming from the ${\bf 16}^{\prime}$, the shift for the weight $-e_i$ 
is $-e_i+e_{\Sigma}$. The ${\bf \bar{5}_{2}}$ weight coming from the ${\bf 10}$ weight of $SO(10)$ remains unchanged. 
These can be also seen explicitly by constructing the effective curves from the generators 
of the relative Mori cone. Then, all the $U(1)$ charges for $B_\Lambda,\;(\Lambda=1,\cdots, 5)$ can be obtained 
from the weights of $SO(10)$ and the formula \eqref{eq:chirality3d} becomes
\begin{align}
&\Theta_{23} = -\chi({\bf 10}), & \qquad &
\Theta_{24} = \tfrac{1}{2} \chi({\bf 10}) + \tfrac{1}{2}(\chi(\bar{{\bf 5}}_{-3})+ \chi(\bar{{\bf 5}}_{2})), &\nn \\
&\Theta_{33} = \chi(\bar{{\bf 5}}_{-3})+ \chi(\bar{{\bf 5}}_{2}),& &\Theta_{44} = -\chi({\bf 10}),   \nn \\
&\Theta_{45}=\tfrac{1}{2}\chi({\bf 10})  - \tfrac{1}{2}\chi(\bar{{\bf 5}}_{-3}) + \tfrac{1}{2}\chi(\bar{{\bf 5}}_{2}), & \label{eq:theta-ex2} \nn 
& \Theta_{55}= - \chi({\bf 10}) + \tfrac{3}{2} \chi(\bar{{\bf 5}}_{-3}) - \chi(\bar{{\bf 5}}_{2}) + \tfrac12 \chi({\bf 1}_{5}) , & \nn \\
&\Theta_{13} = \tfrac{1}{2}\chi({\bf 10}) - \tfrac{1}{2}(\chi(\bar{{\bf 5}}_{-3}) + \chi(\bar{{\bf 5}}_{2})),&
&\Theta_{34} = \tfrac{1}{2}\chi({\bf 10}) - \tfrac{1}{2}(\chi(\bar{{\bf 5}}_{-3}) + \chi(\bar{{\bf 5}}_{2})), \nn  \\
&\Theta_{11} = -\chi({\bf 10}) + \chi(\bar{{\bf 5}}_{-3}) + \chi(\bar{{\bf 5}}_{2}),&&
\Theta_{22} = \chi({\bf 10}) - (\chi(\bar{{\bf 5}}_{-3}) + \chi(\bar{{\bf 5}}_{2})),& 
\end{align}
and the other components are zero. 
From the explicit intersection numbers by using the $G_4$-flux 
\eqref{eq:G4_res}, $\Theta_{\Lambda\Sigma}$ is computed to be
\begin{eqnarray} \label{ev_theta}
\Theta_{23} &=& -3a+2b\, ,\qquad \Theta_{24}= 3a-2b\, , \qquad \Theta_{33}=3a-2b\, ,\\
\Theta_{44} &=& -3a+2b\, , \qquad \Theta_{45} = 24a-8b\, , \qquad \Theta_{55} = 99a+254b\, , \nn
\end{eqnarray}
and the other components are zero. The chirality of \eqref{eq:chi_ex2} 
is precisely reproduced by comparing \eqref{eq:theta-ex2} with the explicit expressions \eqref{ev_theta}.

\subsubsection{Comparison with split spectral cover}

Having obtained the chirality \eqref{eq:chi_ex2} from the 
formula \eqref{eq:chirality2}, we compare the results with 
the ones from the $(4+1)$ split spectral cover model.

As for the $U(1)_X \in SU(5)_{\perp}$, we consider the generator $(1,1,1,1,-4) \in SU(5)_{\perp}$. Then, we have the matter fields localized on the GUT surface $S_b$
\begin{equation}
 {\bf 10}_{1}, \;\; \bar{{\bf 5}}_{-3}, \;\; \bar{{\bf 5}}_{2},
 \label{eq:matters}
\end{equation}
where subscript denotes the charge under the $U(1)_X$. The $\bar{{\bf 5}}_{-3}$ representation 
matter fields are localized along $a_{3,2}=w=0$ and the $\bar{{\bf 5}}_{2}$ matter fields are 
localized along $a_1a_{4,3}-a_{2,1}a_{3,2}=w=0$. The chirality formulas for ${\bf 10}$ and $\bar{{\bf 5}}$ 
matter are \cite{Marsano:2009gv,Blumenhagen:2009yv}
\begin{eqnarray}
\chi_{{\bf 10}} &=& (-\lambda \tilde{\eta} + \frac{1}{4}\zeta) \cdot (\tilde \eta - 4c_{1}(S_{\rm b})),\label{eq:chirality_res10}\\
\chi_{\bar{{\bf 5}}_{-3}} &=& \lambda(-\tilde{\eta}^{2}+6\tilde{\eta}\, c_{1}(S_{\rm b})-8c_{1}^{2}(S_{\rm b})) + \frac{1}{4}\zeta(-3\tilde{\eta}+6c_{1}(S_{\rm b})),\label{eq:chirality_res5bar1}\\
\chi_{\bar{{\bf 5}}_{2}} &=& \lambda(-2\tilde{\eta}\, c_1(S_{\rm b})+8c_{1}^{2}(S_{\rm b})) + \frac{1}{4}\zeta(4\tilde{\eta}-10c_{1}(S_{\rm b})),\label{eq:chirality_res5bar2}
\end{eqnarray}
where $\tilde{\eta} = \eta-c_{1}(S_{\rm b})$ and $\eta$ is related to the first Chern class of the normal bundle \eqref{eq:normal_eta}. $\zeta$ is a flux part on $S$, namely $\zeta \in H^{2}(S_{\rm b},\mathbb{Z})$. In the current example \eqref{eq:toric4}, $\zeta$ can be chosen as 
\begin{equation}
\frac{1}{4} \zeta = (a H + b S)|_{S_{\rm b}}.
\end{equation} 
Then, the chirality formulas \eqref{eq:chirality_res10}--\eqref{eq:chirality_res5bar2} can be computed as
\beq
\chi_{{\bf 10}} = 3a -2b -38\lambda, \qquad  
\chi_{\bar{{\bf 5}}_{-3}} = -21a+6b-22\lambda,\label{eq:res5bar1} \qquad
\chi_{\bar{{\bf 5}}_{2}} = 24a-8b -16\lambda.
\eeq
For the comparison with the results of \eqref{eq:chi_ex2} note that we turn 
on the $G_4$ flux only in the direction of the $U(1)_X$ \eqref{eq:G4_res}. 
This corresponds to the case where $\lambda=0$ in the $(4+1)$ split spectral 
cover model. By putting $\lambda=0$, \eqref{eq:res5bar1} exactly 
reproduce the chirality formulas \eqref{eq:chi_ex2} 
when we identify $\frac{1}{4}\zeta$ with $F_X$.

\section{Conclusions}

In this paper we discussed the determination of the 
net chiral matter spectrum of a four-dimensional F-theory 
compactification on a singular Calabi-Yau manifold $X_4$.
We argued that the description of F-theory as a limit of 
M-theory allows to extract these data on the resolved 
fourfold $\tilde X_4$ with $G_4$ flux. The resolution is 
physical in the effective three-dimensional theory obtained 
from M-theory on $\tilde X_4$, and corresponds to moving to the 
Coulomb branch of the gauge theory. Due to the 
$G_4$ fluxes the resulting theory contains Chern-Simons 
couplings, proportional to $\Theta_{\Lambda \Sigma} A^\Lambda \wedge F^\Sigma$, 
for the $U(1)$ vector fields $A^\Lambda$. 
In contrast, such couplings are not 
induced by a classical circle reduction of a general 
four-dimensional $\cN=1$ theory which arises as the low energy 
limit of F-theory on $X_4$. However, upon reduction to three dimensions
the charged matter becomes massive in the Coulomb branch of 
the gauge theory. This precisely corresponds to the resolution 
process of $X_4$ to $\tilde X_4$. The Chern-Simons couplings are then 
induced by one-loop corrections with the massive charged matter 
running in the loop. Matching the Chern-Simons couplings of the 
fluxed M-theory reduction with the one-loop corrections in the 
F-theory reduction, we argued that the map between $G_4$ fluxes
and net chiral matter can be inferred.   

The study of one-loop corrections in the three-dimensional Chern-Simons 
theory requires the knowledge of the $U(1)$ charges, as well as 
some positivity properties of the scalars in the three-dimensional vector 
multiplets. Geometrically this corresponds to the fact that curves 
associated to the singularity resolution can have positive or formally 
negative volume. This led us to introduce the relative Mori cone 
which contains all effective curves of $\tilde X_4$ 
which shrink to points in $X_4$. 
A detailed map between these curves and the weights of different 
representations of the matter fields allowed a deeper understanding 
of the resolution process at co-dimensions two and three where matter and
Yukawa couplings are localized. With this data at hand the one-loop Chern-Simons 
couplings can be evaluated and matched with the $G_4$ flux 
result $\Theta_{\Lambda \Sigma}$. The expressions manifestly depend on the number of charged 
fermions and led to a computation of the chiral index $\chi({\bf R})$. While we have shown
this for single non-Abelian gauge groups quite generally, 
it would be interesting to find a more
group-theoretic reasoning that the index $\chi({\bf R})$ can always be extracted.

An analysis of the extended relative Mori cone using the $\tilde X_4$-intersection 
numbers resulted in a detailed map between weights and resolution curves. 
We then proposed a formalism to determine the matter surfaces $S_{\bf R}$ which 
extract the chiral index directly from viable $G_4$-fluxes via $\chi({\bf R}) = \int_{S_{\bf R}} G_4$.
We have shown that $S_{\bf R}$ can be constructed for a chosen weight of the 
representation ${\bf R}$. Only the integral $\chi({\bf R})$ is independent of the 
weight and the topological phase of the resolution. From the Chern-Simons analysis 
one realizes that $\chi({\bf R}) $ can be written as $\chi({\bf R})  = t^{\Lambda \Sigma}_{\bf R} \Theta_{\Lambda \Sigma}$, 
where $t^{\Lambda \Sigma}_{\bf R}$ is either determined from the matter surface, or from the charges 
and positivity properties of the curve classes. It would be nice to work out more  
details of the map from the resolution geometry to~$t^{\Lambda \Sigma}_{\bf R}$.

In the last part of the paper we have evaluated the 
net chirality for two specific examples with $SU(5)$ and $SU(5)\times U(1)_X$
gauge group. We have found that in the first example there is a 
single parameter encoding $G_4$ flux which preserves four-dimensional
Poincar\'e invariance and does not break the $SU(5)$ gauge symmetry.
This is consistent with a spectral cover construction. However, 
our construction does not depend on the existence of a globally 
valid spectral cover description. In the second example we only included  
the $U(1)$-flux gauging the $U(1)_X$, and determined the induced net chirality. 
The result was successfully matched with the split spectral cover construction
for states localized on the $SU(5)$ brane. The number of singlets localized 
away from the $SU(5)$-brane can equally determined using our construction. Both the 
Chern-Simons analysis as well as the explicit construction of the matter surfaces 
led to matching answers. It would be interesting to generalize the flux in this 
configuration. It can indeed be checked that there exists a one-parameter family 
of $G_4$ fluxes which do break $SU(5)$ and induce new chiral fields. Matching with 
a global split spectral cover is hard in this case, since this universal flux 
is not entirely localized on the $SU(5)$-brane, and our model has no 
heterotic dual. So while it is straightforward to evaluate the chirality 
using our formalism, there are not many results with which we can compare 
the answer. 
One other issue, which we addressed only briefly, is to determine 
the correct quantization conditions on the parameters determining the flux $G_4$.
It is straightforward to compute the second Chern class for our examples which is required 
to evaluate \eqref{quantization}. The complication lies in the determination of a 
minimal integral basis of $H^4(\tilde X_4,\mathbb{Z})$. It would be interesting to 
do that for both the $SU(5)$ and $SU(5) \times U(1)_X$ model, e.g.~by 
using the explicit hypersurface equation and resolution. One expects also an
interpretation of these quantization conditions in the three-dimensional 
Chern-Simons theory.

An interesting extension of this work would be the systematic study of more 
complicated examples with varying gauge groups and representations. This includes 
cases with multiple non-Abelian factors, to which our formalism has to be extended. 
Furthermore, already the exceptional gauge group cases might yield some interesting 
new properties which can be studied for a given $\tilde X_4$. Much of our formalism can 
be algorithmically implement in a computer search. One of the main challenges 
will remain the systematic implementation of the quantization conditions.

\vspace*{1.2cm}
\noindent
{\bf Acknowledgments}: We would like to thank Federico Bonetti, 
I\~ naki Garc\' ia-Etxebarria, Kenji Hashimoto, Denis Klevers, 
Albrecht Klemm, Seung-Joo Lee, Noppadol Mekareeya, Raffaele Savelli, 
Gary Shiu, Wati Taylor, and Timo Weigand for discussions. HH would like to thank 
the Hong Kong Institute for Advanced Study at HKUST and Max-Planck-Institut f\"ur Physik for hospitality and financial support 
during part of this work. The work  of TG~was supported by a research grant of the 
Max Planck Society.
The work of HH research was supported in part by JSPS Research Fellowships 
for Young Scientists.

\newpage

\appendix

\noindent {\bf \LARGE Appendices}

\section{Six-dimensional matter via five-dimensional loops} \label{5dCS}

In this appendix we describe how the 6d F-theory spectrum can be 
extracted from a 5d compactification 
of M-theory on a resolved Calabi-Yau threefold. Six-dimensional F-theory 
compactifications have been reviewed in detail in \cite{Taylor:2011wt}.\footnote{An incomplete list of recent works on this subject includes \cite{Kumar:2009us,Morrison:2011mb,Park:2011wv}.}
A more complete analysis of the 6d effective action and the 
analysis of the spectrum using loop corrections of 5d Chern-Simons theory
can be found in \cite{BonettiGrimm}.  
If this is done for a resolved threefold $\tilde{X}_3$ the resulting 
5d $\cN=2$ theory will be in its Coulomb branch.  
In the effective theory one will find a Chern-Simons coupling 
of the form 
\beq \label{Chern-Simons}
   S^{(5)}_{\rm CS} =\int_{\mathbb{M}^{4,1}}  \cK_{ABC} A^A  \wedge F^B \wedge F^C
\eeq
where $A^A$ are the 5d vectors which arise by expanding 
the M-theory three-form into a basis of $(1,1)$-forms of $X_3$.
The coefficients $\cK_{ABC} $ in \eqref{Chern-Simons} are precisely 
the intersection numbers of $\tilde{X}_3$. One realizes 
that upon lifting these couplings to an F-theory compactification to 
six dimensions, terms coupling with $\cK_{ijk}$, i.e.~the intersections 
of the exceptional divisors for the resolved singularities, are absent at
in the tree-level effective action. In fact, it is known that these couplings 
precisely arise from a one loop correction to the 5d gauge theory with Coulomb-branch
$U(1)$'s $A^i$. On the gauge theory side these corrections can also 
be determined in terms of the number of hypermultiplets in various representations 
which became massive when going to the Coulomb branch. In other 
words, in the 5d theory one precisely finds the link 
of the intersection numbers $\cK_{ijk}$ with the 
charged matter spectrum.

Five dimensional gauge theories with matter fields have a prepotential of the form \cite{Intriligator:1997pq}
\begin{equation}
\mathcal{F}=\tfrac{1}{2}m_{0}h_{ij}\xi^{i}\xi^{j}+\tfrac{1}{6}c_{{\rm class}}d_{ijk}\xi^{i}\xi^{j}\xi^{k}+\tfrac{1}{12}\Big( \sum_{{\bf R}}|{\bf R} \cdot \xi |^{3} - \sum_{f}\sum_{{\bf w} \in {\bf W}_{f}}|{\bf w}\cdot \xi + m_{f}|^{3} \Big),
\label{eq:prepotential}
\end{equation}
where $h_{ij}={\rm Tr}(T_{i}T_{j})$ and $d_{ijk}=\frac{1}{2}{\rm Tr}T_{i}(T_{j}T_{k}+T_{k}T_{j})$ with $T_{i}$ the Cartan generator of a gauge group $G$. ${\bf R}$ are the roots of $G$, ${\bf W}_{f}$ are the weights of $G$ in the representation of ${\bf r}_{f}$. The first two terms of \eqref{eq:prepotential} are the classical terms and the last two terms of \eqref{eq:prepotential} are the quantum contributions of massive charged vector and matter multiplets. $\xi_{i}$ is a real scalar field in a vector multiplet. Geometrically, $\xi_i$ comes from the dimensional reduction of K\"ahler forms which are Poinca\'e dual to exceptional divisors $\tilde{J} = \xi^i \omega_i$. 
It is not hard to compare the field theoretic expression 
\eqref{eq:prepotential} with the pre-potential $\cF(\xi^i) =\frac16 \cK_{ijk} \xi^{i} \xi^j \xi^k$ arising 
in the  M-theory compactification, and derive the identification of the coefficients $\cK_{ijk}$
with the number of charged matter fields in the various representations.

Let us now see more explicitly how the matter content in five 
dimensional $SU(N)$ gauge theories is determined from the 
triple intersection numbers of the exceptional divisors. 
The prepotential \eqref{eq:prepotential} 
for $SU(N)$ gauge theories can be written as
\begin{equation}
\tfrac{1}{6} \cK_{ijk} \xi^i \xi^j \xi^k = \tfrac{1}{12}\Big( (2-2n_{{\bf Adj}})\sum_{I < J}^{N}(a^{I}-a^{J})^{3} + 2c_{{\rm class}}\sum_{I=1}^{N}(a^{I})^{3} - n_{{\bf AS}} \sum_{I < J}^{N}|a^{I} + a^{J}|^{3} - n_{{\bf F}} \sum_{I=1}^{N}|a^{I}|^{3} \Big),
\label{eq:geometry_gauge}
\end{equation} 
where $a^{I}$'s are vevs of a vector multiplet $\Phi = {\rm diag}(a^{1}, \cdots, a^{N})$ and parameterize the Coulomb branch. Here we include the contribution from matter in the 
adjoint representation, the fundamental representation and the anti-symmetric representation. We suppress the contribution from the 
symmetric representation fields since their contribution is the same as the one from an anti-symmetric tensor and eight fundamentals. Note that to compare this term 
with the expression obtained by the Calabi-Yau reduction 
of M-theory one has to identify the $a^I, I = 1\ldots N$ with the $N-1$ expansion 
coefficients of the rescaled K\"ahler form $J = \xi^i \omega_i$ as 
\bea
a^{1} &=& \xi^{1},\qquad  a^{N}=- \xi^{N-1} ,\\
a^{I}&=&\xi^{I}- \xi^{I-1},\quad I =2,\ldots,N-1
\label{eq:relations1}
\eea

We can write general expressions for the matter content of the fundamental representation matter and anti-symmetric representation matter in $SU(N)$ gauge theories in terms of the triple intersection numbers by exploiting the relation \eqref{eq:geometry_gauge}, . The formulas in the case of $n_{{\rm AS}}=0$ was obtained in \cite{Intriligator:1997pq} purely from geometry. We propose the generalization of the results including anti-symmetric representation matter assuming \eqref{eq:geometry_gauge}. We also assume that there are only fields in the adjoint representation, the fundamental representation and anti-symmetric representation. We propose the generalized formulas are   
\begin{eqnarray}
D_{j}^{2}D_{j+1} + D_{j}D_{j+1}^{2} &=& 2 n_{{\bf Adj}} -2, \label{eq:formula1}\\
D_{j}^{3} &=& 8 - 8 n_{{\bf Adj}} - an_{{\bf AS}} - bn_{{\bf F}}, \label{eq:formula2}\\
D_{i}D_{j}D_{k} &=& 0 \;\; {\rm or} \;\; n_{{\bf AS}}, \label{eq:formula3}
\end{eqnarray}
where $i,j,k$ in \eqref{eq:formula3} are all different. The coefficient of $a$ and $b$ 
in \eqref{eq:formula2} can be determined in the following way. The parameters of 
the Coulomb branch, namely ${\bf w}_{{\bf f}}\cdot\xi = a^{I}$ for a fundamental representation 
weight ${\bf w}_{{\bf f}}$ and ${\bf w}_{{\bf as}}\cdot\xi=a^{I} + a^{J}$ for an anti-symmetric representation 
weight ${\bf w}_{{\bf as}}$, are positive or negative depending on the sub-wedge of the Weyl 
chamber in the Coulomb branch. For each phase, there are some weights whose signs 
change after one subtracts a simple root from them. Then, $a$ and $b$ are determined 
by counting how many simple roots $\alpha_{j}$ corresponding to $D_{j}$ have such a feature. 

For example, let us consider the relative Mori cone for the phase I 
in \eqref{Mori-generators1}.\footnote{Although the 
relative Mori cone for the phase I was determined from the Calabi--Yau 
fourfolds \eqref{eq:toric3}, one can also 
obtain the same relative Mori cone for a Calabi--Yau 
threefold with a similar 
singularity structure.} 
In this phase, $a^{1}, a^{2}$ are positive and $a^{3}, a^{4}, a^{5}$ are negative. 
Then, we have the relation
\begin{equation}
e_{2}\cdot\xi - (e_{2}-e_{3})\cdot\xi = e_{3}\cdot\xi.
\end{equation}
Hence, $b=1$ for $D_{2}^{3}$ where $D_{2}$ corresponds to a simple roots $e_{2}-e_{3}$ 
and $b=0$ for the other $D_{j}^{3}$'s. For the ${\bf 10}$ weights, the sign for the 
Coulomb branch parameters are depicted in Figure \ref{fig:10phase1}. 
The sign change happens from $a^{2}+a^{4}$ to $a^{3}+a^{4}$ and from $a^{2}+a^{5}$ to 
$a^{3}+a^{5}$. Then, we have the relations 
\begin{eqnarray}
(e_{2}+e_{4})\cdot\xi - (e_{2}-e_{3})\cdot\xi &=& (e_{3}+e_{4})\cdot\xi,\\
(e_{2}+e_{5})\cdot\xi - (e_{2}-e_{3})\cdot\xi &=& (e_{3}+e_{5})\cdot\xi. 
\end{eqnarray}
Hence, $a=2$ for $D_{2}^{3}$ and $a=0$ for the other $D_{j}^{3}$'s. Schematically, 
$a$ and $b$ for the triple intersection $D_{j}^{3}$ are written as
\begin{eqnarray}
a &=& \# (e_{j}-e_{j+1})\;\; s.t. \;\; {\bf w_{as}}(+) - (e_{j}-e_{j+1}) = {\bf w_{as}}(-),\\
b &=& \# (e_{j}-e_{j+1})\;\; s.t. \;\; {\bf w_{f}}(+) - (e_{j}-e_{j+1}) = {\bf w_{f}}(-)
\end{eqnarray}
for each phase. Note that the formulas \eqref{eq:formula1}--\eqref{eq:formula3} are 
enough to determine the three unknown parameters $n_{{\bf Adj}}$, $n_{{\bf AS}}$ and $n_{{\bf F}}$. 
We have checked that the proposed formulae \eqref{eq:formula1}--\eqref{eq:formula3} 
correctly reproduce the relation \eqref{eq:geometry_gauge} at least for $SU(5)$, $SU(6)$ and $SU(7)$ gauge theories. 

For completeness, we review how to obtain the matter spectrum in six-dimensional theories from F-theory compactifications. The matter content of six-dimensional theories are constrained from the anomaly cancellation conditions \cite{Sadov:1996zm}. The anomaly cancellation via Green-Schwarz mechanism \cite{Green:1984sg, Sagnotti:1992qw} requires that the anomaly eight-form should be factorized in a particular way. F-theory compactifications require that Calabi--Yau threefolds $X_3$ have an elliptic-fibration over the surface $\mathcal{B}_2$. Then, the matter content from F-theory compactifications on $X_3$ is \cite{Sadov:1996zm}
\begin{eqnarray}
{\rm index}({\bf Ad}_{a}) - \sum_{{\bf r}}{\rm index}({\bf r}_{a})n_{{\bf r}_{a}} &=& 6 (K_{\mathcal{B}_2} \cdot D_{a}),\label{eq:sadov1}\\
y_{{\bf Ad}_{a}} - \sum_{{\bf r}}y_{{\bf r}_{a}} n_{{\bf r}_{a}} &=& -3(D_{a} \cdot D_{a}),\label{eq:sadov2}\\
x_{{\bf Ad}_{a}} - \sum_{{\bf r}}x_{{\bf r}_{a}}n_{{\bf r}_{a}} &=& 0,\label{eq:sadov3}\\  
\sum_{{\bf r}, {\bf r}^{\prime}}{\rm index}({\bf r}_{a}){\rm index}({\bf r}^{\prime}_{b})n_{{\bf r}_{a}{\bf r}_{b}} &=& (D_{a} \cdot D_{b}),\label{eq:sadov4}
\end{eqnarray}
where $D_{a}$ is a curve where a 7-brane wraps on. The matter in the representations ${\bf Ad}_{a},\; {\rm r}_{a}$ are localized on the 7-brane wrapping on the curve $D_a$. $x_{{\bf r}_{a}}, y_{{\bf r}_{a}}$ are defined as the coefficients in the decomposition
\begin{equation}
{\rm tr}_{{\bf r}_{a}}F^{4} = x_{{\bf r}_{a}} {\rm tr}F^{4} + y_{{\bf r}_{a}}({\rm tr}F^{2})^{2},
\label{eq:decom}
\end{equation} 
Here we assume that ${\bf r}_{a}$ has two independent fourth order invariants. If ${\bf r}_{a}$ has only one fourth order invariant, then $x_{{\bf r}_{a}}=0$. Let us see the formulas \eqref{eq:sadov1}--\eqref{eq:sadov4} more explicitly for an $SU(N)$ case. The coefficients in \eqref{eq:decom} are 
\begin{eqnarray}
{\rm tr}_{{\bf Adj}}F^{4} &=& 2N {\rm tr}F^{4} + 6{\rm tr}(F^{2})^{2},\label{eq:decom1}\\
{\rm tr}_{{\bf AS}}F^{4} &=& (N-8) {\rm tr}F^{4} + 3({\rm tr}F^{2})^{2},
\label{eq:decom2}
\end{eqnarray}
for the $SU(N)$ case. 
The index of each representation is 
\begin{equation}
{\rm index}({\bf F}) = 1,\qquad {\rm index}({\bf AS}) = N-2,\qquad {\rm index}({\bf Adj})=2N. \label{eq:index}
\end{equation}
Inserting \eqref{eq:decom1}--\eqref{eq:index} into \eqref{eq:sadov1}--\eqref{eq:sadov3}, we find 
\begin{equation}
n_{{\rm Adj}} = g,\; n_{{\bf AS}} = -(K\cdot D),\; n_{{\bf F}} = -8(K_{\mathcal{B}_2} \cdot D) - N (D \cdot D),
\label{eq:anomaly}
\end{equation}
where $g$ is a genus of the curve $D$ on which the $SU(N)$ 7-branes wrap. 

The formulas \eqref{eq:sadov1}--\eqref{eq:sadov4} determine the spectrum for charged matter fields in six-dimensional theories from F-theory compactifications. They are written by the local geometric data on the base $\mathcal{B}_2$. On the other hand, the matter content of five-dimensional gauge theories are characterized by the intersection between the exceptional divisors of resolved Calabi--Yau threefolds $\tilde{X}_3$. Both spectrum should be the same due to the duality between M-theory and F-theory. For  $SU(N)$ gauge theories, the spectrum obtaining from the formulas \eqref{eq:formula1}--\eqref{eq:formula3} from five-dimensional theories should match the spectrum from the 
six-dimensional theories formulas \eqref{eq:anomaly}. Indeed, we have explicitly checked 
the exact matching of the charged matter spectrum in several examples of $SU(5)$ gauge theories.

\section{Mori cone of hypersurfaces in toric varieties}
\label{sec:mori_cone}

We review how to obtain the generators of the Mori cone systematically by 
following the algorithm presented in \cite{Berglund:1995gd, Berglund:1996uy, Braun:2000hh}. 
We consider a hypersurface in a ambient toric space. Note that this algorithm can 
be also applied to complete intersections in principle. Let the polar polyhedron of the ambient 
toric space be $\Delta^{\ast}$ which is $d$ dimension. For example, $d=5$ for 
Calabi--Yau fourfold hypersurfaces. We denote the vertices of $\Delta^{\ast}$ by $\nu^{\ast (i)}$, where 
$i$ labels the vertices. The origin of  $\Delta^{\ast}$ is chosen to be $\nu^{\ast (0)}$. We first 
derive the generators of the Mori cone for the ambient toric space $\Delta^{\ast}$. Then, 
we have to specify a particular star triangulation. This means that we subdivides 
$\Delta^{\ast}$ into a set of $d$ dimensional simplicies where every simplex 
includes the vertex $\nu^{\ast (0)}$. This triangulation procedure is not unique. 
We have in general many triangulations from one $\Delta^{\ast}$, namely we have 
different Calabi--Yau hypersurfaces from the triangulation.

Let us pick one of the star triangulations. 
The generators of the Mori cone for this particular star triangulation of the ambient 
toric space $\Delta^{\ast}$ can be obtained systematically in the following way \cite{Berglund:1995gd}:
\begin{enumerate}
\item Extend the vertices $\nu^{\ast (i)}$ to $\bar{\nu}^{\ast (i)}=(1,\nu^{\ast (i)})$.
\item Take all the pairs of the $d$ dimensional simplicies $(S_k, S_l)$ which share a $(d-1)$ dimensional simplex
\item For each pair, find a linear relation $\sum_{i} \ell_{i}^{k,l}\bar{\nu}^{\ast (i)}=0$ among the vertices of 
         $S_k \cup S_l$. Furthermore, the coefficients $\ell_{i}^{k,l}$ should be minimal integers and the 
         coefficients of the points $(S_k \cup S_l)\setminus (S_k \cap S_l)$ should be non-negative.
\item Find a set of generators $\ell_i$ which express any $\ell_i^{k,l}$ of all the phases by the linear 
         combination of the elements in the set with positive integer coefficients. The generators 
         $\ell_i$ are the generators of the Mori cone of $\Delta^{\ast}$ of a particular phase.
\end{enumerate}

So far we have determined the generators of the Mori cone for the toric ambient space $\Delta^{\ast}$. 
Let us next turn to the generators of the Mori cone of Calabi--Yau hypersurfaces in the ambient space. 
In general, the Mori cone of an ambient space is different from the Mori cone of a hypersurface 
in the ambient space. In order to see how the Mori cone changes for hypersurfaces, let us see 
how the K\"ahler cone, which is dual to the Mori cone, changes for hypersurfaces first. The different 
K\"ahler cones from the different triangulations of the ambient space are connected through the 
boundaries of the K\"ahler cone. They are related to each other by flop transitions. Let us focus 
on one of the walls of the K\"ahler cone of the ambient space. When one passes the wall of the 
K\"ahler cone, a submanifold of the ambient space in one phase blows down on one side and 
another submanifold of the ambient space in another phase emerges on the other side of the 
wall. By restricting to the hypersurface in the ambient space, the submanifold 
which is flopped may not be contained in the hypersurface. In that case, the wall 
describing the flop of the ambient space is not a part of  
the K\"ahler cone of the hypersurface. Hence, we have to connect the two K\"ahler cones of the 
ambient space. By checking all the walls of the K\"ahler cones of the ambient space and removing 
the walls which are not related to the flops in the hypersurface, one can obtain the enlarged K\"ahler 
cone of the hypersurface. The dual to the union of the K\"ahler 
cones is the intersection of the corresponding Mori cones. Therefore, we consider the intersection 
of the Mori cones which characterizes one phase for the hypersurface in the ambient space.

For the ambient toric space, this procedure can also be computed systematically 
by exploiting the Stanley-Reisner ideal of the ambient toric space \cite{Berglund:1996uy}:
\begin{enumerate}
\item Find all pairs of the triangulations which are 
          related to each other by a flop through the co-dimension 
          one boundary of the K\"ahler cones.
\item For each pair, pick up a generator of the Mori cone which blows down/up through the wall. 
\item Determine a submanifold which contains the curve from the negative intersection numbers. 
         The submanifold can be written by the intersection of the divisors of the ambient toric space.
\item Restrict a Calabi--Yau hypersurface equation on the intersection of the divisors.
\item If the hypersurface equation reduces to a monomial and the reduced defining equation is not compatible with the Stanley-Reisner ideal, then the flop does not exist in the hypersurface. If not, the flop exists in the hypersurface.
\item Try the procedures $3 \sim 5$ for all the pairs. 
\item Take the intersection of the Mori cones which are related by the flops which do not exist in the hypersurface. 
          The generators of the intersection of the Mori cones are the generators of the Mori cone of the hypersurface.
\end{enumerate}
Note that we only use the information of the hypersurface for writing down its defining equation. 
Since we can also write down the defining equation of complete intersections in the ambient toric 
variety, we can in principle apply this procedure to complete intersection Calabi--Yau manifolds.

Actually, there is a shortcut for the above procedures. Since the different phases of Calabi--Yau 
hypersurfaces have distinct intersection numbers, one can find a set of  Mori cones which 
describes one phase of the hypersurface. The intersection 
numbers of hypersurfaces can be computed  by restricting the ambient space 
divisors to the hypersurface. 
If all intersection numbers of the hypersurface are equal for some phases of the 
ambient toric space, they are actually a single phase of the hypersurface. Therefore, 
we can obtain the Mori cone for the hypersurface from the intersection of the Mori 
cones which generate the same intersection 
numbers on the hypersurface.

\section{Resolution from Tate form}
\label{sec:direct_comp}

In the section \ref{sec:degeneration}, we have determined the degeneration of the $\tilde X_4$-fibers 
from the Mori cone of the Calabi--Yau fourfold \eqref{eq:toric3}. In this appendix, we describe the 
degeneration from the direct analysis of the defining equation. 
The procedure applied here essentially follows \cite{Krause:2011xj} but for the case $a_6 \neq 0$.
The defining equation of the elliptically fibered Calabi--Yau fourfold can be written by the Tate form
\begin{equation}
P = \{ y^{2} + a_{1}xyz + a_{3}yz^{3} = x^{3} + a_{2}x^{2}z^{2} + a_{4}xz^{4} + a_{6}z^{6}\}.\label{eq:Tate}
\end{equation}
For the Calabi--Yau fourfold $X_{4}$, the divisors $D_{1,2,3}$ corresponds to $y=0,x=0, z=0$ respectively.\footnote{To be precise, $D_{1,2,3}$ are the divisors defined by the coordinates after the resolution.} If one imposes the $A_{4}$ singularity along the $w=0$, then the coefficients of \eqref{eq:Tate} have to take special forms 
\begin{equation}
a_{1} = a_{1},\; a_{2} = a_{2,1}w,\; a_{3}=a_{3,2}w^{2},\;a_{4}=a_{4,3}w^{3},\;a_{6}=a_{6,5}w^{5}.\label{eq:SU(5)}
\end{equation}
In our example, $w=0$ is the divisor $D_{4}$. Introducing the divisors $D_{9},\cdots D_{12}$ resolve the $A_{4}$ singularity. The resolution process is 
\begin{eqnarray}
&&(x,y,w) \rightarrow (\tilde{x}E_{1}, \tilde{y}E_{1}, \tilde{w}E_{1}),\\
y\rightarrow y_{1}w && (x, y_{1}, w) \rightarrow (\tilde{x}E_{4}, \tilde{y}_{1}E_{4}, \tilde{w}E_{4}),\\
x\rightarrow x_{1}w && (x_{1}, y_{1}, w)\rightarrow (\tilde{x}_{1}E_{2}, \tilde{y}_{1}E_{2}, \tilde{w}E_{2}),\\
y_{1} \rightarrow y_{2}w && (x_{1}, y_{1}, w) \rightarrow (\tilde{x}_{1}E_{3}, \tilde{y}_{2}E_{3}, \tilde{w}E_{3}),
\end{eqnarray}
where $E_{1}=0, E_{2}=0, E_{3}=0, E_{4}=0 $ denote the divisors $D_{9}, D_{11}, D_{12}, D_{10}$ respectively. Note that for each line, we omit tilde when one moves on to the next line. Hence the resolution can be summarized as 
\begin{equation}
(x, y, w) \rightarrow (\tilde{x}E_{1}E_{2}^{2}E_{3}^{2}E_{4}, \tilde{y}E_{1}E_{2}^{2}E_{3}^{3}E_{4}^{2}, \tilde{w}E_{1}E_{2}E_{3}E_{4}).
\label{eq:resolution_Tate}
\end{equation}
By the resolution \eqref{eq:resolution_Tate}, the proper transform of the defining equation \eqref{eq:Tate} of the Calabi--Yau fourfold becomes 
\begin{eqnarray}
\tilde{P} &=& \{ y^{2}E_{3}E_{4} + a_{1}xyz + a_{3,2}yz^{3}E_{0}^{2}E_{1}E_{4} = x^{3}E_{1}E_{2}^{2}E_{3} + a_{2,1}x^{2}z^{2}E_{0}E_{1}E_{2} \nonumber \\ && + a_{4,3}xz^{4}E_{0}^{3}E_{1}^{2}E_{2}E_{4} + a_{6,5}z^{6}E_{0}^{5}E_{1}^{3}E_{2}E_{4}^{2}\}.\label{eq:resolvedTate}
\end{eqnarray}
where we rewrite $w\rightarrow E_{0}$ since the $E_{0}=0$ divisor correspond to a extended node of the extended $A_{4}$ Dynkin diagram.

The curve corresponding to the negative simple roots of the $SU(5)$ can be described by the intersection 
\begin{eqnarray}
e_{1}-e_{5} &=& \hat{S} \cdot D_{\alpha} \cdot D_{\beta},\;\; -e_{1}+e_{2} = B_1 \cdot D_{\alpha} \cdot D_{\beta},\;\;-e_{2}+e_{3} = B_3 \cdot D_{\alpha} \cdot D_{\beta},\nonumber \\
-e_{3}+e_{4} &=& B_4 \cdot D_{\alpha} \cdot D_{\beta},\;\; -e_{4}+e_{5} = B_2 \cdot D_{\alpha} \cdot D_{\beta},\label{eq:simple_roots1}
\end{eqnarray}
in the hypersurface. $D_{\alpha}$ and $D_{\beta}$ also satisfy the condition
\begin{equation}
S \cdot_{\cB} D_{\alpha} \cdot_{\cB} D_{\beta} = 1.
\end{equation}
One can always write down the negative simple roots from the intersection of the divisors due to the intersection \eqref{dynkin_intersect}. The intersection \eqref{eq:simple_roots1} in the hypersurface can be promoted to the intersection in the ambient space by further intersecting with the defining equation \eqref{eq:resolvedTate}. When one restricts on the divisors $D_{i}$ to consider the curves $\mathcal{C}_{-\alpha_i}$, the defining equation \eqref{eq:resolvedTate} is further reduced. Then, \eqref{eq:simple_roots1} become
\begin{eqnarray}
\cC_{e_{1}-e_{5}} &=& E_{0} \cap y^{2}E_{3}E_{4}+a_{1}xyz - x^{3}E_{1}E_{2}^{2}E_{3},\label{eq:P1-0}\\
\cC_{-e_{1}+e_{2}} &=& E_{1} \cap y^{2}E_{3}E_{4} + a_{1}xyz,\label{eq:P1-1}\\
\cC_{-e_{2}+e_{3}} &=& E_{2} \cap y^{2}E_{3}E_{4}+a_{1}xyz+a_{3,2}yz^{3}E_{0}^{2}E_{1}E_{4},\label{eq:P1-2}\\
\cC_{-e_{3}+e_{4}} &=& E_{3} \cap (a_{1}xyz + a_{3,2}yz^{3}E_{0}^{2}E_{1}E_{4} - a_{2,1}x^{2}z^{2}E_{0}E_{1}E_{2}  \nonumber \\
&&- a_{4,3}xz^{4}E_{0}^{3}E_{1}^{2}E_{2}E_{4} - a_{6,5}z^{6}E_{0}^{5}E_{1}^{3}E_{2}E_{4}^{2}),\label{eq:P1-3}\\
\cC_{-e_{4}+e_{5}} &=& E_{4} \cap a_{1}xyz - x^{3}E_{1}E_{2}^{2}E_{3} - a_{2,1}x^{2}z^{2}E_{0}E_{1}E_{2}.\label{eq:P1-4}
\end{eqnarray}
Here we omit the intersection $D_{\alpha}\cdot D_{\beta}$ in \eqref{eq:simple_roots1} since it is common for all the cases. If a divisor is written as $E_{1}$ for example, it should be understood as $E_{1} = 0$. 

Let us focus on the chain of the degeneration $A_{4} \rightarrow D_{5} \rightarrow E_{6}$ and $A_{4} \rightarrow D_{5} \rightarrow D_{6}$ for the phase I. Other chains and other phases can be understood in a similar manner. The $D_{5}$ singularity enhancement locus is characterized by $a_{1}=0$\footnote{To be precise, $a_{1}$ can be written as $a_{1} = a_{1,0} + a_{1.1}w + \cdots$. The {\bf 10} matter curve is obtained by imposing both $a_{1,0}=0$ and $w=0$. The latter condition is the restriction to $S_b$. From these two conditions, the {\bf 10} matter curve can be characterized just by $a_{1}=0$.}. Along the locus $a_{1}=0$, two of the curves of \eqref{eq:P1-0}--\eqref{eq:P1-4} further degenerate into smaller components. The curves along $a_1=0$ are
\begin{eqnarray}
\cC_{e_{1}-e_{5}}&\rightarrow& E_{0} \cap y^{2}E_{4}-x^{3}E_{1}E_{2}^{2},\label{eq:P1-D5-0}\\
\cC_{-e_{1}+e_{2}}&\rightarrow& E_{1} \cap E_{4},\label{eq:P1-D5-1}\\
\cC_{-e_{2}+e_{3}}&\rightarrow& E_{2} \cap E_{4},\; E_{2} \cap yE_{3}+a_{3,2}z^{3}E_{0}^{2}E_{1},\label{eq:P1-D5-2}\\
\cC_{-e_{2}+e_{3}}&\rightarrow& E_{3} \cap (a_{3,2}yzE_{0}E_{4}-a_{2,1}x^{2}E_{2} -a_{4,3}xz^{2}E_{0}^{2}E_{1}E_{2}E_{4}\nonumber \\
&&-a_{6,5}z^{4}E_{0}^{4}E_{1}^{2}E_{2}E_{4}^{2}),\label{eq:P1-D5-3}\\
\cC_{-e_{4}+e_{5}}&\rightarrow& E_{4} \cap E_{1},\;E_{4} \cap E_{2},\;\\&&E_{4} \cap -xE_{2}E_{3} - a_{2,1}z^{2}E_{0}.\label{eq:P1-D5-4}
\end{eqnarray}
In order to obtain this decomposition, we use the Stanley-Reisner ideal \eqref{eq:SR1} for the phase I. Note that the curves corresponding to the weights $-e_{2}+e_{3}$ and $-e_{4}+e_{5}$ further decompose and one can identify them with the weights
\begin{eqnarray}
E_{2} \cap E_{4} &\rightarrow& \mathcal{C}_{-e_{2}-e_{4}},\\
E_{2} \cap yE_{3}+a_{3,2}z^{3}E_{0}^{2}E_{1} &\rightarrow& \mathcal{C}_{e_{3}+e_{4}},\\
E_{4} \cap -xE_{2}E_{3} - a_{2,1}z^{2}E_{0} &\rightarrow& \mathcal{C}_{e_{1}+e_{5}}.
\end{eqnarray}
This is consistent with the decomposition of \eqref{eq:decomposition_D5-1} and \eqref{eq:decomposition_D5-2}. The intersection between the curves of \eqref{eq:P1-D5-0}--\eqref{eq:P1-D5-4} form the extended $D_{5}$ Dynkin diagram as depicted in Figure \ref{fig:Dynkin_phase1}. 

The $E_{6}$ enhancement point is characterized by $a_{1}=0$ and $a_{2,1}=0$. The curves of \eqref{eq:P1-D5-3} and \eqref{eq:P1-D5-4} further degenerate as
\begin{eqnarray}
\cC_{-e_{2}+e_{3}} &\rightarrow& E_{3} \cap E_{4},\; \\
&&E_{3} \cap a_{3,2}yz - a_{4,3}xz^{2}E_{0}E_{1}E_{2} - a_{6,5}z^{4}E_{0}^{3}E_{1}^{2}E_{2}E_{4},\label{eq:P1-E6-1}\\
\cC_{e_{1}+e_{5}} &\rightarrow& E_{4}\cap E_{2},\; E_{4}\cap E_{3}.\label{eq:P1-E6-2}
\end{eqnarray}
The other curves do not change. The identification of the weights is 
\begin{eqnarray}
E_{3} \cap E_{4} &\rightarrow& \cC_{-e_{3}},\\
E_{3} \cap (a_{3,2}yz - a_{4,3}xz^{2}E_{0}E_{1}E_{2} - a_{6,5}z^{4}E_{0}^{3}E_{1}^{2}E_{2}E_{4}) &\rightarrow& \cC_{e_{4}}.\\
\end{eqnarray}
Therefore, we have six curves altogether 
\begin{equation}
\cC_{e_{1}-e_{5}},\; \cC_{-e_{1}+e_{2}},\; \cC_{-e_{2}-e_{4}},\; \cC_{e_{3}+e_{4}},\; \cC_{-e_{3}},\; \cC_{e_{4}}
\end{equation}
at the $E_6$ singularity enhancement points. They form the $E_{6}$ Dynkin diagram as depicted in Figure \ref{fig:Dynkin_phase1}. 

The $D_{6}$ enhancement point, on the other hand, is characterized by $a_{1}=0$ and $a_{3,2}=0$. Then, \eqref{eq:P1-D5-3} decomposes as 
\begin{eqnarray}
\cC_{-e_{2}+e_{3}} &\rightarrow& E_{3} \cap E_{2},\; E_{3} \cap \alpha_{1} x + \beta_{1}z^{2}E_{0}^{2}E_{1}E_{4},\\
&& E_{3} \cap \alpha_{2} x + \beta_{2}z^{2}E_{0}^{2}E_{1}E_{4},
\label{eq:D6_degeneration_Tate}
\end{eqnarray}
where $\alpha_{1,2},\beta_{1,2}$ are constants which satisfy $\alpha_{1}\alpha_{2}=-a_{2,1},\;\alpha_{1}\beta_{2}+\alpha_{2}\beta_{1}=-a_{4,3},\; \beta_{1}\beta_{2} = -a_{6,5}$. The identification of the components in \eqref{eq:D6_degeneration_Tate} with the weights is 
\begin{eqnarray}
E_{3} \cap E_{2} &\rightarrow& \cC_{e_{3}+e_{4}},\\
E_{3} \cap \alpha_{1} x + \beta_{1}z^{2}E_{0}^{2}E_{1}E_{4} &\rightarrow& \cC_{-e_{3}},\\
E_{3} \cap \alpha_{2} x + \beta_{2}z^{2}E_{0}^{2}E_{1}E_{4} &\rightarrow& \cC_{-e_{3}^{\prime}}.
\end{eqnarray}
Hence we have seven curves altogether 
\begin{equation}
\cC_{e_{1}-e_{5}},\; \cC_{e_{1}+e_{5}},\; \cC_{-e_{1}+e_{2}},\; \cC_{-e_{2}-e_{4}},\; \cC_{e_{3}+e_{4}},\; \cC_{-e_{3}},\; \cC_{-e_{3}^{\prime}}
\end{equation}
at the $D_6$ singularity enhancement points. They form the extended $D_{6}$ Dynkin diagram as depicted in Figure \ref{fig:Dynkin_phase1}. 

\begin{figure}[ht]
\begin{center}
\begin{tabular}{c}
\includegraphics[width=100mm]{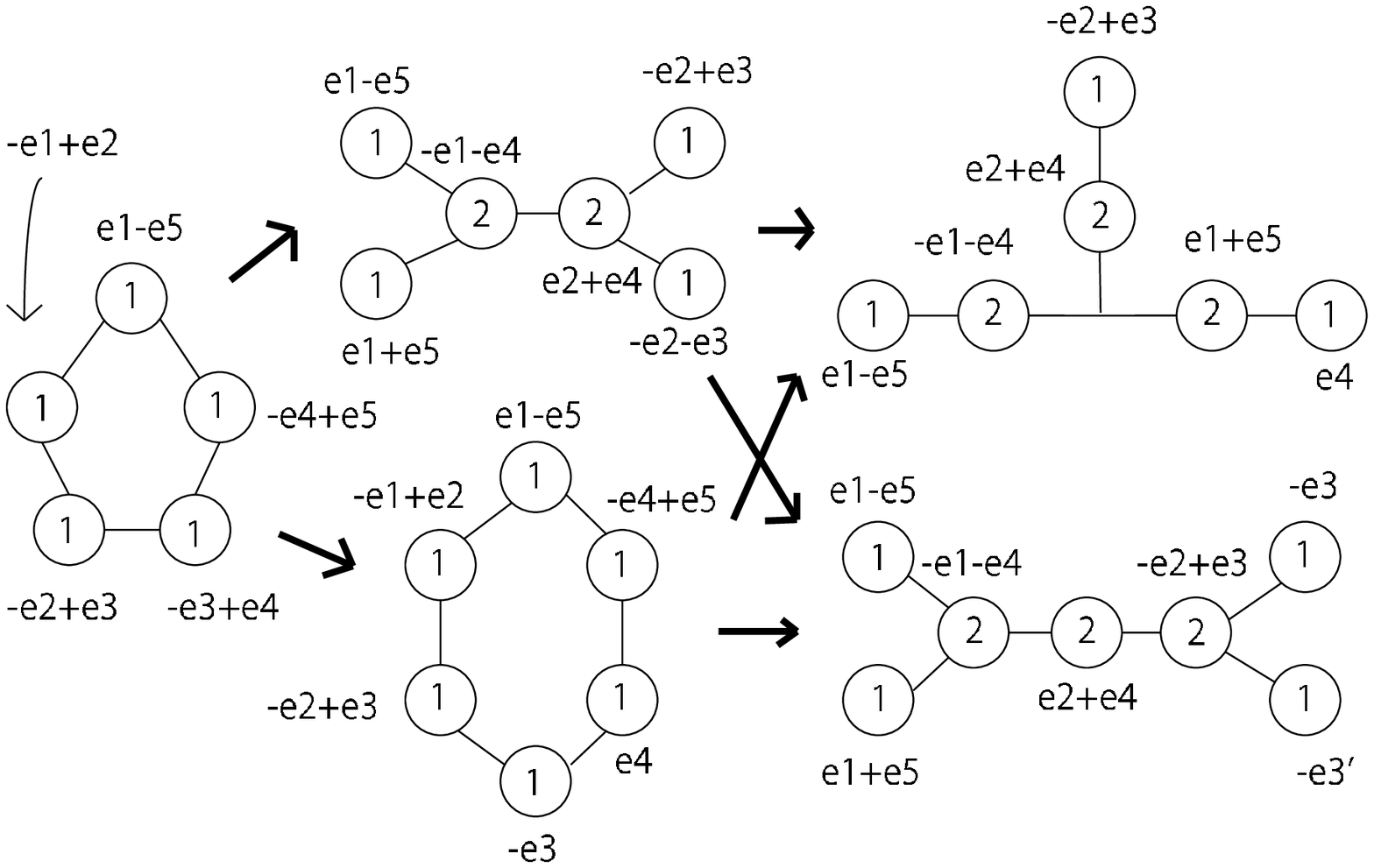} \\ 
\end{tabular}
\caption{The chain of the Dynkin diagrams for the phase II. The number in the nodes denotes the multiplicity.}
\label{fig:Dynkin_phase2}
\end{center}

\begin{center}
\begin{tabular}{c}
\includegraphics[width=100mm]{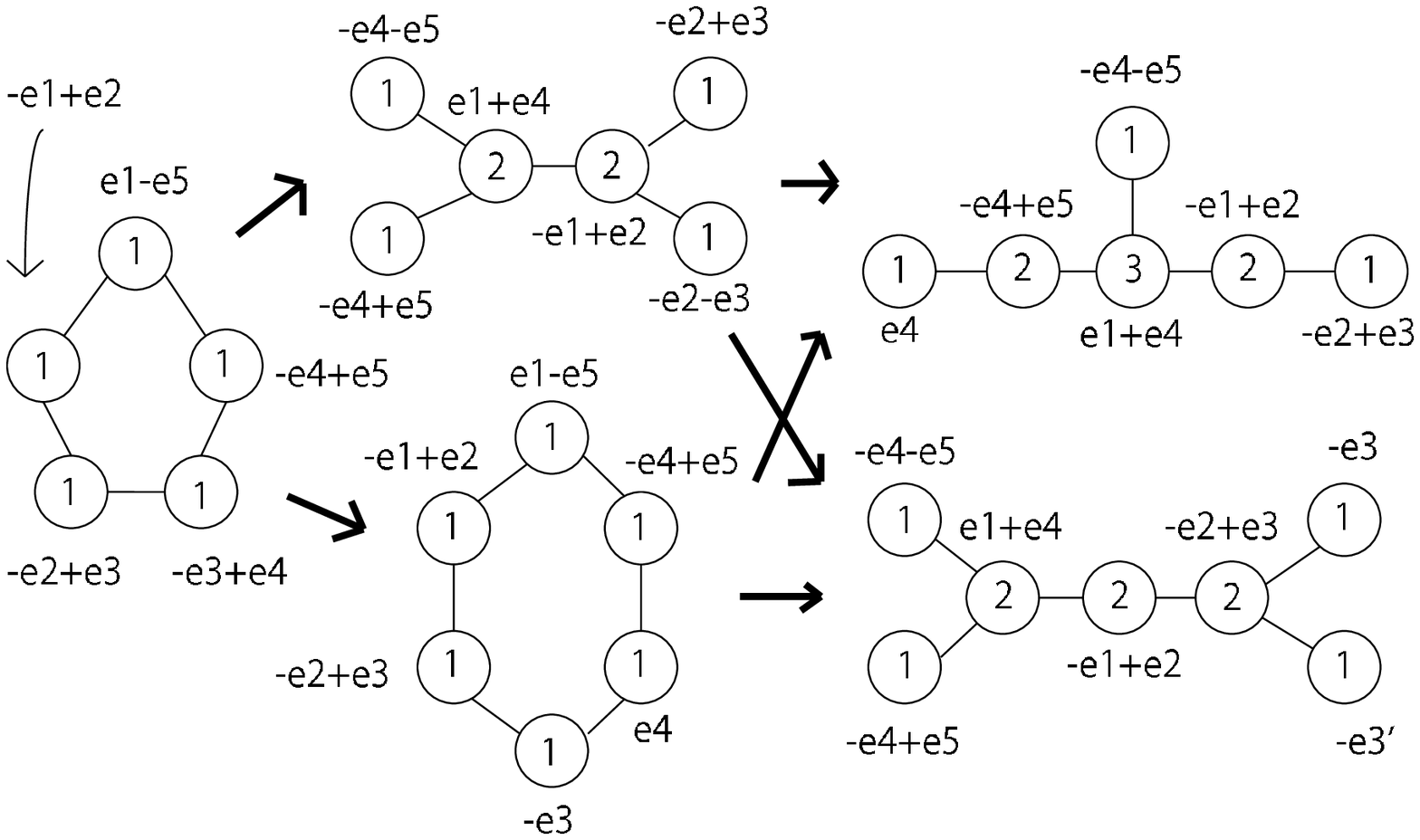} \\ 
\end{tabular}
\caption{The chain of the Dynkin diagrams for the phase III. The number in the node denotes the multiplicity.}
\label{fig:Dynkin_phase3}
\end{center}
\end{figure}

One can also do the same computation for the other phases. The chains of the Dynkin diagrams for the phase II are depicted in Figure \ref{fig:Dynkin_phase2} and the ones for the phase III are depicted in Figure \ref{fig:Dynkin_phase3}. For the phase II, the $E_{6}$ enhancement points generate a $T^{-}_{3,3,3}$ diagram, not the $E_{6}$ Dynkin diagram. The two ${\bf 10}$ weights and one $\overline{{\bf 10}}$ weights form the junction-type intersection in the $T^{-}_{3,3,3}$ in Figure \ref{fig:Dynkin_phase2}. The phase II is the only phase where three ${\bf 10}$ or ${\bf \overline{10}}$ weights appear in the generators of the Mori cone. This may be a criterion to see whether the degeneration generates the $E_{6}$ Dynkin diagram or $T^{-}_{3,3,3}$ diagram from the generators of the relative Mori cone.

\end{document}